\documentclass[epsfig,12pt]{article}
\usepackage{makeidx}
\usepackage{amsmath}
\usepackage{amsfonts}
\usepackage{amssymb}
\usepackage{graphicx}
\usepackage{accents}

\input epsf.sty
\newtheorem{theorem}{Theorem}
\newtheorem{acknowledgement}[theorem]{Acknowledgement}

\makeatletter \@addtoreset{equation}{section}
\renewcommand{\theequation}{\thesection.\arabic{equation}}
\input{tcilatex}
\begin{document}

\title{\vspace{-1.99cm}%
\rightline{\mbox{\small
{Lab/UFR-HEP0806/GNPHE/0806}}}\vspace{0.5cm}\textbf{\ Computing the Scalar
Field Couplings in }\\
\textbf{6D Supergravity}}
\author{ El Hassan Saidi\thanks{%
h-saidi@fsr.ac.ma} \\
{\small 1. Lab/UFR-Physique des Hautes Energies, Fac Sciences, Rabat,
Morocco,}\\
{\small 2. GNPHE, point focal, Lab/UFR-PHE, Fac Sciences, Rabat, Morocco.} \\
{\small 3. Coll\`{e}ge SPC, Acad\'{e}mie Hassan II des Sciences et
Techniques, Rabat, Morocco} }
\maketitle

\begin{abstract}
Using non chiral supersymmetry in \emph{6D }space time, we compute the
explicit expression of the metric the scalar manifold $SO\left( 1,1\right)
\times \frac{SO\left( 4,20\right) }{SO\left( 4\right) \times SO\left(
20\right) }$ of the ten dimensional type IIA superstring on generic K3. We
consider as well the scalar field self-couplings in the general case where
the non chiral \emph{6D} supergravity multiplet is coupled to generic $n$
vector supermultiplets with moduli space $SO\left( 1,1\right) \times \frac{%
SO\left( 4,n\right) }{SO\left( 4\right) \times SO\left( n\right) }$. We also
work out a dictionary giving a correspondence between hyperKahler geometry
and the Kahler geometry of the Coulomb branch of \emph{10D} type IIA on
Calabi-Yau threefolds. Others features are also discussed.\newline
\end{abstract}


\section{Introduction}

At Planck scale, ten dimensional type IIA superstring compactification on K3
is described by non chiral $\mathcal{N}=2$ supergravity in six dimensions 
\textrm{\cite{P}-\cite{SG}}. There, the dynamical degrees of freedom come
into two kinds of supersymmetric multiplets:\newline
(\textbf{1}) the gravity multiplet $\mathcal{G}_{6D}^{N=2}$ consisting of 
\emph{32} bosonic and \emph{32} fermionic propagating degrees of freedom.
The propagating bosonic fields are:%
\begin{equation*}
\begin{tabular}{llllllll}
$g_{\mu \nu }\left( x\right) $ & , & $\mathcal{B}_{\mu \nu }\left( x\right) $
& , & $\mathcal{A}_{\mu }^{a}\left( x\right) $ & , & $\sigma \left( x\right) 
$ & .%
\end{tabular}%
\end{equation*}%
They describe respectively the gravity field $g_{\mu \nu }$, the
antisymmetric gauge field $\mathcal{B}_{\mu \nu }$, \emph{four} Maxwell type
gauge fields $\mathcal{A}_{\mu }^{a}$, $a=1,2,3,4$, and the\ dilaton $\sigma 
$.\newline
(\textbf{2}) \emph{twenty} Maxwell type supermultiplets $\left\{ \mathcal{V}%
_{6D,N=2}^{I}\right\} _{1\leq I\leq 20}$ having $\emph{20}\times $\emph{8}
bosonic and $\emph{20}\times $\emph{8} fermionic propagating degrees of
freedom. The \emph{6D} bosonic fields content of these supermultiplets
consists of%
\begin{equation*}
\begin{tabular}{llll}
$\mathcal{A}_{\mu }^{I}\left( x\right) $ & , & $\phi ^{aI}\left( x\right) $
& ,%
\end{tabular}%
\end{equation*}%
that is \emph{twenty} gauge fields $\mathcal{A}_{\mu }^{I}$, capturing a
local $U^{20}\left( 1\right) $ gauge invariance,%
\begin{equation}
\begin{tabular}{llll}
$\mathcal{A}_{\mu }^{I}$ & $\longrightarrow $ & $\mathcal{A}_{\mu
}^{I}+\partial _{\mu }\varphi ^{I}$ & ,%
\end{tabular}
\label{sj}
\end{equation}%
with gauge parameters $\varphi ^{I}$; and \emph{eighty} real scalar $\left\{
\phi ^{aI}\right\} $ parameterizing the real eighty dimensional manifold 
\begin{equation*}
\begin{tabular}{llllllll}
$\boldsymbol{Q}_{80}$ & $=$ & $\frac{SO\left( 4,20\right) }{SO\left(
4\right) \times SO\left( 20\right) }$ & , & $\dim \boldsymbol{Q}_{80}$ & $=$
& $4\times 20$ & .%
\end{tabular}%
\end{equation*}%
The dynamical scalar fields of the non chiral \emph{6D}$\ \mathcal{N}=2$
supergravity theory that we will deal with are then $\left\{ \sigma ,\phi
^{aI}\right\} $; they transform in the following particular representations
of the $SO\left( 4\right) \times SO\left( 20\right) $ isotropy symmetry 
\textrm{\cite{BDSS,S1}}, 
\begin{equation*}
\begin{tabular}{llllllll}
$\sigma $ & $\sim $ & $\left( \underline{1},\underline{1}\right) $ & , & $%
\phi ^{aI}$ & $\sim $ & $\left( \underline{4},\underline{20}\right) $ & .%
\end{tabular}%
\end{equation*}%
Generally, instead of $\phi ^{aI}\sim \left( \underline{4},\underline{20}%
\right) $, these scalar fields may be thought of as $\phi ^{aI}\sim \left( 
\underline{4},\underline{n}\right) $ parameterizing, together with $\sigma $%
, the following typical moduli space family \textrm{\cite{S2} }involving a
generic number $n$ of Maxwell supermultiplets, 
\begin{equation*}
\begin{tabular}{llll}
$\boldsymbol{M}_{6D}^{\mathcal{N}=2}$ & $=$ & $SO\left( 1,1\right) \times 
\boldsymbol{Q}_{4n}$ & .%
\end{tabular}%
\end{equation*}%
The real dimension of $\boldsymbol{M}_{6D}^{\mathcal{N}=2}$ is equal to $%
\left( 1+4n\right) $; the case of \emph{10D} type IIA superstring on K3
corresponds obviously to $n=20$. For generic \emph{6D}$\ \mathcal{N}=2$
supergravity models, the integer $n$ can however be any positive number; $%
n\geq 1$. The leading term of the family is particularly remarkable since,
as we will show, corresponds to the well known real four dimensional
Taub-NUT geometry.

In this paper, we freeze the dilaton $\sigma $ ($d\sigma =0$) and study the
interacting dynamics of the real \emph{4n} scalars $\phi ^{aI}$ that
parameterize the \emph{4n }real dimensional scalar manifold 
\begin{equation*}
\begin{tabular}{llll}
$\boldsymbol{Q}_{4n}$ & $=$ & $\frac{SO\left( 4,n\right) }{SO\left( 4\right)
\times SO\left( n\right) }$ & ,%
\end{tabular}%
\end{equation*}%
with generic integers $n\geq 1$. We use rigid non chiral supersymmetry in $%
6D $ space time to determine the \emph{explicit} expression of the scalar
field couplings of the non linear sigma model that governs the dynamics of
the scalars fields $\phi ^{aI}$. Besides the hyperKahler geometry of the
underlying non linear sigma model, the knowledge of the scalar fields self-
couplings associated with $\boldsymbol{Q}_{4n}$ is important for the study
of the BPS and non BPS attractors in non chiral \emph{6D} supergravity 
\textrm{\cite{BDSS,S1,S2,S3}}. The interacting dynamics of the dilaton $%
\sigma $ with the scalar field $\phi ^{aI}$ is recovered as usual; it will
be implemented at the end of this work. \newline
Having introduced the basic ingredients and the main objective of this
study; the question that we have to answer is how to get the $\phi ^{aI}$
self-interactions. To that purpose, we shall proceed in three main steps as
follows:\newline
\textbf{(a)} \emph{Introduce a} \emph{complex representation to deal with }$%
\phi ^{aI}$\newline
Instead of working with the \emph{4n} real coordinates $\phi ^{aI}$, we use
rather \emph{2n} complex fields given by the doublets 
\begin{equation*}
\begin{tabular}{llllllll}
$\mathrm{f}^{iA}\left( x\right) $ & $\sim $ & $\left( 2,n\right) $ & , & $%
\overline{\mathrm{f}}_{iA}\left( x\right) $ & $\sim $ & $\left( \overline{2},%
\overline{n}\right) $ & ,%
\end{tabular}%
\end{equation*}%
and transforming in the fundamentals of the group $SU\left( 2\right) \times
U\left( n\right) $. In using this complex representation, the above family
of real \emph{4n} manifold $\boldsymbol{Q}_{4n}$ gets replaced by the
complex \emph{2n} manifold family 
\begin{equation}
\begin{tabular}{llll}
$\boldsymbol{H}_{2n}$ & $=$ & $\frac{U\left( 2,n\right) }{U\left( 2\right)
\times U\left( n\right) }$ & .%
\end{tabular}
\label{hn}
\end{equation}%
In addition to the power of complex analysis, this representation allows to
exhibit manifestly the $U^{n}\left( 1\right) $ gauge symmetry (\ref{sj}) by
performing phases change in the complex fields $\mathrm{f}^{iA}$. The $\phi
^{aI}$ are real since they describe matter in adjoint representation of the
gauge symmetry (adjoint matter for short).\newline
(\textbf{b}) \emph{Supersymmetry as a basic invariance}\newline
Besides fermions, the \emph{6D} $\mathcal{N}=2$ Maxwell multiplet $\mathcal{V%
}_{6D,N=2}$ has, in addition to the gauge field $\mathcal{A}_{\mu }$, the
four real scalars $\phi ^{a}$, which now on should be thought of as, 
\begin{equation*}
\begin{tabular}{llll}
$\phi ^{a}$ & $\equiv $ & $\left( \mathrm{f}^{i},\overline{\mathrm{f}}%
_{i}\right) $ & .%
\end{tabular}%
\end{equation*}%
To study the geometry (scalar fields self-couplings) of the scalar manifold
parameterized by these scalars, it is interesting to split the gauge
supermultiplet $\mathcal{V}_{6D,\mathcal{N}=2}$ in terms of $\mathcal{N}=1$
supermultiplets as given below%
\begin{equation*}
\begin{tabular}{llllll}
$\mathcal{V}_{6D,\mathcal{N}=2}$ & $=$ & $\mathcal{V}_{6D,\mathcal{N}=1}$ & $%
\oplus $ & $\mathcal{H}_{6D,\mathcal{N}=1}$ & ,%
\end{tabular}%
\end{equation*}%
where $\mathcal{V}_{6D,\mathcal{N}=1}$ is the \emph{6D} $\mathcal{N}=1$
Maxwell multiplet and $\mathcal{H}_{6D,N=1}$ is the hypermultiplet. Notice
in passing that the same approach is used in dealing with the Kahler
geometry of the Coulomb branch of the \emph{4D} $\mathcal{N}=2$\
supergravity theory.\newline
(\textbf{c}) \emph{HSS method to get the explicit expression of the metric}%
\newline
In the \emph{H}armonic \emph{S}uper\emph{S}pace (\emph{HSS}) method \textrm{%
\cite{HS}}-\textrm{\cite{4}}, the \emph{n} hypermultiplets $\left\{ \mathcal{%
H}_{6D,N=1}^{I}\right\} _{1\leq I\leq 20}$ are adequately described by the
superfields $\Phi ^{+A}$ and their conjugate $\tilde{\Phi}_{A}^{+}$,%
\begin{equation*}
\begin{tabular}{llll}
$\Phi ^{+A}=\Phi ^{+A}\left( x,\theta ^{+},u^{\pm }\right) $ & , & $\tilde{%
\Phi}_{A}^{+}=\tilde{\Phi}_{A}^{+}\left( x,\theta ^{+},u^{\pm }\right) $ & ,%
\end{tabular}%
\end{equation*}%
where $x$, $\theta ^{\pm }=u_{i}^{\pm }\theta ^{i}$ and $u_{i}^{\pm }$ stand
for the space time coordinates, the Grassmann variables and harmonic
variables respectively. The \emph{HSS} superfields $\Phi ^{+A}$ and $\tilde{%
\Phi}_{A}^{+}$ transform in the fundamental representations of the $U\left(
n\right) $ isotropy symmetry of (\ref{hn}), 
\begin{equation*}
\begin{tabular}{llllllll}
$\Phi ^{+A}$ & $\sim $ & \underline{$n$} & , & $\tilde{\Phi}_{A}^{+}$ & $%
\sim $ & \underline{$\overline{n}$} & .%
\end{tabular}%
\end{equation*}%
Notice in passing that the $\Phi ^{+}$ and $\tilde{\Phi}^{+}$\ description
of hypermultiplets as well as their general self-interactions are well
established in literature on harmonic superspace \textrm{\cite{5}; }see also%
\textrm{\ \cite{z}-\cite{y} }for related matters. The \emph{HSS}\ superfield
action describing hypermultiplet interactions has the typical form $\mathcal{%
S}=\int d^{6}xL\left( x\right) $ with%
\begin{equation}
\begin{tabular}{llll}
$L\left( x\right) $ & $=$ & $\int_{S^{2}}du\mathcal{L}\left( x,u\right) $ & ,
\\ 
$\mathcal{L}\left( x,u\right) $ & $=$ & $\int d^{4}\theta ^{+}\left[ \tilde{%
\Phi}^{+}D^{++}\Phi ^{+}-\mathcal{L}_{int}^{+4}\left( \tilde{\Phi}^{+},\Phi
^{+},u^{\pm }\right) \right] $ & ,%
\end{tabular}
\label{ls}
\end{equation}%
where $D^{++}$ is the usual harmonic derivative. The first term of the right
hand side of the second relation may be thought of as the Kinetic term and $%
\mathcal{L}_{int}^{+4}$ stands for the hypermultiplet self interactions. 
\newline
Here, we will use known results on \emph{HSS} method and the prepotential
derived in \textrm{\cite{S3}} to deal with the interacting dynamics
associated with the manifold $\boldsymbol{H}_{2n}$ (\ref{hn}). This dynamics
is given by the Lagrangian super-density $\mathcal{L}_{n}^{+4}$,%
\begin{equation}
\begin{tabular}{llll}
$\mathcal{L}_{n}$ & $=$ & $\int d^{4}\theta ^{+}\left( \dsum\limits_{A=1}^{n}%
\tilde{\Phi}_{A}^{+}D^{++}\Phi ^{+A}+\mathcal{L}_{n}^{+4}\right) $ & ,%
\end{tabular}
\label{ln}
\end{equation}%
with 
\begin{equation}
\begin{tabular}{llll}
$\mathcal{L}_{n}^{+4}$ & $=$ & $\frac{1}{2}\dsum\limits_{I,J=1}^{n}\lambda
_{IJ}T^{++I}T^{++J}$ & , \\ 
$T_{I}^{++}$ & $=$ & $\frac{1}{i}Tr\left( \tilde{\Phi}^{+}H_{I}\Phi
^{+}\right) $ & .%
\end{tabular}
\label{qp}
\end{equation}%
The real symmetric matrix $\lambda _{IJ}$ describes the superfield coupling
constants and the $H^{I}$'s are the Cartan generators of the $U\left(
n\right) $ isotropy group of eq(\ref{hn}). Notice that for the particular $%
n=1$ case, eq(\ref{ln}) gets reduced to 
\begin{equation}
\mathcal{L}_{1}=\int d^{4}\theta ^{+}\left[ \tilde{\Phi}^{+}D^{++}\Phi ^{+}-%
\frac{\lambda }{2}\left( \tilde{\Phi}^{+}\Phi ^{+}\right) ^{2}\right]
\label{t}
\end{equation}%
which is nothing but the \emph{HSS}\ hypermultiplet model that describe the
real 4 dimensional Taub-NUT geometry \textrm{\cite{TN}}. The successive
integration of eq(\ref{t}) first with respect to the Grassmann $\theta ^{+}$
and then with respect to the harmonic $u^{\pm }$ variables lead to%
\begin{equation}
L_{1}\left( \mathrm{f,}\overline{\mathrm{f}}\right) =\overline{g}%
_{ij}\partial _{\mu }\mathrm{f}^{i}\partial ^{\mu }\mathrm{f}%
^{j}+g^{ij}\partial _{\mu }\overline{\mathrm{f}}_{i}\partial _{\mu }%
\overline{\mathrm{f}}_{j}+2h_{i}^{j}\partial _{\mu }\mathrm{f}^{i}\partial
^{\mu }\overline{\mathrm{f}}_{j}  \label{l1}
\end{equation}%
with%
\begin{equation}
\begin{tabular}{llll}
$\overline{g}_{ij}$ & $=$ & $\frac{\lambda }{2}\frac{\left( 2+\lambda 
\mathrm{f}\overline{\mathrm{f}}\right) }{\left( 1+\lambda \mathrm{f}%
\overline{\mathrm{f}}\right) }\overline{\mathrm{f}}_{i}\overline{\mathrm{f}}%
_{j}$ & , \\ 
$g^{_{ij}}$ & $=$ & $\frac{\lambda }{2}\frac{\left( 2+\lambda \mathrm{f}%
\overline{\mathrm{f}}\right) }{\left( 1+\lambda \mathrm{f}\overline{\mathrm{f%
}}\right) }\mathrm{f}^{i}\mathrm{f}^{j}$ & , \\ 
$h_{i}^{j}$ & $=$ & $\delta _{i}^{j}\left( 1+\lambda \mathrm{f}\overline{%
\mathrm{f}}\right) -\frac{\lambda }{2}\frac{\left( 2+\lambda \mathrm{f}%
\overline{\mathrm{f}}\right) }{\left( 1+\lambda \mathrm{f}\overline{\mathrm{f%
}}\right) }\mathrm{f}^{j}\overline{\mathrm{f}}_{i}$ & ,%
\end{tabular}
\label{tn}
\end{equation}%
where $\lambda $ is a real coupling constant. For generic $n\geq 1$; the
Lagrangian density (\ref{l1}) extends as 
\begin{equation}
\mathcal{L}_{n}\left( \mathrm{f,}\overline{\mathrm{f}}\right) =\overline{g}%
_{iAjB}\partial _{\mu }\mathrm{f}^{iA}\partial ^{\mu }\mathrm{f}%
^{jB}+g^{iAjB}\partial _{\mu }\overline{\mathrm{f}}_{iA}\partial _{\mu }%
\overline{\mathrm{f}}_{jB}+2h_{iA}^{jB}\partial _{\mu }\mathrm{f}%
^{iA}\partial ^{\mu }\overline{\mathrm{f}}_{jB}.  \label{mb}
\end{equation}%
The main purpose of this paper is to first compute explicitly the metric
components 
\begin{equation}
\begin{tabular}{llll}
$\overline{g}_{iAjB}$ & $=$ & $\overline{g}_{iAjB}\left( \mathrm{f,}%
\overline{\mathrm{f}}\right) $ & , \\ 
$g^{iAjB}$ & $=$ & $g^{iAjB}\left( \mathrm{f,}\overline{\mathrm{f}}\right) $
& , \\ 
$h_{iA}^{jB}$ & $=$ & $h_{iA}^{jB}\left( \mathrm{f,}\overline{\mathrm{f}}%
\right) $ & .%
\end{tabular}
\label{mc}
\end{equation}%
We also give a dictionary drawing the correspondence between the Kahler
geometry of \emph{10D} type IIA superstring on Calabi-Yau threefolds and the
hyperKahler geometry of \emph{10D} type IIA on K3. \newline
The organization\textrm{\ }of this paper is as follows. In section $2$, we
describe some basic tools. In section $3$, we study the 4D\ hyperKahler
Taub-NUT geometry as it is the leading term of the family $SO\left(
1,1\right) \times \boldsymbol{H}_{2n}$. In section $4$, we study the
quaternionic 2- form and derive the HSS prepotential. In section $5$, we
consider the real \emph{4n} dimensional generalization of the Taub-NUT
supersymmetric model and\ in section $6$ we derive the hyperKahler metric
with $\ U^{n}\left( 1\right) $ abelian symmetry. In section $7$, we give the
conclusion and make a discussion concerning the correspondence between
Kahler and hyperKahler geometries. In sections $8$ and $9$, we give two
appendices A and B where technical computations are presented.

\section{Basic tools}

In this section, we describe the three following points: (\textbf{1}) The
mapping from the real fields $\phi _{I}^{a}$ to the complex $\mathrm{f}^{iA}$
and $\overline{\mathrm{f}}_{iA}$. (\textbf{2}) Supersymmetric
representations in \emph{6D }\textrm{\cite{HST}}\emph{\ }and reduction down
to \emph{4D}. (\textbf{3}) Harmonic superspace method (\emph{HSS}).

\subsection{From real $\protect\phi _{I}^{a}$ to the complex $\left( \mathrm{%
f}^{iA},\overline{\mathrm{f}}_{iA}\right) $}

With the objective to use \emph{HSS} method to get eq(\ref{mc}), it is
interesting to work with the complex field coordinates $\mathrm{f}^{iA}$ and 
$\overline{\mathrm{f}}_{iA}$ rather than the real fields $\phi _{I}^{a}$.
Below we show how this mapping can be obtained. \newline
\emph{First} notice that $\phi _{I}^{a}$ is in the $\left( \underline{4},%
\underline{n}\right) $ bi-fundamental representation of $SO\left( 4\right)
\times SO\left( n\right) $ group. Using the property $SO\left( 4\right) \sim
SU\left( 2\right) \times SU\left( 2\right) $ and the usual Pauli $2\times 2$
matrices $\mathcal{\sigma }_{ij}^{a}$, we can put $\phi _{I}^{a}$ in the
equivalent form $\phi _{I}^{ij}$ with $\phi _{I}^{a}=\sum_{i,j=1}^{2}%
\mathcal{\sigma }_{ij}^{a}\phi _{I}^{ij}$.\newline
\emph{Second}, using $\phi _{I}^{ij}$, the mapping from these real scalars
to the complex fields $\mathrm{f}^{iA}$ and $\overline{\mathrm{f}}_{iA}$ is
given by the following relation \textrm{\cite{S3}}, 
\begin{equation}
\begin{tabular}{llllll}
$\phi _{k}^{iI}$ & $=$ & $Tr\left( \left[ \overline{\mathrm{f}}_{k}H_{I}%
\mathrm{f}^{i}\right] \right) $ & , & $I=1,...,n$ & .%
\end{tabular}
\label{ch}
\end{equation}%
where $Tr\left( \left[ \overline{\mathrm{f}}_{k}H_{I}\mathrm{f}^{i}\right]
\right) $ stands for 
\begin{equation}
\begin{tabular}{llllll}
$\sum\limits_{A,B=1}^{n}\overline{\mathrm{f}}_{kA}\left( H_{I}\right)
_{B}^{A}\mathrm{f}^{iB}$ & $\equiv $ & $\overline{\mathrm{f}}_{k}H_{I}%
\mathrm{f}^{i}$ & , & $I=1,...,n$ & .%
\end{tabular}%
\end{equation}%
In this relation, the complex field coordinates $\mathrm{f}^{iA}$ and $%
\overline{\mathrm{f}}_{iA}$ are in the bi-fundamentals of the isotropy group 
$SU\left( 2\right) \times U\left( n\right) $ of the moduli space of the
Coulomb branch of the supergravity theory. The $n\times n$ matrices $\left\{
H^{I}\right\} $ are the Cartan generators of the $U\left( n\right) $ group
satisfying the usual properties,%
\begin{equation}
\begin{tabular}{llllllll}
$\left[ H^{I},H^{J}\right] $ & $=$ & $0$ & , & $\left[ H,H^{J}\right] $ & $=$
& $0$ & , \\ 
$\left( H^{I}\right) ^{\dagger }$ & $=$ & $H^{I}$ & , & $\left( H\right)
^{\dagger }$ & $=$ & $H$ & .%
\end{tabular}
\label{h}
\end{equation}%
The $n\times n$ hermitian matrix $H$ stands for $H=\sum_{I=1}^{n}\varphi
_{I}H^{I}$ with $\varphi _{I}\in \mathbb{R}$, where the real functions $%
\varphi _{I}$ are the abelian $U^{n}\left( 1\right) $ group parameters. The
reality condition of the adjoint matter, $\overline{\left( \phi
_{k}^{iI}\right) }=\phi _{i}^{kI}$, follows directly from%
\begin{equation}
\begin{tabular}{llll}
$\overline{\left( \mathrm{f}^{iA}\right) }=\overline{\mathrm{f}}_{iA}$ & , & 
$\left( H_{I}\right) ^{\dagger }=H_{I}$ & .%
\end{tabular}%
\end{equation}%
\emph{Third}, it is interesting to notice that the change from the real
field coordinates $\phi _{I}^{a}$ to the complex $\mathrm{f}^{iA}$ and $%
\overline{\mathrm{f}}_{iA}$ is not uniquely defined. Indeed under the change 
\begin{equation}
\begin{tabular}{llll}
$\mathrm{f}\longrightarrow \mathrm{q}=e^{iH}\mathrm{f}$ & , & $\overline{%
\mathrm{f}}\longrightarrow \overline{\mathrm{q}}=\overline{\mathrm{f}}%
e^{-iH} $ & ,%
\end{tabular}
\label{gs}
\end{equation}%
or more explicitly by exhibiting the indices, 
\begin{equation}
\begin{tabular}{ll}
$\mathrm{f}^{kA}\longrightarrow \mathrm{q}^{kA}=\left( e^{iH}\right) _{B}^{A}%
\mathrm{f}^{kB}$ & , \\ 
$\overline{\mathrm{f}}_{kA}\longrightarrow \overline{\mathrm{q}}_{kA}=%
\overline{\mathrm{f}}_{kB}\left( e^{-iH}\right) _{A}^{B}$ & ,%
\end{tabular}%
\end{equation}%
where $H$ is as in (\ref{h}), the mapping (\ref{ch}) remains invariant%
\begin{equation}
\begin{tabular}{llll}
$Tr\left( \overline{\mathrm{q}}_{k}H_{I}\mathrm{q}^{i}\right) $ & $=$ & $%
Tr\left( \overline{\mathrm{f}}_{k}H_{I}\mathrm{f}^{i}\right) $ & .%
\end{tabular}%
\end{equation}%
Therefore the field change (\ref{ch}) has a $U^{n}\left( 1\right) $ gauge
symmetry which can be promoted to the local gauge symmetry (\ref{sj}) of the
Coulomb branch of the non chiral \emph{6D} $\mathcal{N}=2$ supergravity
theory. Since, we are not interested here by the gauge-hypermultiplet
interactions, we then restrict our attention below to global invariance.

\subsection{Supersymmetry}

The moduli space of non chiral $6D$ $\mathcal{N}=2$ supergravity multiplet
coupled $n$ Maxwell gauge supermultiplets has the form 
\begin{equation}
\begin{tabular}{llll}
$\frac{SO\left( 4,n\right) }{SO\left( 4\right) \times SO\left( n\right) }$ & 
$\times $ & $SO\left( 1,1\right) $ & ,%
\end{tabular}%
\end{equation}%
where the factor $SO\left( 1,1\right) $ is parameterized by $e^{\sigma }$
with $\sigma $\ standing for the dilaton. The real \emph{4n} moduli $\phi
^{aI}$ describe the \emph{vevs} of the scalars of the $6D$ $\mathcal{N}=2$
vector multiplets $\mathcal{V}_{6D,N=2}^{I}$.

$6D$ $\mathcal{N}=1$ \emph{formalism} \newline
In the language of $6D$ $\mathcal{N}=1$ supersymmetric representations, the
Maxwell supermultiplet $\mathcal{V}_{6D,\mathcal{N}=2}=\left( 1,\frac{1}{2}%
^{2},0^{4}\right) _{{\small 6D}}$, split into a vector $\mathcal{V}_{6D,%
\mathcal{N}=1}$ and a hypermultiplet $\mathcal{H}_{6D,\mathcal{N}=1}$. We
have%
\begin{equation}
\begin{tabular}{llllllll}
$\mathcal{V}_{6D,\mathcal{N}=2}^{I}$ & $=$ & $\mathcal{V}_{6D,\mathcal{N}%
=1}^{I}$ & $\oplus $ & $\mathcal{H}_{6D,\mathcal{N}=1}^{I}$ & , & $I=1,...,n$
& ,%
\end{tabular}
\label{213}
\end{equation}%
with the following fields content%
\begin{equation}
\begin{tabular}{llll}
$\mathcal{V}_{6D,\mathcal{N}=2}$ & $=$ & $\left( 1,\frac{1}{2}\right) _{%
{\small 6D}}$ & , \\ 
$\mathcal{H}_{6D,\mathcal{N}=1}$ & $=$ & $\left( \frac{1}{2},0^{4}\right) _{%
{\small 6D}}$ & .%
\end{tabular}
\label{hy}
\end{equation}%
As we see, the vector supermultiplets $\mathcal{V}_{6D,\mathcal{N}=2}^{I}$
have no scalars. The real \emph{4n} scalar fields are all of them in the
hypermultiplets $\mathcal{H}_{6D,N=1}^{I},$ $I=1,...,n$.

\emph{4D} $\mathcal{N}=2$\ \emph{formalism }\newline
A more convenient way to deal with $6D$ $\mathcal{N}=1$ hypermultiplets is
to use $4D$ $\mathcal{N}=2$ superspace. In this $4D$ $\mathcal{N}=2$
language, the hypermultiplet fields content decomposed as follows, 
\begin{equation}
\begin{tabular}{llll}
$\mathcal{H}_{6D,\mathcal{N}=1}=\left( \frac{1}{2},0^{4}\right) _{{\small 6D}%
}$ & $\rightarrow $ & $\mathcal{H}_{4D,\mathcal{N}=2}=\left( \frac{1}{2}%
^{2},0^{4}\right) _{{\small 4D}}$ & .%
\end{tabular}
\label{215}
\end{equation}%
A similar relation is valid for $\mathcal{V}_{6D,\mathcal{N}=1}$ which
decomposes like 
\begin{equation}
\begin{tabular}{llll}
$\mathcal{V}_{6D,\mathcal{N}=1}=\left( 1,\frac{1}{2}\right) _{{\small 6D}}$
& $\rightarrow $ & $\mathcal{V}_{4D,\mathcal{N}=2}=\left( 1,\frac{1}{2}%
^{2},0^{2}\right) _{{\small 4D}}$ & .%
\end{tabular}%
\end{equation}%
This reduction is obtained by decomposing $6D$ vectors as $4D$ vectors plus
2 scalars. The $6D$ spinors; say $\theta ^{\widehat{\alpha }i}$, split
equality into a $4D$ Weyl spinor $\theta ^{ai}$ and its complex conjugate $%
\overline{\theta }_{\dot{a}i}$ like,%
\begin{equation}
\begin{tabular}{llll}
$\left( \theta ^{\widehat{\alpha }i}\right) _{1\leq \widehat{\alpha }\leq 4}$
& $\longrightarrow $ & $\left( 
\begin{array}{c}
\theta ^{ai} \\ 
\overline{\theta }_{\dot{a}i}%
\end{array}%
\right) _{a=1,2}$ & .%
\end{tabular}%
\end{equation}%
Hypermultiplets couplings in $6D$ $\mathcal{N}=1$\ supersymmetric gauge
theory can be then conveniently studied in the framework of the \emph{4D} $%
\mathcal{N}=2$ \emph{HSS} formalism \textrm{\cite{HS}} where several results
have been obtained. Below, we give a brief description of the \emph{4D} $%
\mathcal{N}=2$ \emph{HSS} and make comments regarding our purposes.

\subsection{General on HSS in \emph{4D}}

In the \emph{HSS} formulation of $4D$ $\mathcal{N}=2$ hypermultiplet theory,
the ordinary superspace with $SU_{R}\left( 2\right) $ R-symmetry, 
\begin{equation}
\begin{tabular}{llllllll}
$z^{M}$ & $=$ & $\left( x^{\mu },\theta _{a}^{i},\overline{\theta }_{\dot{a}%
}^{i}\right) $ & , & $i=1,2$ & , & $a,\text{ }\dot{a}=1,2$ & ,%
\end{tabular}%
\end{equation}%
gets mapped into the harmonic superspace $z^{M}=\left( Y^{m},\theta _{a}^{-},%
\overline{\theta }_{\dot{a}}^{-},u_{i}^{\pm }\right) $, with an analytic
sub-superspace parameterized by the super-coordinates%
\begin{equation}
Y^{m}=\left( y^{\mu },\theta _{a}^{+},\overline{\theta }_{\dot{a}%
}^{+}\right) ,
\end{equation}%
and 
\begin{equation}
\begin{tabular}{llll}
$y^{\mu }$ & $=$ & $x^{\mu }+i\left( \theta ^{+}\sigma ^{\mu }\overline{%
\theta }^{-}+\theta ^{-}\sigma ^{\mu }\overline{\theta }^{+}\right) $ & , \\ 
$\theta _{a}^{+}$ & $=$ & $u_{i}^{+}\theta _{a}^{i}$ , $a=1,2$ & , \\ 
$\overline{\theta }_{\dot{a}}^{+}$ & $=$ & $u_{i}^{+}\overline{\theta }_{%
\dot{a}}^{i}$ , $\dot{a}=1,2$ & ,%
\end{tabular}%
\end{equation}%
where $u_{i}^{\pm }$ are the harmonic variables satisfying the relations $%
u^{+i}u_{i}^{-}=1$ and $u^{\pm i}u_{i}^{\pm }=0$.\newline
The hypermultiplets are described by an analytic \emph{HSS} function $\Phi
^{+}=\Phi ^{+}\left( Y,u\right) $,%
\begin{equation}
\begin{tabular}{llll}
$D_{a}^{+}\Phi ^{+}=0$ & , & $\overline{D}_{\dot{a}}^{+}\Phi ^{+}=0$ & ,%
\end{tabular}%
\end{equation}%
with covariant spinor derivatives as $D_{a}^{\pm }=u_{i}^{\pm }D_{a}^{i}$, $%
D_{a}^{+}=\frac{\partial }{\partial \theta ^{-a}}$ and $\overline{D}_{\dot{a}%
}^{+}=\frac{\partial }{\partial \overline{\theta }^{-\dot{a}}}$. The
superfield $\Phi ^{+}$ satisfy as well the property%
\begin{equation}
\begin{tabular}{llll}
$\left[ D^{0},\Phi ^{+}\right] $ & $=$ & $\Phi ^{+}$ & .%
\end{tabular}%
\end{equation}%
In this relation $D^{0}$ is a $U\left( 1\right) $ charge operator given by%
\begin{equation}
\begin{tabular}{llll}
$D^{0}$ & $=$ & $\partial ^{0}+\left( \theta ^{+}\frac{\partial }{\partial
\theta ^{+}}+\overline{\theta }^{+}\frac{\partial }{\partial \overline{%
\theta }^{+}}\right) -\left( \theta ^{-}\frac{\partial }{\partial \theta ^{-}%
}+\overline{\theta }^{-}\frac{\partial }{\partial \overline{\theta }^{-}}%
\right) $ & .%
\end{tabular}%
\end{equation}%
It generates, together with the two following operators 
\begin{equation}
\begin{tabular}{llll}
$D^{++}$ & $=$ & $u^{+i}\frac{\partial }{\partial u^{-i}}-2i\theta
^{+}\sigma ^{\mu }\overline{\theta }^{+}\partial _{\mu }-2i\theta ^{+2}\frac{%
\partial }{\partial \overline{\tau }}-2i\overline{\theta }^{+2}\frac{%
\partial }{\partial \tau }$ & , \\ 
&  &  &  \\ 
$D^{--}$ & $=$ & $u^{-i}\frac{\partial }{\partial u^{+i}}-2i\theta
^{-}\sigma ^{\mu }\overline{\theta }^{-}\partial _{\mu }-2i\theta ^{-2}\frac{%
\partial }{\partial \overline{\tau }}-2i\overline{\theta }^{-2}\frac{%
\partial }{\partial \tau }$ & , \\ 
&  &  & 
\end{tabular}%
\end{equation}%
where we have set $\tau =x^{4}+ix^{5}$, the SU$_{R}\left( 2\right) $
symmetry. In particular, we have the usual commutation relations of the $%
su\left( 2\right) $ algebra, 
\begin{equation}
\begin{tabular}{llll}
$\left[ D^{0},D^{++}\right] $ & $=$ & $+2D^{++}$ & , \\ 
$\left[ D^{0},D^{--}\right] $ & $=$ & $-2D^{--}$ & , \\ 
$\left[ D^{++},D^{--}\right] $ & $=$ & $D^{0}$ & ,%
\end{tabular}%
\end{equation}%
These operators play a crucial role in the \emph{HSS} formulation of $4D$ $%
\mathcal{N}=2$ supersymmetric field theory and obey the \emph{twild}\textrm{%
\footnote{%
The twild $\left( \sim \right) $ is an automorphism combining the usual
complex conjugation $\left( -\right) $ and the conjugation $\left( \ast
\right) $ of the charge of the $U\left( 1\right) $ Cartan sub-symmetry of $%
SU_{R}\left( 2\right) $.}} reality property 
\begin{equation}
\begin{tabular}{llllll}
$\tilde{D}^{++}=D^{++}$ & , & $\tilde{D}^{0}=D^{0}$ & , & $\tilde{D}%
^{--}=D^{--}$ & .%
\end{tabular}%
\end{equation}%
The $\theta ^{+}$- expansion of the \emph{HSS}\ hypermultiplet superfield $%
\Phi ^{+}$ reads as 
\begin{equation}
\begin{tabular}{llll}
$\Phi ^{+}\left( Y,u\right) $ & $=$ & $q^{+}+\theta ^{+2}F^{-}+\overline{%
\theta }^{+2}G^{-}+i\theta ^{+a}\overline{\theta }^{+\dot{a}}B_{a\dot{a}%
}^{-}+\theta ^{+2}\overline{\theta }^{+2}\Delta ^{---}$ & ,%
\end{tabular}
\label{sp}
\end{equation}%
where we have ignored fermions for simplicity. Notice that the components, 
\begin{equation}
\begin{tabular}{llll}
$F^{-}$ & $=$ & $F^{-}\left( x,u\right) $ & , \\ 
$G^{-}$ & $=$ & $G^{-}\left( x,u\right) $ & , \\ 
$B_{a\dot{a}}^{-}$ & $=$ & $B_{a\dot{a}}^{-}\left( x,u\right) $ & ,%
\end{tabular}
\label{ax}
\end{equation}%
are auxiliary fields scaling as a mass squared; i.e $\left( \emph{mass}%
\right) ^{2}$. The extra remaining one, 
\begin{equation}
\Delta ^{---}=\Delta ^{---}\left( x,u\right) ,  \label{ay}
\end{equation}%
is also an auxiliary field; but scaling as $\left( \emph{mass}\right) ^{3}$.
All these auxiliary fields are needed to have off shell supersymmetry; in
particular for the computation of eqs(\ref{mc}). We also have 
\begin{equation}
\begin{tabular}{llll}
$\tilde{\Phi}^{+}\left( Y,u\right) $ & $=$ & $\tilde{q}^{+}+\theta ^{+2}%
\tilde{G}^{-}+\overline{\theta }^{+2}\tilde{F}^{-}+i\theta ^{+a}\overline{%
\theta }^{+\dot{a}}\tilde{B}_{a\dot{a}}^{-}+\theta ^{+2}\overline{\theta }%
^{+2}\tilde{\Delta}^{---}$ & ,%
\end{tabular}
\label{ps}
\end{equation}%
where $\left( \sim \right) =\left( \overline{\ast }\right) $ stands for the
twild conjugation preserving\ the harmonic analiticity \cite{HS}.\ Moreover,
the component fields $\mathcal{F}^{q}=\mathcal{F}^{q}\left( x,u\right) $,
with Cartan charge $q$, can be also expanded in a harmonic series as follows:%
\begin{equation}
\begin{tabular}{llll}
$\mathcal{F}^{q}\left( y,u\right) $ & $=$ & $\sum_{n=0}^{\infty
}u_{(i_{1}...i_{n+q}}^{+n+q}u_{j_{1}}^{-n}..._{j_{n})}\mathcal{F}^{\left(
i_{1}...i_{n+q}j_{1}...j_{n}\right) }\left( y\right) $ & ,%
\end{tabular}%
\end{equation}%
where we have taken $q\geq 0$ and set for convenience 
\begin{equation}
\begin{tabular}{llll}
$u_{(i_{1}...i_{n+q}}^{+\left( n+q\right) }u_{j_{1}}^{-n}..._{j_{n})}$ & $%
\equiv $ & $%
u_{(i_{1}}^{+}u_{i_{2}}^{+}...u_{i_{n+q}}^{+}u_{j_{1}}^{-}...u_{j_{n})}^{-}$
& .%
\end{tabular}%
\end{equation}

\emph{HSS hypermultiplet action}\newline
Following \textrm{\cite{HS}}, the \emph{HSS} action $\mathcal{S}$ describing
the dynamics of interacting hypermultiplets $\Phi ^{+A}$ and their conjugate 
$\tilde{\Phi}_{A}^{+}$ has the form 
\begin{equation}
\mathcal{S}_{n}=\int d^{4}x\left( \int_{S^{2}}du\left[ \int d^{4}\theta ^{+}%
\mathcal{L}_{n}^{4+}\left( \Phi ^{+},\tilde{\Phi}^{+},u^{\pm }\right) \right]
\right) ,  \label{sl}
\end{equation}%
where $d^{4}\theta ^{+}=d^{2}\theta ^{+}d^{2}\overline{\theta }^{+}$ should
be understood in the usual way; that is as the derivatives $d^{4}\theta
^{+}\sim \left( D^{-a}D_{a}^{-}\right) \left( \overline{D}_{\dot{a}}^{-}%
\overline{D}^{-a}\right) $. This integral measure captures \emph{four}
negative charges. As such, the $SU\left( 2\right) $ invariance of $\mathcal{S%
}$ requires the Lagrangian super-density to carry \emph{four} positive
Cartan charges and reads as%
\begin{equation}
\mathcal{L}^{4+}=\tilde{\Phi}_{A}^{+}D^{++}\Phi ^{+A}+\mathcal{L}_{\text{%
{\small int}}}^{4+},
\end{equation}%
with hypermultiplet self interactions $\mathcal{L}_{\text{{\small int}}%
}^{4+} $ given by%
\begin{equation}
\mathcal{L}_{\text{{\small int}}}^{4+}=-\frac{\lambda }{2}\left[ Tr\left( 
\tilde{\Phi}^{+}H^{I}\Phi ^{+}\right) \right] g_{IJ}\left[ Tr\left( \tilde{%
\Phi}^{+}H^{J}\Phi ^{+}\right) \right] \text{ ,}
\end{equation}%
where $H^{I}$ as in eqs(\ref{h}). The coupling constant matrix $\lambda
_{IJ} $ has been factorized as $\lambda g_{IJ}$. The scale $\lambda $ can be
interpreted in terms of the black hole horizon radius. For $n=20$, the
matrix $g_{IJ}$ can be interpreted as the intersection matrix of the 2-
cycles of the real second homology of K3.

\section{U$\left( 1\right) $ supersymmetric model}

In this section, we study the scalar field self- couplings for the simplest
case of \emph{6D} $\mathcal{N}=2$ supergravity with one (\emph{n=1}) Maxwell
supermultiplet $\mathcal{V}_{6D,N=2}$. This study has been first considered
in \textrm{\cite{TN}}; but here it will be used as a first step towards the
derivation the U$^{n}\left( 1\right) $ extension of the Taub-NUT geometry.
We also take this opportunity to give a geometric interpretation of the
harmonic superspace\ prepotential%
\begin{equation}
\mathcal{L}_{\text{{\small taub-NUT}}}^{4+}=-\frac{\lambda }{2}\left( \tilde{%
\Phi}^{+}\Phi ^{+}\right) ^{2},
\end{equation}%
in the framework of \emph{10D} type IIA superstring compactification on
complex surfaces.\newline
As noted earlier, the vector supermultiplet $\mathcal{V}_{6D,N=2}$ has,
besides fermions, the following bosonic fields:\newline
(\textbf{1}) A gauge field $\mathcal{A}_{\mu }$ with the abelian gauge
symmetry (\ref{sj}),\newline
(\textbf{2}) \emph{Four} real scalars $\phi ^{a}$ parameterizing the real
scalar manifold $Q_{4}=\frac{SO\left( 4,1\right) }{SO\left( 4\right) }$. 
\newline
In the complex coordinates $\left( \mathrm{f}^{i}\mathrm{,}\overline{\mathrm{%
f}}_{i}\right) $, the manifold $Q_{4}$ gets mapped to the complex surface 
\begin{equation}
\boldsymbol{H}_{2}=\frac{SU\left( 2,1\right) }{SU\left( 2\right) \times
U\left( 1\right) }.  \label{2}
\end{equation}%
To deal with the underlying geometry of $\boldsymbol{H}_{2}$, it is useful
to freeze the dynamics of the gauge field $\mathcal{A}_{\mu }$ and use the 
\emph{6D} $\mathcal{N}=1$ supersymmetric formalism. There, the real four
scalars are all of them in the hypermultiplet $\mathcal{H}_{6D,\mathcal{N}%
=1} $ (\ref{hy}) which in turn is conveniently described in the $4D$ $%
\mathcal{N}=2$ harmonic superspace formalism where several results have been
obtained.\newline
In \emph{HSS}, the hypermultiplet is represented by the superfield $\Phi
^{+} $ (\ref{sp}-\ref{ps}); and its self-coupling is described by the action%
\begin{equation}
\begin{tabular}{llll}
$\mathcal{S}_{1}$ & $=$ & $\int d^{4}x\left( \int_{S^{2}}du\left[ \int
d^{4}\theta ^{+}\mathcal{L}_{1}^{4+}\left( \Phi ^{+},\tilde{\Phi}^{+}\right) %
\right] \right) $ & .%
\end{tabular}
\label{ac}
\end{equation}%
The Lagrangian super-density $\mathcal{L}_{1}^{4+}$ is given by the
supersymmetric Taub-NUT model 
\begin{equation}
\begin{tabular}{llll}
$\mathcal{L}_{1}^{4+}$ & $=$ & $\tilde{\Phi}^{+}D^{++}\Phi ^{+}-\frac{%
\lambda }{2}\left( \tilde{\Phi}^{+}\Phi ^{+}\right) ^{2}$ & ,%
\end{tabular}
\label{la}
\end{equation}%
where $\lambda $ is a coupling constant to be interpreted later in terms of
the mass $M$ of the Taub-NUT black hole ($\lambda \sim M^{-2}$). The \emph{%
HSS}\ Lagrangian density (\ref{la}) is invariant under the abelian global $%
U\left( 1\right) $ symmetry (\ref{gs})%
\begin{equation}
\begin{tabular}{llll}
$\Phi ^{+\prime }=e^{i\Lambda }\Phi ^{+}$ & , & $\tilde{\Phi}^{+\prime
}=e^{-i\Lambda }\tilde{\Phi}^{+}$ & ,%
\end{tabular}
\label{u1}
\end{equation}%
with super- parameter $\Lambda $ constrained as $D^{++}\Lambda =0$. As noted
before, this symmetry can be promoted to a local gauge invariance 
\begin{equation}
D^{++}\Lambda \neq 0
\end{equation}
by coupling the hypermultiplet $\Phi ^{+}$ to a \emph{4D} $\mathcal{N}=2$
Maxwell gauge superfield $V^{++}$ with the abelian gauge symmetry 
\begin{equation}
V^{++\prime }=V^{++}-D^{++}\Lambda \text{ }.
\end{equation}%
Below, we shall not develop this issue; and focus just on the U$\left(
1\right) $ global gauge invariance of $\mathcal{L}_{1}^{4+}\left( \Phi ^{+},%
\tilde{\Phi}^{+}\right) $.\newline
To get the explicit component field expression of the action, we have to
integrate eq(\ref{ac}) with respect to the Grassmann variables $\theta ^{+}$%
, then eliminate the auxiliary fields through their eqs of motion and
finally integrate with respect to the harmonic variables. These technical
steps are a little bit cumbersome; they are collected in appendix A. \newline
Using the results obtained in appendix A, we can put the superfield action (%
\ref{ac}) into the following component field one,%
\begin{equation}
\mathcal{S}_{1}=\frac{-1}{2}\int d^{4}x\left( \bar{g}_{ij}\partial _{\mu }%
\mathrm{f}^{i}\partial ^{\mu }\mathrm{f}^{j}+g^{_{ij}}\partial _{\mu }%
\overline{\mathrm{f}}_{i}\partial ^{\mu }\overline{\mathrm{f}}%
_{j}+2h_{i}^{j}\partial _{\mu }\mathrm{f}^{i}\partial ^{\mu }\overline{%
\mathrm{f}}_{j}\right) ,  \label{lag}
\end{equation}%
with,%
\begin{equation}
\begin{tabular}{llll}
$\bar{g}_{ij}$ & $=$ & $\frac{\lambda }{2}\frac{\left( 2+\lambda \mathrm{f}%
\overline{\mathrm{f}}\right) }{\left( 1+\lambda \mathrm{f}\overline{\mathrm{f%
}}\right) }\overline{\mathrm{f}}_{i}\overline{\mathrm{f}}_{j}$ & , \\ 
$g^{_{ij}}$ & $=$ & $\frac{\lambda }{2}\frac{\left( 2+\lambda \mathrm{f}%
\overline{\mathrm{f}}\right) }{\left( 1+\lambda \mathrm{f}\overline{\mathrm{f%
}}\right) }\mathrm{f}^{i}\mathrm{f}^{j}$ & , \\ 
$h_{i}^{j}$ & $=$ & $\delta _{i}^{j}\left( 1+\lambda \mathrm{f}\overline{%
\mathrm{f}}\right) -\frac{\lambda }{2}\frac{\left( 2+\lambda \mathrm{f}%
\overline{\mathrm{f}}\right) }{\left( 1+\lambda \mathrm{f}\overline{\mathrm{f%
}}\right) }\mathrm{f}^{j}\overline{\mathrm{f}}_{i}$ & .%
\end{tabular}
\label{tnm}
\end{equation}%
Before proceeding ahead, let us make \emph{three} comments: (\textbf{1})
Using the following variables change mapping the complex coordinates to the
real ones $\left( r,\theta ,\psi ,\varphi \right) $,%
\begin{equation}
\begin{tabular}{llll}
$\mathrm{f}^{1}$ & $=$ & $\rho e^{i\left( \mathrm{\psi +\varphi }\right)
}\cos \frac{\mathrm{\theta }}{2}$ & , \\ 
$\mathrm{f}^{2}$ & $=$ & $\rho e^{i\left( \mathrm{\psi -\varphi }\right)
}\sin \frac{\mathrm{\theta }}{2}$ & ,%
\end{tabular}%
\end{equation}%
with%
\begin{equation}
\begin{tabular}{llll}
$\rho ^{2}=2\left( \mathrm{r}-\mathrm{M}\right) \mathrm{M}$ & , & $\mathrm{%
r>M}=\frac{1}{2\sqrt{\lambda }}$ & ,%
\end{tabular}
\label{lam}
\end{equation}%
the Taub-NUT metric%
\begin{equation}
ds^{2}=2h_{i}^{j}d\mathrm{f}^{i}d\overline{\mathrm{f}}_{j}+g_{ij}d\mathrm{f}%
^{i}d\mathrm{f}^{j}+\overline{g}^{_{ij}}d\overline{\mathrm{f}}_{i}d\overline{%
\mathrm{f}}_{j}\text{ },  \label{me}
\end{equation}%
becomes%
\begin{eqnarray}
\mathrm{ds}^{2} &=&\frac{\left( \mathrm{r+M}\right) }{2\left( \mathrm{r-M}%
\right) }\mathrm{dr}^{2}+2\frac{\left( \mathrm{r-M}\right) }{\left( \mathrm{%
r+M}\right) }\left( \mathrm{d\psi }+\cos \mathrm{\theta d\varphi }\right)
^{2}  \notag \\
&&+\frac{\left( \mathrm{r}^{2}\mathrm{-M}^{2}\right) }{2}\left( \mathrm{%
d\theta }^{2}+\sin ^{2}\mathrm{\theta d\varphi }^{2}\right) .
\end{eqnarray}%
This expression of the metric is precisely the standard form of the Taub-NUT
metric where the singularity is manifestly exhibited in real coordinates 
\textrm{\cite{EH}}. Moreover, from (\ref{lam}), we learn that the coupling
constant $\lambda $ is proportional to the mass $M$ of the Taub-NUT black
hole with horizon at $r=M$. Notice that the origin of the conic field
variable $\rho =0$ corresponds exactly to the singularity $r=M$. So, the
field modulus $\rho $ can be interpreted as describing fluctuations near the
Taub-NUT horizon.\newline
(\textbf{2}) The metric (\ref{me}) can be rewritten in an other equivalent
form as follows:%
\begin{equation}
\begin{tabular}{ll}
$\mathrm{ds}^{2}=G_{i\alpha ,j\beta }\mathrm{d\xi }^{i\alpha }\mathrm{d\xi }%
^{j\beta }$ & ,%
\end{tabular}%
\end{equation}%
with $\mathrm{\xi }^{i\alpha }$ standing for the $SU\left( 2\right) \times
SU\left( 2\right) $ doublet $\left( \mathrm{f}^{i},\overline{\mathrm{f}}%
_{i}\right) $ and where the tensor $G_{i\alpha ,j\beta }$ is given by the $%
4\times 4$ matrix,%
\begin{equation}
G_{i\alpha j\beta }=\left( 
\begin{array}{cc}
\overline{g}_{ij} & h_{i}^{j} \\ 
h_{i}^{j} & g^{_{ij}}%
\end{array}%
\right) .
\end{equation}%
This way of writing the metric $G_{i\alpha ,j\beta }$ is interesting since
it allows to express it in terms of vielbeins $E_{i\alpha }^{k\gamma }$ as $%
G_{i\alpha j\beta }=E_{i\alpha }^{k\gamma }E_{j\beta }^{l\delta }\varepsilon
_{kl}\varepsilon _{\gamma \delta }$ with%
\begin{equation}
E_{i\alpha }^{k\gamma }=\left( 
\begin{array}{cc}
\frac{\delta _{i}^{k}\left( 2+\lambda \mathrm{f}\overline{\mathrm{f}}\right)
-\lambda \mathrm{f}_{i}\overline{\mathrm{f}}^{k}}{2\sqrt{1+\lambda \mathrm{f}%
\overline{\mathrm{f}}}} & \frac{\lambda \overline{\mathrm{f}}_{i}\overline{%
\mathrm{f}}^{k}}{2\sqrt{1+\lambda \mathrm{f}\overline{\mathrm{f}}}} \\ 
-\frac{\lambda \mathrm{f}_{i}\mathrm{f}^{k}}{2\sqrt{1+\lambda \mathrm{f}%
\overline{\mathrm{f}}}} & \frac{\delta _{i}^{k}\left( 2+\lambda \mathrm{f}%
\overline{\mathrm{f}}\right) +\lambda \overline{\mathrm{f}}_{i}\mathrm{f}^{k}%
}{2\sqrt{1+\lambda \mathrm{f}\overline{\mathrm{f}}}}%
\end{array}%
\right) .
\end{equation}%
Following \textrm{\cite{TN}}, the hyperKahler 2- form $\Omega ^{\left(
kl\right) }$ reads as, 
\begin{equation}
\begin{tabular}{llll}
$\Omega ^{\left( kl\right) }$ & $=$ & $\frac{1}{1+\lambda \mathrm{\rho }^{2}}%
\left( \varepsilon _{\gamma \delta }E_{i\alpha }^{(k\gamma }E_{i\beta
}^{l)\delta }\right) d\mathrm{\xi }^{i\alpha }\wedge d\mathrm{\xi }^{j\beta
} $ & .%
\end{tabular}%
\end{equation}%
As we see, this 2-form is given by the irreducible isotriplet factor of the
following reducible quaternionic 2- form 
\begin{equation}
\begin{tabular}{llll}
$\Omega ^{kl}$ & $=$ & $d\mathrm{\xi }^{i\alpha }\wedge d\mathrm{\xi }%
^{j\beta }\left( \frac{1}{1+\lambda \mathrm{\rho }^{2}}\varepsilon _{\gamma
\delta }E_{i\alpha }^{k\gamma }E_{i\beta }^{l\delta }\right) $ & .%
\end{tabular}%
\end{equation}%
The extra irreducible term, namely the isosinglet 
\begin{equation*}
\Omega ^{0}=\Omega ^{kl}\varepsilon _{kl},
\end{equation*}%
can be also written as $d\mathrm{\xi }^{i\alpha }\wedge d\mathrm{\xi }%
^{j\beta }\left( \frac{1}{1+\lambda \mathrm{\rho }^{2}}G_{i\alpha j\beta
}\right) $. As we will see in the discussion section, this term may be
interpreted in terms of the flux of the \emph{NS-NS} antisymmetric B- field.

\section{Quaternionic 2- form}

In this section, we want to give a geometric interpretation of the Taub-NUT
geometry discussed in the above section in terms of the periods $\phi ^{a}$
of the quaternionic form $\Omega ^{a}$ like,%
\begin{equation}
\begin{tabular}{llll}
$\phi ^{a}$ & $=$ & $\int_{C_{2}}\Omega ^{a}$ & ,%
\end{tabular}%
\end{equation}%
where the real 2-cycle $C_{2}$ will be specified later on.\newline
Using the homomorphism $SO\left( 4\right) \simeq SU\left( 2\right) \times
SU\left( 2\right) $, we can rewrite the field moduli $\phi ^{a}$ and the
quaternionic 2- form $\Omega ^{a}$ like,%
\begin{equation}
\begin{tabular}{llll}
$\phi _{j}^{i}$ & $=$ & $\sum\limits_{a=1}^{4}\left( \sigma ^{a}\right)
_{j}^{i}\phi ^{a}$ & , \\ 
$\Omega _{j}^{i}$ & $=$ & $\sum\limits_{a=1}^{4}\left( \sigma ^{a}\right)
_{j}^{i}\Omega ^{a}$ & .%
\end{tabular}%
\end{equation}%
The analysis to be given in this section can be also viewed as a first step
towards the study of the $U^{n}\left( 1\right) $ supersymmetric model based
on the moduli space (\ref{hn}). The non linear $U^{n}\left( 1\right) $
supersymmetric sigma model in six dimension and the underlying hyperKahler
metric (\ref{me}) will be studied in the next sections.\newline
To that purpose, we first study the quaternionic 2- form by borrowing
methods from Kahler geometry and type II superstring compactification on
Calabi-Yau threefolds.\newline
Then, we consider the derivation of the \emph{HSS}\ potential (\ref{qp}).

\subsection{Quaternionifying the hyperKahler form}

We begin by describing the \emph{complexified} Kahler 2- form in type IIA
superstring on Calabi-Yau threefolds. Then we study the case of \emph{10D}
type IIA superstring on K3.

\emph{Complexified Kahler 2- form}\newline
In \emph{10D} type IIA superstring on Calabi-Yau threefolds X$_{3}$, the
usual Kahler form on X$_{3}$%
\begin{equation}
K=\Omega ^{\left( 1,1\right) }\text{ ,}
\end{equation}%
gets complexified by the implementation of the NS-NS B- field as follows%
\begin{equation}
\begin{tabular}{llll}
$J$ & $=$ & $\mathcal{B}_{NS}+iK$ & .%
\end{tabular}%
\end{equation}%
The Kahler moduli $z^{a}$ capturing the Kahler deformations of the
Calabi-Yau threefold are given by the periods%
\begin{equation}
z^{a}=\int_{C_{2}^{a}}J\text{ },
\end{equation}%
where $C_{2}^{a}$ is a real 2-cycle basis of the second homology of X$_{3}$.
The holomorphic prepotential $F\left( z\right) $ is given by%
\begin{equation}
\begin{tabular}{llll}
$F\left( z\right) $ & $=$ & $\int_{X_{3}}J\wedge J\wedge J$ & ,%
\end{tabular}
\label{hol}
\end{equation}%
which, up on using the tri-intersection tensor $d_{abc}$, gives the well
known relation 
\begin{equation}
\begin{tabular}{llll}
$F\left( z\right) $ & $=$ & $\sum\limits_{a,b,c=1}^{h_{X_{3}}^{\left(
1,1\right) }}d_{abc}z^{a}z^{b}z^{c}$ & .%
\end{tabular}%
\end{equation}%
Below, we will show that the analog of this relation in the case of \emph{10D%
} type IIA superstring on K3 is precisely given by eq(\ref{qp}).

\emph{From HyperKahler 2- form to quaternionic }$\Omega ^{a}$\emph{\ }%
\newline
To begin recall that in the case of the complex surface K3, the complex 2-
forms 
\begin{equation}
\begin{tabular}{llll}
$\Omega ^{+}=\Omega ^{\left( 2,0\right) }$ & , & $\Omega ^{-}=\Omega
^{\left( 0,2\right) }$ & ,%
\end{tabular}%
\end{equation}%
and the Kahler 2- form 
\begin{equation}
\begin{tabular}{ll}
$\Omega ^{0}=\Omega ^{\left( 1,1\right) }$ & ,%
\end{tabular}%
\end{equation}%
are in the same cohomology class $H^{2}\left( K3\right) $. This property
reflects the fact that K3 has a hyperKahler structure described by the
isotriplet%
\begin{equation}
\begin{tabular}{llll}
$\Omega ^{\left( ij\right) }$ & $\equiv $ & $\left( 
\begin{array}{c}
\Omega ^{+} \\ 
\Omega ^{0} \\ 
\Omega ^{-}%
\end{array}%
\right) $ & .%
\end{tabular}%
\end{equation}%
In \emph{10D} type IIA superstring on K3, the hyperKahler 2- form $\Omega
^{\left( ij\right) }$ gets quaternionified as follows%
\begin{equation}
\begin{tabular}{llll}
$\Omega ^{ij}$ & $=$ & $\mathcal{B}_{NS}\mathcal{\varepsilon }^{ij}+\Omega
^{\left( ij\right) }$ & ,%
\end{tabular}%
\end{equation}%
where the singlet $\mathcal{B}_{NS}$ stands for the NS-NS antisymmetric
2-form B- field of the non chiral \emph{6D} $\mathcal{N}=2$ supergravity
theory. The above relation can be also put in the equivalent form%
\begin{equation}
\begin{tabular}{llll}
$\Omega ^{a}$ & $=$ & $\sum\limits_{a=1}^{4}\mathcal{\sigma }_{ij}^{a}\Omega
^{ij}$ & ,%
\end{tabular}
\label{sa}
\end{equation}%
where $\mathcal{\sigma }_{ij}^{a}$ are the usual Pauli $2\times 2$ matrices.

\subsection{Periods}

The scalar fields $\phi ^{aI}$ of the hypermultiplets $\mathcal{H}_{6D,N=1}$
(\ref{213}-\ref{215}) have a geometric interpretation in terms of periods of
the above quaternionic 2-form $\Omega ^{a}$. We have%
\begin{equation}
\begin{tabular}{llll}
$\phi ^{aI}$ & $=$ & $\int_{C_{2}^{I}}\Omega ^{a}$ & ,%
\end{tabular}%
\end{equation}%
where $C_{2}^{I}$ is a generic real 2-cycle of the $H^{\left( 1,1\right)
}\left( K3\right) $ Dalbeault homology of K3. For a given 2- cycle C$_{2}$,
the above relation simplifies as 
\begin{equation}
\begin{tabular}{llll}
$\phi ^{a}$ & $=$ & $\int_{C_{2}}\Omega ^{a}$ & ,%
\end{tabular}
\label{c2}
\end{equation}%
and is associated with Taub-NUT geometry. Indeed, using eq(\ref{sa}), we can
rewrite the above relation like,%
\begin{equation}
\begin{tabular}{llll}
$\overline{\mathrm{f}}^{i}\mathrm{f}^{j}$ & $=$ & $\frac{1}{i}%
\int_{C_{2}}\Omega ^{ij}$ & ,%
\end{tabular}%
\end{equation}%
where we have used the complex coordinates $\phi ^{ij}=i$ $\overline{\mathrm{%
f}}^{i}\mathrm{f}^{j}$. Multiplying both sides of this relation by the
harmonic variables $u_{k}^{+}u_{l}^{+}$, we can put the it in the form,%
\begin{equation}
\begin{tabular}{llll}
$\mathrm{\tilde{f}}^{+}\mathrm{f}^{+}$ & $=$ & $\frac{1}{i}%
\int_{C_{2}}\Omega ^{++}$ & ,%
\end{tabular}
\label{ff}
\end{equation}%
with $\mathrm{w}^{++}=\mathrm{\tilde{f}}^{+}\mathrm{f}^{+}$ and%
\begin{equation}
\begin{tabular}{llll}
$\Omega ^{++}$ & $=$ & $u_{k}^{+}u_{l}^{+}\Omega ^{kl}$ & , \\ 
$\mathrm{w}^{++}$ & $=$ & $u_{k}^{+}u_{l}^{+}\mathrm{\tilde{f}}^{k}\mathrm{f}%
^{l}$ & ,%
\end{tabular}%
\end{equation}%
satisfying the obvious identity 
\begin{equation}
\begin{tabular}{llll}
$u^{+i}\frac{\partial }{\partial u^{-i}}\Omega ^{++}$ & $=$ & $\partial
^{++}\Omega ^{++}=0$ & , \\ 
$u^{+i}\frac{\partial }{\partial u^{-i}}\mathrm{w}^{++}$ & $=$ & $\partial
^{++}\mathrm{w}^{++}=0$ & ,%
\end{tabular}%
\end{equation}%
which should be associated with the conservation law of the \emph{HSS}
current (\ref{csl}). Thinking about eq(\ref{ff}) as the leading $\theta ^{+}$%
- component of $\mathrm{\tilde{\Phi}}^{+}\mathrm{\Phi }^{+}$ (\ref{ccs}), we
can promote it to the following superfield relation%
\begin{equation}
\begin{tabular}{llll}
$T^{++}$ & $=$ & $\int_{C_{2}}\mathcal{J}^{++}$ & , \\ 
$T^{++}$ & $=$ & $i\mathrm{\tilde{\Phi}}^{+}\mathrm{\Phi }^{+}$ & ,%
\end{tabular}
\label{fa}
\end{equation}%
satisfying 
\begin{equation}
\begin{tabular}{llll}
$D^{++}T^{++}=0$ & , & $D^{++}\mathcal{J}^{++}=0$ & .%
\end{tabular}
\label{fb}
\end{equation}%
Moreover, denoting by $\mathbf{\omega }$ the real 2-form which is dual to
the 2-cycle $C_{2}$ involved in (\ref{c2}); 
\begin{equation}
\begin{tabular}{llll}
$\int_{C_{2}}\mathbf{\omega }$ & $=$ & $1$ & ,%
\end{tabular}%
\end{equation}%
with the normalization%
\begin{equation}
\begin{tabular}{llll}
$\int_{{\small CY2}}\mathbf{\omega \wedge \omega }$ & $=$ & $1$ & ,%
\end{tabular}
\label{no}
\end{equation}%
then we have 
\begin{equation}
\begin{tabular}{ll}
$\mathcal{J}^{++}=T^{++}\mathbf{\omega }=i\mathrm{\tilde{\Phi}}^{+}\mathrm{%
\Phi }^{+}\mathbf{\omega }$ & .%
\end{tabular}%
\end{equation}%
Now computing the\ analog of (\ref{hol}), it is not difficult to see that
the \emph{HSS} Lagrangian (\ref{t}) may be defined as 
\begin{equation}
\begin{tabular}{ll}
$\mathcal{L}_{1}^{+4}=\frac{\lambda }{2}\int_{CY2}\mathcal{J}^{++}\wedge 
\mathcal{J}^{++}$ & .%
\end{tabular}
\label{hy1}
\end{equation}%
Substituting $\mathcal{J}^{++}$ by its expression $T^{++}\mathbf{\omega }$
and using the normalization (\ref{no}), we obtain precisely the \emph{HSS}
potential of the Taub-NUT model namely 
\begin{equation}
\begin{tabular}{llllll}
$\mathcal{L}_{1}^{+4}$ & $=$ & $\frac{\lambda }{2}\left( T^{++}\right) ^{2}$
& $=$ & $-\frac{\lambda }{2}\left( \mathrm{\tilde{\Phi}}^{+}\mathrm{\Phi }%
^{+}\right) ^{2}$ & .%
\end{tabular}
\label{hy2}
\end{equation}%
Now we turn to study the generic case.

\subsection{Deriving the HSS prepotential (\protect\ref{qp})}

The above analysis extend naturally to the case of \emph{10D} type IIA
superstring on K3. There, eq(\ref{fa}) and the hypermultiplet coupling $%
\mathcal{L}_{1}^{+4}$ (\ref{hy1}-\ref{hy2}) generalizes as follows:\newline
\textbf{(i)} Instead of one super- current $T^{++}$, we have twenty \emph{HSS%
}\ conserved currents $T^{++I}$, 
\begin{equation}
\begin{tabular}{llll}
$D^{++}T^{++I}=0$ & , & $I=1,...,20$ & ,%
\end{tabular}%
\end{equation}%
given by 
\begin{equation}
\begin{tabular}{llll}
$T^{++I}$ & $=$ & $iTr\left( \mathrm{\tilde{\Phi}}^{+}H^{I}\mathrm{\Phi }%
^{+}\right) $ & .%
\end{tabular}
\label{car}
\end{equation}%
They are expressed in terms of the \emph{twenty} hypermultiplets%
\begin{equation}
\begin{tabular}{llllll}
$\Phi ^{+A}$ & , & $\tilde{\Phi}_{A}^{+}$ & , & $A=1,...,20$ & ,%
\end{tabular}%
\end{equation}%
and the Cartan generators $\left\{ H^{I}\right\} $ of the $U\left( 20\right) 
$ isotropy group of the scalar manifold $\frac{SU\left( 2,20\right) }{S\left[
U\left( 2\right) \times U\left( 20\right) \right] }$. Similarly as in eq(\ref%
{fa}), we also have%
\begin{equation}
\begin{tabular}{llll}
$T^{++I}=\int_{C_{2}^{I}}\mathcal{J}^{++}$ & , & $I=1,...,20$ & ,%
\end{tabular}
\label{fc}
\end{equation}%
where $C_{2}^{I}$ is a generic real 2-cycle of $\mathcal{H}^{\left(
1,1\right) }\left( K3\right) $. Using the duality relation,%
\begin{equation}
\begin{tabular}{ll}
$\int_{C_{2}^{I}}\mathbf{\omega }_{K}=\delta _{K}^{I}$ & ,%
\end{tabular}%
\end{equation}%
and the intersections, 
\begin{equation}
\begin{tabular}{ll}
$\int_{K3}\mathbf{\omega }_{I}\wedge \mathbf{\omega }_{J}=g_{IJ}$ & ,%
\end{tabular}%
\end{equation}%
with real intersection matrix $g_{IJ}=g_{JI}$, we can rewrite the \emph{HSS}
2-form $\mathcal{J}^{++}$ like%
\begin{equation}
\begin{tabular}{ll}
$\mathcal{J}^{++}=\sum\limits_{I=1}^{20}T^{++I}\mathbf{\omega }%
_{I}=\sum\limits_{I=1}^{20}\mathbf{\omega }_{I}$Tr$\left( \mathrm{\tilde{\Phi%
}}^{+}H^{I}\mathrm{\Phi }^{+}\right) $ & .%
\end{tabular}
\label{jp}
\end{equation}%
Furthermore, using (\ref{hy1}), we can compute the HSS prepotential

\begin{equation}
\begin{tabular}{llll}
$\mathcal{L}_{20}^{4+}$ & $=$ & $\frac{\lambda }{2}\int_{K3}\mathcal{J}%
^{++}\wedge \mathcal{J}^{++}$ & .%
\end{tabular}%
\end{equation}%
Substituting $\mathcal{J}^{++}$ by its expression eq(\ref{jp}), we obtain
the following HSS Lagrangian density,%
\begin{equation}
\begin{tabular}{llll}
$\mathcal{L}_{20}^{4+}$ & $=$ & $\frac{\lambda }{2}\left(
\sum\limits_{I,J=1}^{20}g_{KL}T^{++K}T^{++L}\right) $ & ,%
\end{tabular}%
\end{equation}%
or equivalently%
\begin{equation}
\begin{tabular}{llll}
$\mathcal{L}_{20}^{4+}$ & $=$ & -$\frac{\lambda }{2}\sum\limits_{I,J=1}^{20}%
\left[ Tr\left( \mathrm{\tilde{\Phi}}^{+}H^{K}\mathrm{\Phi }^{+}\right) %
\right] g_{KL}\left[ Tr\left( \mathrm{\tilde{\Phi}}^{+}H^{L}\mathrm{\Phi }%
^{+}\right) \right] $ & .%
\end{tabular}
\label{hsc}
\end{equation}%
In the next section, we compute the metric associated with this \emph{HSS}\
Lagrangian density. \newline
Below, we relax the above hypermultiplets self- coupling (\ref{hsc}) to
generic integers $n\geq 1$ dealing with the scalar manifold (\ref{hn}). The
corresponding \emph{HSS} prepotential will be denoted as $\mathcal{L}%
_{n}^{4+}$.

\section{$U^{n}\left( 1\right) $\ supersymmetric model}

In the $\mathcal{N}=1$ formalism of the Coulomb branch of the non chiral 
\emph{6D} $\mathcal{N}=2$ supergravity with generic $n$ Maxwell
supermultiplets (\ref{213}), the self- couplings of the hypermultiplets $%
\left\{ \Phi ^{+A}\right\} $ is given by the Lagrangian density,%
\begin{equation}
\begin{tabular}{ll}
$L_{n}\left( x\right) =\int_{S^{2}}du$ $\mathcal{L}_{n}\left( x,u\right) $ & 
,%
\end{tabular}%
\end{equation}%
with%
\begin{equation}
\begin{tabular}{llll}
$\mathcal{L}_{n}\left( x,u\right) $ & $=$ & $\int d^{4}\theta ^{+}\left(
\dsum\limits_{A=1}^{n}\tilde{\Phi}_{A}^{+}D^{++}\Phi ^{+A}+\frac{\lambda }{2}%
\dsum\limits_{K,L=1}^{n}g_{KL}T^{++K}T^{++L}\right) $ & ,%
\end{tabular}
\label{lan}
\end{equation}%
where the $H^{I}$'s are the generators of the \emph{U}$\left( n\right) $
group and $T^{++I}=-i$Tr$\left( \Phi ^{+}H^{I}\tilde{\Phi}^{+}\right) $.

\subsection{Symmetries}

The \emph{HSS} Lagrangian density (\ref{lan}) has the following continuous
symmetries:\newline
(\textbf{1}) It has a manifest \emph{4D} $\mathcal{N}=2$ (or equivalently 
\emph{6D} $\mathcal{N}=1$ ) supersymmetry captured by the superfield
formulation.\newline
(\textbf{2}) It has a manifest $SU_{R}\left( 2\right) $ symmetry captured by
the charges of the harmonic variables $u^{\pm }$. The total charge of $%
\mathcal{L}_{n}$ should be zero knowing that the charge of the measure is $%
Q_{U_{C}\left( 1\right) }\left( d^{4}\theta ^{+}\right) =-4$. This charge is
balanced by the charge of the HSS prepotential; i.e: $Q_{U_{C}\left(
1\right) }\left( \mathcal{L}_{n}^{+4}\right) =+4$ since%
\begin{equation}
\begin{tabular}{ll}
$\left[ D^{0},\mathcal{L}_{n}^{+4}\right] =4\mathcal{L}_{n}^{+4}$ & .%
\end{tabular}%
\end{equation}%
The $SU_{R}\left( 2\right) $ invariance will be explicitly exhibited after
integration with respect to the harmonic variables $u^{\pm }$.\newline
(\textbf{3}) It has a manifest $U^{n}(1)$ global invariance acting by
changing the phases of the \emph{HSS} superfields as follows,%
\begin{equation}
\begin{tabular}{llll}
$\Phi ^{+\prime }=e^{iH}\Phi ^{+}$ & , & $\tilde{\Phi}^{+\prime }=e^{-iH}%
\tilde{\Phi}^{+}$ & ,%
\end{tabular}
\label{uu}
\end{equation}%
with%
\begin{equation}
\begin{tabular}{llll}
$H=\sum_{I=1}^{n}\Lambda _{I}H^{I}$ & , & $\left[ H^{I},\Phi _{A}^{+}\right]
=q_{A}^{I}\Phi _{A}^{+}$ & .%
\end{tabular}%
\end{equation}%
To be more explicit, we choose the charges $q_{A}^{I}$ of the
hypermultiplets with respect to the $U\left( 1\right) $ generators $H^{I}$
as follows:%
\begin{equation}
\begin{tabular}{llll}
$\left( q_{A}^{I}\right) $ & $=$ & $\left( 
\begin{array}{ccccccc}
{\small 1} & {\small -1} & {\small 0} & {\small \cdots } & {\small 0} & 
{\small 0} & {\small 0} \\ 
{\small 0} & {\small 1} & {\small -1} & {\small \cdots } & {\small 0} & 
{\small 0} & {\small 0} \\ 
{\small 0} & {\small 0} & {\small 0} & {\small \ddots } & {\small 0} & 
{\small 0} & {\small 0} \\ 
{\small 0} & {\small 0} & {\small 0} & {\small 0} & {\small 1} & {\small -1}
& {\small 0} \\ 
{\small 1} & {\small 1} & {\small 1} & {\small 1} & {\small 1} & {\small 1}
& {\small 1}%
\end{array}%
\right) $ & .%
\end{tabular}%
\end{equation}%
(\textbf{4}) It has a manifest global $U\left( n\right) $ invariance acting
by $n\times n$ unitary matrices $U$ and $U^{\dagger }$ as given below%
\begin{equation}
\begin{tabular}{llll}
$\Phi ^{+\prime }=U$ $\Phi ^{+}$ & , & $\tilde{\Phi}^{+\prime }=\tilde{\Phi}%
^{+}U^{\dagger }$ & ,%
\end{tabular}%
\end{equation}%
with $U^{\dagger }U=I$. Before going ahead notice the two following:\newline
First, the group $U^{n}(1)$ of eq(\ref{uu}) is the maximal abelian
subsymmetry of the $U\left( n\right) $ group. Its local version ($%
D^{++}\Lambda _{I}=0$) is associated with the Coulomb branch of the \emph{6D}
$\mathcal{N}=2$ supergravity theory.\newline
Second, the $U\left( n\right) $ invariance allows to extend eq(\ref{lan}) to
the more general relation%
\begin{equation}
\begin{tabular}{lll}
$\mathcal{L}_{\text{non abelian}}^{+4}=$ & $-\frac{\lambda }{2}%
\dsum\limits_{K,L}g_{KL}T^{++K}T^{++L}-\frac{\lambda }{2}\dsum\limits_{K,%
\alpha }g_{K\alpha }T^{++K}T^{++\alpha }$ &  \\ 
& $-\frac{\lambda }{2}\dsum\limits_{\alpha ,\beta }g_{\alpha \beta
}T^{++\alpha }T^{++\beta }$ & ,%
\end{tabular}%
\end{equation}%
where $g_{K\alpha }$ and $g_{\alpha \beta }$ are coupling constants, $%
T^{++K} $ associated with the Cartan basis as in eq(\ref{car}) and where%
\begin{equation}
\begin{tabular}{ll}
$T^{++\alpha }=Tr\left( \Phi ^{+}E^{\alpha }\tilde{\Phi}^{+}\right) $ & .%
\end{tabular}%
\end{equation}%
In this relation, the $E^{\alpha }$ matrices stand for generic step
operators of the $U\left( n\right) $ isotropy group. Recall in passing that
the set $\left\{ H^{I},E^{\alpha }\right\} $ define the $n+\left(
n^{2}-n\right) =n^{2}$ generators of the $u\left( n\right) $ algebra.
Invariance under $U\left( n\right) $ follows from the trace property 
\begin{equation}
Tr\left( U\Phi ^{+}E^{\alpha }\tilde{\Phi}^{+}U^{\dagger }\right) =Tr\left(
\Phi ^{+}E^{\alpha }\tilde{\Phi}^{+}\right) .
\end{equation}%
Below, we focus our attention on eq(\ref{lan}); i.e $g_{K\alpha }=g_{\alpha
\beta }=0$; but notice that the general case where $g_{K\alpha }\neq 0,$ $%
g_{\alpha \beta }\neq 0$ is also an interesting issue as it concerns the non
abelian extension.

\emph{Equations of motion}\newline
The equations of motion of the hypermultiplets $\Phi ^{+A}$ and $\tilde{\Phi}%
_{A}^{+}$ following from the variation of the Lagrangian density (\ref{lan}%
), can be put in the form,%
\begin{equation}
\begin{tabular}{llll}
$\left( D^{++}-\lambda T^{++}\right) \Phi ^{+}$ & $=$ & $0$ & $,$ \\ 
$\left( D^{++}+\lambda T^{++}\right) \tilde{\Phi}^{+}$ & $=$ & $0$ & $,$%
\end{tabular}
\label{em}
\end{equation}%
with%
\begin{equation}
\begin{tabular}{llll}
$T^{++}$ & $=$ & $\dsum\limits_{I=1}^{n}T_{I}^{++}H^{I}=\dsum%
\limits_{I=1}^{n}T^{++I}H_{I}$ & ,%
\end{tabular}
\label{vp}
\end{equation}%
and where we have set%
\begin{equation}
\begin{tabular}{llllll}
$H_{I}=g_{IJ}H^{J}$ & , & $T_{I}^{++}=g_{IJ}T^{++J}$ & , & $%
g_{IJ}g^{JK}=\delta _{I}^{K}$ & .%
\end{tabular}%
\end{equation}%
Expanding $\Phi ^{+}$ and $\tilde{\Phi}^{+}$ as in eqs(\ref{sp},\ref{ps}),
we can write down the component field eqs of motion. To that purpose, it is
interesting to expand $T^{++}$ (\ref{vp}) as 
\begin{equation}
\frac{1}{i}T^{++}=\mathrm{w}^{++}+\theta ^{+2}\mathrm{M}+\overline{\theta }%
^{+2}\mathrm{N}+i\theta ^{+}\sigma ^{\mu }\overline{\theta }^{+}\mathrm{A}%
_{\mu }+\theta ^{+2}\overline{\theta }^{+2}\mathrm{P}^{--},
\end{equation}%
where, for simplicity, the fermionic contribution have been dropped out and
where 
\begin{equation}
\begin{tabular}{llllllll}
$\mathrm{w}^{++}$ & $=$ & $\dsum\limits_{I=1}^{n}w^{++I}H_{I}$ & , & $%
\mathrm{A}_{\mu }$ & $=$ & $\dsum\limits_{I=1}^{n}A_{\mu }^{I}H_{I}$ & , \\ 
$\mathrm{M}$ & $=$ & $\dsum\limits_{I=1}^{n}M^{I}H_{I}$ & , & $\mathrm{N}$ & 
$=$ & $\dsum\limits_{I=1}^{n}N^{I}H_{I}$ & ,%
\end{tabular}
\label{w1}
\end{equation}%
with 
\begin{equation}
\begin{tabular}{lll}
$w^{++I}=$ & $\tilde{q}^{+}H^{I}q^{+}$ & , \\ 
$M^{I}=$ & $\left( \tilde{q}^{+}H^{I}F^{-}-\tilde{G}^{-}H^{I}q^{+}\right) $
& , \\ 
$N^{I}=$ & $\left( \tilde{q}^{+}H^{I}G^{-}-\tilde{F}^{-}H^{I}\phi
^{+}\right) $ & , \\ 
$A_{\mu }^{I}=$ & $\left( \tilde{q}^{+}H^{I}B_{\mu }^{-}+\tilde{B}_{\mu
}^{-}H^{I}\phi ^{+}\right) $ & ,%
\end{tabular}
\label{w2}
\end{equation}%
as well as $\partial ^{++}w^{++I}=0$ and $\partial ^{++}A_{\mu
}^{I}=2i\partial _{\mu }\left( w^{++I}\right) $. Then put these relations
back into eq(\ref{em}), we obtain the component field eqs of motion: The
leading term in $\theta ^{+}$ gives, 
\begin{equation}
\begin{tabular}{ll}
$\left( D^{++}-\lambda w^{++}\right) q^{+}=0$ & $,$%
\end{tabular}
\label{del}
\end{equation}%
where $w^{++}$, valued in the Cartan subalgebra of the $U\left( n\right) $
group, is as in eqs(\ref{w1}-\ref{w2}). The term in $\theta ^{+}\sigma ^{\mu
}\overline{\theta }^{+}$ gives,%
\begin{equation}
\begin{tabular}{lll}
$\left[ \partial ^{++}-\lambda w^{++}\right] \mathrm{B}_{\mu }^{-}-\lambda
A_{\mu }\mathrm{q}^{+}=$ & $2\partial _{\mu }\mathrm{q}^{+}$ & $,$%
\end{tabular}
\label{self}
\end{equation}%
where $w^{++}$ and $A_{\mu }$ are given by eqs(\ref{w1}-\ref{w2}). The terms
in $\theta ^{+2}$ and $\overline{\theta }^{+2}$ give the equations of motion
of the auxiliary fields $\mathrm{F}^{-}$ and $\mathrm{G}^{-}$, 
\begin{equation}
\begin{tabular}{llll}
$\left[ \partial ^{++}-\lambda w^{++}\right] \mathrm{F}^{-}-\lambda M\mathrm{%
q}^{+}$ & $=$ & $=2\partial _{\tau }\mathrm{q}^{+}$ & $,$ \\ 
$\left[ \partial ^{++}-\lambda w^{++}\right] \mathrm{G}^{-}-\lambda N\mathrm{%
q}^{+}$ & $=$ & $=2\partial _{\bar{\tau}}\mathrm{q}^{+}$ & $,$%
\end{tabular}%
\end{equation}%
with $\tau =\left( x^{5}+ix^{6}\right) $. These fields are irrelevant for
the determination (\ref{mc}). They will be ignored below.\newline
The last relation corresponds to the term $\theta ^{+2}\overline{\theta }%
^{+2}$; it gives the space time dynamics of the propagating scalars; this
equation is also not needed for the determination of (\ref{mc}).

\subsection{Solving the constraint eqs(\protect\ref{del}-\protect\ref{self})}

The working of the solution of eqs(\ref{del}) and (\ref{self}) is very
technical. For simplicity, we will focus below on the main steps and focus
on the results. The details of the computations are presented in the
appendix B.\newline
The solution of eq(\ref{del}) expressing $\mathrm{q}^{+}\left( x,u\right) $
in terms of $\left( \mathrm{f}^{i}\left( x\right) ,\overline{\mathrm{f}}%
_{i}\left( x\right) \right) $ and the harmonics $u_{i}^{\pm }$ reads as%
\begin{equation}
\begin{tabular}{llll}
$\mathrm{q}^{+}$ & $=$ & $u_{i}^{+}\mathrm{f}^{i}\exp \left( \frac{\lambda }{%
2}\sum\limits_{k,l=1}^{2}u_{(k}^{+}u_{l)}^{-}\sum\limits_{I=1}^{n}\left[
Tr\left( \overline{\mathrm{f}}^{k}H_{I}\mathrm{f}^{l}\right) \right]
H^{I}\right) $ & , \\ 
&  &  &  \\ 
$\mathrm{\tilde{q}}^{+}$ & $=$ & $u_{i}^{+}\overline{\mathrm{f}}^{i}\exp
\left( \frac{-\lambda }{2}\sum\limits_{k,l=1}^{2}u_{(k}^{+}u_{l)}^{-}\sum%
\limits_{I=1}^{n}\left[ Tr\left( \overline{\mathrm{f}}^{k}H_{I}\mathrm{f}%
^{l}\right) \right] H^{I}\right) $ & .%
\end{tabular}
\label{a524}
\end{equation}%
In the limit where the coupling constant $\lambda \longrightarrow 0$, we
recover the free fields $\mathrm{q}^{+}=u_{i}^{+}\mathrm{f}^{i}\left(
x\right) $.\newline
To get the solution of eq(\ref{self}), we need several steps (see appendix B
for details): First use the $U^{n}\left( 1\right) $ symmetry to make the
change 
\begin{equation}
\begin{tabular}{llll}
$\mathrm{B}_{\mu }^{-}=e^{\lambda w}\mathrm{C}_{\mu }^{-}$ & , & $\mathrm{%
\tilde{B}}_{\mu }^{-}=e^{-\lambda w}\mathrm{\tilde{C}}_{\mu }^{-}$ & ,%
\end{tabular}
\label{a25}
\end{equation}%
where $w$ is as in eq(\ref{wh}) and where $\mathrm{C}_{\mu }^{-}$ is the new
auxiliary field satisfying the differential equation,%
\begin{eqnarray}
\partial ^{++}\mathrm{C}_{\mu }^{-}-\lambda A_{\mu }\mathrm{f}^{+}
&=&2\nabla _{\mu }\mathrm{f}^{+},  \notag \\
\partial ^{++}\mathrm{C}_{\mu }^{-}+\lambda A_{\mu }\overline{\mathrm{f}}%
^{+} &=&2\overline{\nabla }_{\mu }\overline{\mathrm{f}}^{+},  \label{acp}
\end{eqnarray}%
with%
\begin{equation}
\begin{tabular}{llll}
$\nabla _{\mu }\mathrm{f}^{+}$ & $=$ & $\left[ \partial _{\mu }+\lambda
\left( \partial _{\mu }w\right) \right] \mathrm{f}^{+}$ & , \\ 
$\overline{\nabla }_{\mu }\overline{\mathrm{f}}^{+}$ & $=$ & $\left[
\partial _{\mu }-\lambda \left( \partial _{\mu }w\right) \right] \overline{%
\mathrm{f}}^{+}$ & .%
\end{tabular}%
\end{equation}%
We also have the decomposition $A_{\mu }=\sum_{I=1}^{n}A_{\mu }^{I}H_{I}$
with 
\begin{equation}
\begin{tabular}{llll}
$A_{\mu }^{I}$ & $=$ & $\mathrm{\tilde{C}}_{\mu }^{-}H^{I}\mathrm{f}^{+}+%
\mathrm{\tilde{f}}^{+}H^{I}\mathrm{C}_{\mu }^{-}$ & .%
\end{tabular}
\label{ajc}
\end{equation}%
The next step is to use the identity $\lambda A_{\mu }\mathrm{f}^{+}=\lambda
\partial ^{++}\left( \vartheta _{\mu }\mathrm{f}^{-}+\partial _{\mu }w^{--}%
\mathrm{f}^{+}\right) $ to solve eq(\ref{acp}) like,%
\begin{equation}
\begin{tabular}{llll}
$\mathrm{C}_{\mu }^{-}$ & $=$ & $2\partial _{\mu }\mathrm{f}^{-}+\lambda
\vartheta _{\mu }\mathrm{f}^{-}+\lambda \left( \partial _{\mu }w^{--}\right) 
\mathrm{f}^{+}$ & , \\ 
$\mathrm{\tilde{C}}_{\mu }^{-}$ & $=$ & $2\partial _{\mu }\mathrm{\tilde{f}}%
^{-}-\lambda \vartheta _{\mu }\mathrm{\tilde{f}}^{-}-\lambda \left( \partial
_{\mu }w^{--}\right) \mathrm{\tilde{f}}^{+}$ & .%
\end{tabular}
\label{apa}
\end{equation}%
To determine the quantity $\vartheta _{\mu }$, we have to compute the term $%
\mathrm{\tilde{f}}^{+}H^{I}\mathrm{C}_{\mu }^{-}+\mathrm{\tilde{C}}_{\mu
}^{-}H^{I}\mathrm{f}^{+}$ by using eqs(\ref{apa}) and derive a constraint
equation that allows us to fix $\vartheta _{\mu }$. We have%
\begin{equation}
\begin{tabular}{llll}
$\mathrm{\tilde{f}}^{+}H^{I}\mathrm{C}_{\mu }^{-}$ & $=$ & $2\mathrm{\tilde{f%
}}^{+}H^{I}\partial _{\mu }\mathrm{f}^{-}+\lambda \vartheta _{\mu J}\left( 
\mathrm{\tilde{f}}^{+}H^{I}H^{J}\mathrm{f}^{-}\right) +2\lambda \partial
_{\mu }w_{J}^{--}\left( \mathrm{\tilde{f}}^{+}H^{I}H^{J}\mathrm{f}%
^{+}\right) $ & , \\ 
$\mathrm{\tilde{C}}_{\mu }^{-}H^{I}\mathrm{f}^{+}$ & $=$ & $2\partial _{\mu }%
\mathrm{\tilde{f}}^{-}H^{I}\mathrm{f}^{+}-\lambda \vartheta _{\mu J}\left( 
\mathrm{\tilde{f}}^{-}H^{J}H^{I}\mathrm{f}^{+}\right) -2\lambda \partial
_{\mu }w_{J}^{--}\left( \mathrm{\tilde{f}}^{+}H^{J}H^{I}\mathrm{f}%
^{+}\right) $ & .%
\end{tabular}
\label{anv}
\end{equation}%
For simplicity of the equations, it is convenient to introduce the following
conventional notations: 
\begin{equation}
\begin{tabular}{ll}
$\mathrm{Q}_{I}^{\pm A}=u_{i}^{\pm }\mathrm{Q}_{I}^{iA}\equiv \left( \mathrm{%
f}^{\pm }H_{I}\right) ^{A}$ & , \\ 
$\mathrm{\tilde{Q}}_{B}^{\pm I}=u_{i}^{\pm }\overline{\mathrm{Q}}%
_{B}^{iI}\equiv \left( \mathrm{\tilde{f}}^{\pm }H^{I}\right) _{B}$ & ,%
\end{tabular}
\label{aqq}
\end{equation}%
with 
\begin{equation}
\begin{tabular}{ll}
$\mathrm{Q}_{I}^{iA}=\left( H_{I}\right) _{C}^{A}\mathrm{f}^{iC}$ & , \\ 
$\overline{\mathrm{Q}}_{jB}^{I}=\overline{\mathrm{f}}_{jD}\left(
H^{I}\right) _{B}^{D}$ & , \\ 
$R_{B}^{A}=\overline{\mathrm{Q}}_{iB}^{I}\mathrm{Q}_{I}^{iA}$ & .%
\end{tabular}
\label{aqqb}
\end{equation}%
Using these fields, one can build the following composites%
\begin{equation}
\begin{tabular}{llllllll}
$\overline{\mathrm{Q}}_{iB}^{I}\mathrm{Q}_{I}^{iA}$ & , & $\mathrm{Q}%
_{I}^{iA}\overline{\mathrm{Q}}_{jB}^{J}$ & , & $\mathrm{Q}_{I}^{iA}\mathrm{Q}%
_{J}^{kC}$ & , & $\overline{\mathrm{Q}}_{jB}^{I}\overline{\mathrm{Q}}%
_{lD}^{J}$ & .%
\end{tabular}
\label{aQ}
\end{equation}%
For $n=1$, the unique Cartan generator reduces to the identity operator, $%
H_{1}=I$, the fields $\mathrm{Q}_{I}^{iA}$\ reduce down to $\mathrm{f}^{i}$
and eqs(\ref{aQ}) to%
\begin{equation}
\begin{tabular}{llll}
$\overline{\mathrm{Q}}_{iB}^{I}\mathrm{Q}_{I}^{iA}\rightarrow \overline{%
\mathrm{f}}_{i}\mathrm{f}^{i}$ & , & $\mathrm{Q}_{I}^{iA}\overline{\mathrm{Q}%
}_{jB}^{J}\rightarrow \mathrm{f}^{i}\overline{\mathrm{f}}_{j}$ & , \\ 
$\mathrm{Q}_{I}^{iA}\mathrm{Q}_{J}^{kC}\rightarrow \mathrm{f}^{i}\mathrm{f}%
^{k}$ & , & $\overline{\mathrm{Q}}_{jB}^{I}\overline{\mathrm{Q}}%
_{lD}^{J}\rightarrow \overline{\mathrm{f}}_{j}\overline{\mathrm{f}}_{l}$ & .%
\end{tabular}
\label{aQQ}
\end{equation}%
Using the field moduli $\mathrm{Q}^{\pm }$ and $\mathrm{\tilde{Q}}^{\pm }$,
we can rewrite eqs(\ref{anv}) like,%
\begin{equation}
\begin{tabular}{ll}
$\mathrm{\tilde{Q}}^{+I}\mathrm{C}_{\mu }^{-}=2\mathrm{\tilde{Q}}%
^{+I}\partial _{\mu }\mathrm{f}^{-}+\lambda \vartheta _{\mu J}\left( \mathrm{%
\tilde{Q}}^{+I}\mathrm{Q}^{-J}\right) +\lambda \partial _{\mu
}w_{J}^{--}\left( \mathrm{\tilde{Q}}^{+I}\mathrm{Q}^{+J}\right) $ & , \\ 
$\mathrm{\tilde{C}}_{\mu }^{-}\mathrm{Q}^{+I}=2\partial _{\mu }\mathrm{%
\tilde{f}}^{-}\mathrm{Q}^{+I}-\lambda \vartheta _{\mu J}\left( \mathrm{%
\tilde{Q}}^{-J}\mathrm{Q}^{+I}\right) -2\lambda \partial _{\mu }\varphi
_{J}^{--}\left( \mathrm{\tilde{Q}}^{+J}\mathrm{Q}^{+I}\right) $ & .%
\end{tabular}%
\end{equation}%
Next, adding the two relations and using eq(\ref{ajc}), we get 
\begin{equation}
\mathcal{E}_{J}^{I}\vartheta _{\mu }^{J}=\upsilon _{\mu }^{I}\text{ },
\label{at1}
\end{equation}%
with%
\begin{equation}
\begin{tabular}{ll}
$\upsilon _{\mu }^{I}=\left( \mathrm{Q}^{iAI}\partial _{\mu }\overline{%
\mathrm{f}}_{iA}-\overline{\mathrm{Q}}_{iA}^{I}\partial _{\mu }\mathrm{f}%
^{iA}\right) $ & , \\ 
$\mathcal{E}_{J}^{I}=\left[ \delta _{J}^{I}+\lambda \overline{\mathrm{Q}}%
_{iAJ}\mathrm{Q}^{iAI}\right] $ & .%
\end{tabular}
\label{at2}
\end{equation}%
Using eq(\ref{aqq}), these relations can be also put in the equivalent form%
\begin{equation*}
\begin{tabular}{ll}
$\upsilon _{\mu }^{I}=\left( \mathrm{f}^{i}H^{I}\partial _{\mu }\overline{%
\mathrm{f}}_{i}-\overline{\mathrm{f}}_{i}H^{I}\partial _{\mu }\mathrm{f}%
^{i}\right) $ & , \\ 
$\mathcal{E}_{J}^{I}=\left[ \delta _{J}^{I}+\lambda \overline{\mathrm{Q}}%
_{iAJ}\mathrm{Q}^{iAI}\right] $ & .%
\end{tabular}%
\end{equation*}%
Then, the solution of $\vartheta _{\mu }^{I}$ reads as, 
\begin{equation}
\begin{tabular}{llll}
$\vartheta _{\mu }^{J}=\mathcal{F}_{I}^{J}v_{\mu }^{I}$ & , & $\mathcal{F}%
_{I}^{J}\mathcal{E}_{K}^{I}=\delta _{K}^{I}$ & .%
\end{tabular}
\label{at3}
\end{equation}%
Notice that for the leading case $n=1$, eqs(\ref{at2}-\ref{at3}) reduce to%
\begin{equation*}
\begin{tabular}{llll}
$\upsilon _{\mu }^{I}\rightarrow \upsilon _{\mu }=\left( \mathrm{f}%
^{i}\partial _{\mu }\overline{\mathrm{f}}_{i}-\overline{\mathrm{f}}%
_{i}\partial _{\mu }\mathrm{f}^{i}\right) $ & , & $\mathcal{E}%
_{J}^{I}\rightarrow \mathcal{E}=\left[ 1+\lambda \overline{\mathrm{f}}_{i}%
\mathrm{f}^{i}\right] $ & ,%
\end{tabular}%
\end{equation*}%
and%
\begin{equation}
\begin{tabular}{llll}
$\mathcal{F}_{I}^{J}\rightarrow \mathcal{F}=\frac{1}{\left[ 1+\lambda 
\overline{\mathrm{f}}_{i}\mathrm{f}^{i}\right] }$ & , & $\mathcal{EF}=1$ & .%
\end{tabular}%
\end{equation}%
Notice moreover that because of the property $\overline{\mathrm{f}}%
_{i}H_{J}H^{I}\mathrm{f}^{i}=\overline{\mathrm{f}}_{i}H^{I}H_{J}\mathrm{f}%
^{i}$, we have the identity $\overline{\mathrm{Q}}_{iAJ}\mathrm{Q}^{iAI}=%
\overline{\mathrm{Q}}_{iA}^{I}\mathrm{Q}_{J}^{iA}$. \newline
The solution $\mathrm{C}_{\mu }^{-A}\left( x,u\right) $ and $\mathrm{\tilde{C%
}}_{\mu B}^{-}\left( x,u\right) $ read, in terms of $\mathrm{Q}_{J}^{\pm }$,
as%
\begin{equation}
\begin{tabular}{lll}
$\mathrm{C}_{\mu }^{-A}=$ & $2\partial _{\mu }\mathrm{f}^{-A}+\lambda 
\mathcal{F}_{I}^{J}\upsilon _{\mu }^{I}\mathrm{Q}_{J}^{-A}+\lambda \mathrm{Q}%
_{J}^{+A}\mathrm{\tilde{Q}}_{B}^{-J}\left( \partial _{\mu }\mathrm{f}%
^{-B}\right) $ &  \\ 
& $+\lambda \mathrm{Q}_{J}^{+A}\mathrm{Q}^{-BJ}\left( \partial _{\mu }%
\mathrm{\tilde{f}}_{B}^{-}\right) $ & ,%
\end{tabular}
\label{aca}
\end{equation}%
and%
\begin{equation}
\begin{tabular}{llll}
$\mathrm{\tilde{C}}_{\mu A}^{-}$ & $=$ & $2\partial _{\mu }\mathrm{\tilde{f}}%
_{A}^{-}-\lambda \mathcal{F}_{I}^{J}\upsilon _{\mu }^{I}\mathrm{\tilde{Q}}%
_{AJ}^{-}-\lambda \mathrm{\tilde{Q}}_{AJ}^{+}\mathrm{Q}^{-BJ}\left( \partial
_{\mu }\mathrm{\tilde{f}}_{B}^{-}\right) $ &  \\ 
&  & $-\lambda \mathrm{\tilde{Q}}_{AJ}^{+}\mathrm{\tilde{Q}}_{B}^{-J}\left(
\partial _{\mu }\mathrm{f}^{-B}\right) $ & .%
\end{tabular}
\label{acb}
\end{equation}%
The harmonic dependence of the fields $\mathrm{C}_{\mu }^{-A}=\mathrm{C}%
_{\mu }^{-A}\left( x,u\right) $ and $\mathrm{\tilde{C}}_{\mu A}^{-}=\mathrm{%
\tilde{C}}_{\mu A}^{-}\left( x,u\right) $ is as follows 
\begin{equation}
\mathrm{C}_{\mu }^{-A}\left( x,u\right) =u_{i}^{-}\mathcal{C}_{\mu
}^{iA}\left( x\right) +u_{(i}^{-}u_{j}^{-}u_{k)}^{+}\mathcal{C}_{\mu
}^{\left( ijk\right) A}\left( x\right) ,
\end{equation}%
with%
\begin{equation}
\begin{tabular}{lll}
$\mathcal{C}_{\mu }^{iA}=$ & $2\partial _{\mu }\mathrm{f}^{iA}+\lambda 
\mathcal{F}_{I}^{J}\upsilon _{\mu }^{I}\mathrm{Q}_{J}^{iA}$ &  \\ 
& $+\frac{\lambda }{3}\mathrm{Q}_{J}^{jA}\overline{\mathrm{Q}}%
_{jB}^{J}\left( \partial _{\mu }\mathrm{f}^{iB}\right) +\frac{\lambda }{3}%
\overline{\mathrm{Q}}_{B}^{iJ}\mathrm{Q}_{J}^{jA}\left( \partial _{\mu }%
\mathrm{f}_{j}^{B}\right) $ &  \\ 
& $+\frac{\lambda }{3}\mathrm{Q}_{J}^{jA}\mathrm{Q}_{j}^{BJ}\left( \partial
_{\mu }\overline{\mathrm{f}}_{B}^{i}\right) +\frac{\lambda }{3}\mathrm{Q}%
^{iBJ}\mathrm{Q}_{J}^{jA}\left( \partial _{\mu }\overline{\mathrm{f}}%
_{jB}\right) $ & .%
\end{tabular}%
\end{equation}%
Analogous relations for $\overline{\mathcal{C}}_{\mu iA}$ and $\mathcal{C}%
_{\mu }^{\left( ijk\right) A}$, $\overline{\mathcal{C}}_{\mu \left(
ijk\right) A}$ are given in appendix B.

\section{Computing the metric}

Starting from the superfield relation (\ref{lan}) and performing the
integration with respect to the Grassmann variables $\theta ^{+}$ and $%
\overline{\theta }^{+}$, we obtain the following component field action,%
\begin{equation}
\mathcal{S}_{n}=\frac{1}{2}\int d^{4}x\left[ \int_{S^{2}}du\sum_{A=1}^{n}%
\left( \mathrm{B}_{\mu }^{-A}\partial ^{\mu }\mathrm{\tilde{q}}_{A}^{+}-%
\mathrm{\tilde{B}}_{\mu A}^{-}\partial ^{\mu }\mathrm{q}^{+A}\right) \right]
.  \label{aty}
\end{equation}%
This action still depends on the auxiliary fields $\mathrm{B}_{\mu }^{-A}$
and the harmonic variables. To get the space time field action,%
\begin{equation}
\mathcal{S}_{n}=\frac{1}{2}\int d^{4}x\left( 2h_{iA}^{jB}\partial _{\mu }%
\mathrm{f}^{iA}\partial ^{\mu }\overline{\mathrm{f}}_{jB}+\overline{g}%
_{iAjB}\partial _{\mu }\mathrm{f}^{iA}\partial ^{\mu }\mathrm{f}%
^{jB}+g^{_{iAjB}}\partial _{\mu }\overline{\mathrm{f}}_{iA}\partial ^{\mu }%
\overline{\mathrm{f}}_{jB}\right) \text{ },
\end{equation}%
we have to eliminate the $\mathrm{B}_{\mu }^{-A}$'s and integrate with
respect the harmonic variables $u^{\pm }$. Substituting $\mathrm{B}_{\mu
}^{-A}$ and $\mathrm{q}^{+A}$ by of their expressions in terms of $\mathrm{C}%
_{\mu }^{-A}$ and $\mathrm{f}^{+A}$ given by eqs(\ref{a524},\ref{a25},\ref%
{aca},\ref{acb}), we can put $\mathcal{S}_{n}$ as%
\begin{equation}
\mathcal{S}=\frac{1}{2}\int d^{4}x\left[ L_{n1}\left( x\right) +L_{n2}\left(
x\right) \right]
\end{equation}%
with 
\begin{equation}
\begin{tabular}{llll}
$L_{n1}\left( x\right) $ & $=$ & $\int_{S^{2}}du\left( \mathrm{C}_{\mu
}^{-A}\partial ^{\mu }\mathrm{\tilde{f}}_{A}^{+}-\mathrm{\tilde{C}}_{\mu
A}^{-}\partial ^{\mu }\mathrm{f}^{+A}\right) $ & , \\ 
$L_{n2}\left( x\right) $ & $=$ & $-\lambda \int_{S^{2}}du\left[ \left( 
\mathrm{\tilde{Q}}_{IA}^{+}\mathrm{C}_{\mu }^{-A}+\mathrm{\tilde{C}}_{\mu
A}^{-}\mathrm{Q}_{I}^{+A}\right) \left( \partial ^{\mu }w^{I}\right) \right] 
$ & ,%
\end{tabular}%
\end{equation}%
and $w^{I}$ as in eq(\ref{x5}). \newline
Substituting $\mathrm{C}_{\mu }^{-A}$ and $\mathrm{\tilde{C}}_{\mu A}^{-}$
by their expressions in terms of the propagating fields and integrating with
respect to the harmonic variables, we find, after some algebra given in
appendix B, subsection B2, that $L_{n1}$ reads as 
\begin{equation}
\begin{tabular}{lll}
$L_{n1}=$ & $-2\partial _{\mu }\mathrm{f}^{iA}\partial ^{\mu }\overline{%
\mathrm{f}}_{iA}+\frac{\lambda }{2}\mathcal{N}_{kC}^{lD}\partial _{\mu }%
\mathrm{f}^{kC}\partial ^{\mu }\overline{\mathrm{f}}_{lD}$ &  \\ 
& $-\frac{\lambda }{2}\mathcal{U}^{kClD}\partial _{\mu }\overline{\mathrm{f}}%
_{kC}\partial ^{\mu }\overline{\mathrm{f}}_{lD}-\frac{\lambda }{2}\overline{%
\mathcal{U}}_{kClD}\partial _{\mu }\mathrm{f}^{lD}\partial ^{\mu }\mathrm{f}%
^{kC}$ & ,%
\end{tabular}
\label{ala1}
\end{equation}%
with%
\begin{equation}
\begin{tabular}{llll}
$\mathcal{N}_{kC}^{lD}$ & $=$ & $\mathcal{F}_{I}^{J}\mathrm{Q}_{J}^{lD}%
\overline{\mathrm{Q}}_{kC}^{I}+\mathcal{F}_{I}^{J}\overline{\mathrm{Q}}_{kCJ}%
\mathrm{Q}^{lDI}$ &  \\ 
&  & $+\left( \overline{\mathrm{Q}}_{C}^{lJ}\mathrm{Q}_{kJ}^{D}\right)
-\left( \overline{\mathrm{Q}}_{iC}^{J}\mathrm{Q}_{J}^{iD}\right) \delta
_{k}^{l}$ & , \\ 
&  &  &  \\ 
$\overline{\mathcal{U}}_{kC,lD}$ & $=$ & $\mathcal{F}_{I}^{J}\overline{%
\mathrm{Q}}_{lDJ}\overline{\mathrm{Q}}_{kC}^{I}$ &  \\ 
&  & $+\frac{1}{2}\left( \overline{\mathrm{Q}}_{kD}^{J}\overline{\mathrm{Q}}%
_{lCJ}\right) -\frac{1}{2}\left( \overline{\mathrm{Q}}_{iD}^{J}\overline{%
\mathrm{Q}}_{CJ}^{i}\right) \varepsilon _{kl}$ & , \\ 
&  &  &  \\ 
$\mathcal{U}^{kC,lD}$ & $=$ & $\mathcal{F}_{I}^{J}\mathrm{Q}_{J}^{kC}\mathrm{%
Q}^{lDI}$ &  \\ 
&  & $+4\mathrm{\xi }\left( \mathrm{Q}^{lCJ}\mathrm{Q}_{J}^{kD}\right) -4%
\mathrm{\xi }\left( \mathrm{Q}_{i}^{CJ}\mathrm{Q}_{J}^{iD}\right)
\varepsilon ^{kl}$ & .%
\end{tabular}
\label{aal1}
\end{equation}%
In the particular case where $n=1$, these quantities reduce to%
\begin{equation}
\begin{tabular}{llll}
$\mathcal{N}_{k}^{l}$ & $=$ & $\frac{2\mathrm{f}^{l}\overline{\mathrm{f}}_{k}%
}{1+\lambda \overline{\mathrm{f}}\mathrm{f}}+\overline{\mathrm{f}}^{l}%
\mathrm{f}_{k}-\delta _{k}^{l}\overline{\mathrm{f}}_{i}\mathrm{f}^{i}$ & ,
\\ 
&  &  &  \\ 
$\overline{\mathcal{U}}_{kl}$ & $=$ & $\frac{\overline{\mathrm{f}}_{l}%
\overline{\mathrm{f}}_{k}}{1+\lambda \overline{\mathrm{f}}\mathrm{f}}+\frac{1%
}{2}\overline{\mathrm{f}}_{k}\overline{\mathrm{f}}_{l}$ & , \\ 
&  &  &  \\ 
$\mathcal{U}^{kl}$ & $=$ & $\frac{\mathrm{f}^{k}\mathrm{f}^{l}}{1+\lambda 
\overline{\mathrm{f}}\mathrm{f}}+\frac{1}{2}\mathrm{f}^{l}\mathrm{f}^{k}$ & ,%
\end{tabular}
\label{anu}
\end{equation}%
A similar analysis shows that \emph{the term }$L_{n2}$\emph{\ }eq(\ref{l2})
has the form 
\begin{equation}
\begin{tabular}{lll}
$L_{n2}=$ & $-\frac{\lambda }{2}\widehat{\mathcal{U}}^{kC,lD}\partial _{\mu }%
\overline{\mathrm{f}}_{kC}\partial ^{\mu }\overline{\mathrm{f}}_{lD}-\frac{%
\lambda }{2}\widehat{\overline{\mathcal{U}}}_{kC,lD}\partial _{\mu }\mathrm{f%
}^{lD}\partial ^{\mu }\mathrm{f}^{kC}$ &  \\ 
& $+\frac{\lambda }{2}\mathcal{\hat{N}}_{kC}^{lD}\partial _{\mu }\mathrm{f}%
^{kC}\partial ^{\mu }\overline{\mathrm{f}}_{lD}$ & ,%
\end{tabular}
\label{ala2}
\end{equation}%
with%
\begin{equation}
\begin{tabular}{lll}
$\mathcal{\hat{N}}_{kC}^{lD}=$ & $\lambda \left( \overline{\mathrm{Q}}%
_{CI}^{l}\mathrm{Q}_{k}^{DI}-\overline{\mathrm{Q}}_{iCI}\mathrm{Q}%
^{iDI}\delta _{k}^{l}\right) $ & , \\ 
$\widehat{\overline{\mathcal{U}}}_{kC,lD}=$ & $\frac{\lambda }{2}\left( 
\overline{\mathrm{Q}}_{lCI}\overline{\mathrm{Q}}_{kD}^{I}-\overline{\mathrm{Q%
}}_{CI}^{i}\overline{\mathrm{Q}}_{iD}^{I}\varepsilon _{kl}\right) $ & , \\ 
$\widehat{\mathcal{U}}^{kC,lD}=$ & $\frac{\lambda }{2}\left( \mathrm{Q}%
_{I}^{lC}\mathrm{Q}^{kDI}-\mathrm{Q}_{iI}^{C}\mathrm{Q}^{iDI}\varepsilon
^{kl}\right) $ & .%
\end{tabular}
\label{aal2}
\end{equation}%
In the case $n=1$, these tensors reduce to%
\begin{equation}
\begin{tabular}{llllll}
$\mathcal{\hat{N}}_{k}^{l}=\lambda \left( \overline{\mathrm{f}}^{l}\mathrm{f}%
_{k}-\delta _{k}^{l}\overline{\mathrm{f}}\mathrm{f}\right) $ & , & $\widehat{%
\overline{\mathcal{U}}}_{kl}=\frac{\lambda }{2}\overline{\mathrm{f}}_{l}%
\overline{\mathrm{f}}_{k}$ & , & $\widehat{\mathcal{U}}^{kl}=\frac{\lambda }{%
2}\mathrm{f}^{l}\mathrm{f}^{k}$ & .%
\end{tabular}%
\end{equation}

\emph{the U}$^{n}\left( 1\right) $\emph{\ hyperKahler metric}\newline
Adding eqs(\ref{ala1}-\ref{aal1}) and eqs(\ref{ala2}-\ref{aal2}), we get the
total Lagrangian density%
\begin{equation}
\begin{tabular}{lll}
$L_{n}=$ & $+g^{kc,ld}\partial _{\mu }\overline{\mathrm{f}}_{kc}\partial
^{\mu }\overline{\mathrm{f}}_{ld}+\overline{g}_{kc,ld}\partial _{\mu }%
\mathrm{f}^{ld}\partial ^{\mu }\mathrm{f}^{kc}$ &  \\ 
& $+2h_{kc}^{ld}\partial _{\mu }\mathrm{f}^{kc}\partial ^{\mu }\overline{%
\mathrm{f}}_{ld}$ & ,%
\end{tabular}%
\end{equation}%
with%
\begin{equation}
\begin{tabular}{lll}
$2h_{kC}^{lD}=$ & $-2\delta _{k}^{l}\delta _{C}^{D}+\frac{\lambda }{2}\left( 
\mathcal{N}_{kC}^{lD}+\mathcal{\hat{N}}_{kC}^{lD}\right) $ & , \\ 
$g^{kC,lD}=$ & $-\frac{\lambda }{2}\left( \mathcal{U}^{kC,lD}+\widehat{%
\mathcal{U}}^{kC,lD}\right) $ & , \\ 
$\overline{g}_{kC,lD}=$ & $-\frac{\lambda }{2}\left( \overline{\mathcal{U}}%
_{kC,lD}+\widehat{\overline{\mathcal{U}}}_{kC,lD}\right) $ & .%
\end{tabular}%
\end{equation}%
Substituting $\mathcal{N}_{kC}^{lDd}$ and $\mathcal{\hat{N}}_{kC}^{lD}$ by
their expressions (\ref{aal1}-\ref{aal2}) and using the relation $\mathcal{F}%
_{I}^{J}\mathcal{E}_{J}^{K}=\mathcal{\delta }_{I}^{K}$, we get the following
explicit field relation of the metric components: 
\begin{equation}
\begin{tabular}{lll}
$h_{kC}^{lD}=$ & $-\delta _{k}^{l}\left( \delta _{C}^{D}+\frac{\lambda }{2}%
\overline{\mathrm{Q}}_{iCI}\mathrm{Q}^{iDI}\right) $ &  \\ 
& $+\frac{\lambda }{2}\mathcal{F}_{I}^{J}\left( \left[ \overline{\mathrm{Q}}%
_{kC}^{I}\mathrm{Q}_{J}^{lD}+\mathcal{E}_{J}^{K}\overline{\mathrm{Q}}%
_{CI}^{l}\mathrm{Q}_{k}^{DK}\right] \right) $ & ,%
\end{tabular}%
\end{equation}%
and%
\begin{equation}
\begin{tabular}{lll}
$g^{kC,lD}=$ & $-\frac{\lambda }{2}\mathcal{F}_{I}^{J}\left[ \mathrm{Q}%
_{J}^{kC}\mathrm{Q}^{lDI}+\mathcal{E}_{J}^{K}\mathrm{Q}_{I}^{lC}\mathrm{Q}%
^{kDK}+\mathcal{E}_{J}^{K}\mathrm{Q}_{iI}^{C}\mathrm{Q}^{iDK}\varepsilon
^{kl}\right] $ & , \\ 
&  &  \\ 
$\overline{g}_{kC,lD}=$ & $-\frac{\lambda }{2}\mathcal{F}_{I}^{J}\left( 
\overline{\mathrm{Q}}_{lDJ}\overline{\mathrm{Q}}_{kC}^{I}+\mathcal{E}_{J}^{K}%
\overline{\mathrm{Q}}_{lCI}\overline{\mathrm{Q}}_{kD}^{K}+\mathcal{E}_{J}^{K}%
\overline{\mathrm{Q}}_{CI}^{i}\overline{\mathrm{Q}}_{iD}^{K}\varepsilon
_{kl}\right) $ & .%
\end{tabular}%
\end{equation}%
In the special case $n=1$, these relations reduce to, 
\begin{equation}
\begin{tabular}{lll}
$h_{k}^{l}=$ & $-2\delta _{k}^{l}\left( 1+\lambda \overline{\mathrm{f}}%
\mathrm{f}\right) +\lambda \left( \frac{1+\left( 1+\lambda \overline{\mathrm{%
f}}\mathrm{f}\right) }{1+\lambda \overline{\mathrm{f}}\mathrm{f}}\right) 
\mathrm{f}^{l}\overline{\mathrm{f}}_{k}$ & , \\ 
$\overline{g}_{kl}=$ & $-\frac{\lambda }{2}\frac{1+\left( 1+\lambda 
\overline{\mathrm{f}}\mathrm{f}\right) }{1+\lambda \overline{\mathrm{f}}%
\mathrm{f}}\overline{\mathrm{f}}_{k}\overline{\mathrm{f}}_{l}$ & , \\ 
$g^{kl}=$ & $-\frac{\lambda }{2}\frac{1+\left( 1+\lambda \overline{\mathrm{f}%
}\mathrm{f}\right) }{1+\lambda \overline{\mathrm{f}}\mathrm{f}}\mathrm{f}^{k}%
\mathrm{f}^{l}$ & ,%
\end{tabular}%
\end{equation}%
where we have used the identity $\overline{\mathrm{f}}^{l}\mathrm{f}_{k}=%
\mathrm{f}^{l}\overline{\mathrm{f}}_{k}-\delta _{k}^{l}\left( \overline{%
\mathrm{f}}\mathrm{f}\right) $. Comparing this expression with eq(\ref{tnm}%
), we recover exactly the Taub-NUT metric.\newline
Furthermore substituting $\mathrm{Q}_{I}^{iA}$ and $\overline{\mathrm{Q}}%
_{jB}^{I}$ by their expressions in terms of $\mathrm{f}^{iA}$, $\overline{%
\mathrm{f}}_{iA}$ and the Cartan matrices $H_{I}$, we can rewrite the metric
components $h_{kC}^{lD}$, $g^{kClD}$ and $\overline{g}_{kClD}$ as follows :%
\newline
(\textbf{i}) \emph{component} $h_{kC}^{lD}$%
\begin{equation}
\begin{tabular}{lll}
$h_{kC}^{lD}=$ & $+\delta _{k}^{l}\left( \delta _{C}^{D}+\frac{\lambda }{2}%
\overline{\mathrm{f}}_{iA}\mathrm{f}^{iB}g_{IJ}\left( H^{I}\right)
_{C}^{A}\left( H^{J}\right) _{B}^{D}\right) $ &  \\ 
& $-\frac{\lambda }{2}\mathcal{F}_{I}^{J}\left( \left[ \overline{\mathrm{f}}%
_{kA}\mathrm{f}^{lB}g_{JL}\left( H^{I}\right) _{C}^{A}\left( H^{L}\right)
_{B}^{D}\right] \right) $ &  \\ 
& $-\frac{\lambda }{2}\mathcal{F}_{I}^{J}\left[ \left( \mathcal{E}_{J}^{K}%
\overline{\mathrm{f}}_{A}^{l}\mathrm{f}_{k}^{D}g_{IL}\left( H^{L}\right)
_{C}^{A}\left( H^{K}\right) _{B}^{D}\right) \right] $ & .%
\end{tabular}
\label{a}
\end{equation}%
where the $n\times n$ matrices $\mathcal{E}_{I}^{J}$ and $\mathcal{F}%
_{J}^{K} $ are given by 
\begin{equation}
\begin{tabular}{llll}
$\mathcal{E}_{I}^{J}=\left[ \delta _{I}^{J}+\lambda \text{Tr}\left( 
\overline{\mathrm{f}}H_{I}H^{J}\mathrm{f}\right) \right] $ & , & $\mathcal{E}%
_{I}^{J}\mathcal{F}_{J}^{K}=\delta _{I}^{K}$ & .%
\end{tabular}%
\end{equation}%
(ii) \emph{component }$g^{kClD}$%
\begin{equation}
\begin{tabular}{lll}
$g^{kClD}=$ & $+\frac{\lambda }{2}\mathcal{F}_{I}^{J}\left[ \mathrm{f}^{kA}%
\mathrm{f}^{lB}\left( g_{JL}\left( H^{L}\right) _{A}^{C}\left( H^{I}\right)
_{B}^{D}\right) \right] $ &  \\ 
& $+\frac{\lambda }{2}\mathcal{F}_{I}^{J}\left( \mathrm{f}^{lA}\mathrm{f}%
^{kB}\left[ \mathcal{E}_{J}^{K}g_{IL}\left( H^{L}\right) _{A}^{C}\left(
H^{K}\right) _{B}^{D}\right] \right) $ &  \\ 
& $+\frac{\lambda }{2}\mathcal{F}_{I}^{J}\left[ \mathrm{f}^{jA}\mathrm{f}%
^{ik}\left( \mathcal{E}_{J}^{K}g_{IL}\left( H^{L}\right) _{A}^{C}\left(
H^{K}\right) _{B}^{D}\right) \varepsilon ^{kl}\varepsilon _{ij}\right] $ & .%
\end{tabular}
\label{b}
\end{equation}%
(iii) \emph{component }$\overline{g}_{kClD}$%
\begin{equation}
\begin{tabular}{lll}
$\overline{g}_{kClD}=$ & $+\frac{\lambda }{2}\mathcal{F}_{I}^{J}\left[ 
\overline{\mathrm{f}}_{lB}\overline{\mathrm{f}}_{kA}^{I}g_{JL}\left(
H^{L}\right) _{D}^{B}\left( H^{I}\right) _{C}^{A}\right] $ &  \\ 
& $+\frac{\lambda }{2}\mathcal{F}_{I}^{J}\left( \overline{\mathrm{f}}_{lA}%
\overline{\mathrm{f}}_{kB}\left[ \mathcal{E}_{J}^{K}g_{IL}\left(
H^{L}\right) _{C}^{A}\left( H^{K}\right) _{D}^{B}\right] \right) $ &  \\ 
& $+\frac{\lambda }{2}\mathcal{F}_{I}^{J}\left[ \overline{\mathrm{f}}_{jA}%
\overline{\mathrm{f}}_{iB}\left( \mathcal{E}_{J}^{K}g_{IL}\left(
H^{L}\right) _{C}^{A}\left( H^{K}\right) _{D}^{B}\right) \varepsilon
_{kl}\varepsilon ^{ij}\right] $ & ,%
\end{tabular}
\label{c}
\end{equation}%
which just the hermitian conjugate of $g^{kClD}$. In the limit $\lambda
\rightarrow 0$, one recover the free field theory and for $n=1$ the Taub-NUT
geometry.

\section{Conclusion and discussion}

Freezing the dynamics of the dilaton $\sigma $ and using rigid harmonic
superspace (\emph{HSS}) method, we have computed in this paper the explicit
expression of the metric of the real \emph{eighty} dimensional moduli space $%
\boldsymbol{M}_{6D}^{N=2}$ of the \emph{10D} type IIA superstring on K3.
This hyperkahler metric has the following form%
\begin{equation}
\begin{tabular}{llll}
$ds_{80}^{2}$ & $=$ & $\sum\limits_{k,l=1}^{2}\left(
\sum\limits_{A,B=1}^{20}2h_{kB}^{lA}d\mathrm{f}^{kB}d\overline{\mathrm{f}}%
_{lA}+g^{kBlA}d\overline{\mathrm{f}}_{kB}d\overline{\mathrm{f}}_{lA}+%
\overline{g}_{kBlA}d\mathrm{f}^{lA}d\mathrm{f}^{kB}\right) $ & ,%
\end{tabular}
\label{mm}
\end{equation}%
and should be put in correspondence with the Kahler metric of the Coulomb
branch of the moduli space of \emph{10D} type IIA superstring on Calabi-Yau
threefolds,%
\begin{equation}
\begin{tabular}{llllll}
$ds_{\text{{\small Kahler}}}^{2}$ & $=$ & $\sum\limits_{a,b=1}^{n_{v}}g_{a%
\bar{b}}dz^{a}d\bar{z}^{b}$ & , & $g_{a\bar{b}}=\frac{\partial ^{2}\mathcal{K%
}}{\partial z^{a}\partial \bar{z}^{b}}$ & .%
\end{tabular}
\label{wt}
\end{equation}%
We have also shown that the metric (\ref{mm}) is a particular member of a
family of hyperKahler metrics (\ref{a}-\ref{c}) of \emph{4n} dimensional
manifolds with $U^{n}\left( 1\right) $ gauge symmetry whose leading term is
given by the well known real \emph{four} dimensional Taub-NUT geometry
associated with eqs(\ref{l1}-\ref{tn}).\newline
The dynamics of the dilaton $\sigma $ can be directly implemented in the
supergravity field action along the line given in \textrm{\cite{S2}. }In
particular we have for the moduli space $SO\left( 1,1\right) \times \frac{%
U\left( 2,20\right) }{S\left( U\left( 2\right) \times U\left( 20\right)
\right) }$ the following result,%
\begin{equation}
\begin{tabular}{llll}
$ds_{81}^{2}$ & $=$ & $\left( d\sigma \right) ^{2}-e^{-2\sigma }ds_{80}^{2}$
& .%
\end{tabular}%
\end{equation}%
With the analysis given in this study, we have learnt as well that
supersymmetry plays a central role in the metric building of the moduli
spaces $\boldsymbol{M}_{4D}^{N=2}$ and $\boldsymbol{M}_{6D}^{N=2}$ of the
Coulomb branches of the \emph{4D} $\mathcal{N}=2$ and non chiral \emph{6D} $%
\mathcal{N}=2$ supergravity theories. We also learnt that the constructions
are quite similar. \newline
To explicitly exhibit the similarities between the ways to get the two kinds
of metrics, we give below a comment regarding this issue. First, we recall
the geometric set up of the moduli spaces $\boldsymbol{M}_{4D}^{N=2}$ and $%
\boldsymbol{M}_{6D}^{N=2}$ of the two supergravity theories. Then, we
describe the way the metrics of the scalar manifolds $\boldsymbol{M}%
_{4D}^{N=2}$ and $\boldsymbol{M}_{6D}^{N=2}$ can be engineered by combining
geometry and supersymmetry in \emph{4D} and \emph{6D} respectively.\newline

\textbf{(1)} \emph{Geometry of} \emph{scalar spaces} \newline
First recall that moduli space of \emph{10D} type II superstring on
Calabi-Yau threefolds has two branches: a Kahler branch with scalar manifold 
$\boldsymbol{M}_{4D}^{N=2}$ and complex a one with a hyperKahler structure 
\textrm{\cite{RO1,RO2}}. Notice also that in \emph{10D} type IIA superstring
on K3, the scalar manifold $\boldsymbol{M}_{6D}^{N=2}$ is hyperKahler since
complex and Kahler deformations of the metric combine to make a hyperKahler
structure. The general picture giving the \emph{4D/6D}\ correspondence is
schematized in the following table:

\begin{equation*}
\begin{tabular}{|llll|}
\hline
\underline{\emph{4D} $\mathcal{N}=2$ sugra} & $\longleftrightarrow \qquad $
&  & \underline{\emph{6D} $\mathcal{N}=2$ sugra} \\ 
&  &  &  \\ 
{\small Type II on CY3} & \ \ \ - &  & {\small Type IIA on K3} \\ 
&  &  &  \\ 
$%
\begin{array}{c}
\text{{\small Kahler\ }} \\ 
\text{{\small complex}}%
\end{array}%
$ & \ \ \ - &  & {\small hyperKahler }${\small \equiv }\left\{ 
\begin{array}{c}
\text{{\small Kahler\ }} \\ 
\text{{\small complex}}%
\end{array}%
\right. $ \\ 
&  &  &  \\ 
$%
\begin{array}{c}
\text{{\small Kahler 2- form:} $\Omega _{2}^{\left( 1,1\right) }$ } \\ 
\text{{\small holomorphic form:} }{\scriptsize \QTR{sc}{\Omega }_{3}^{\left(
3,0\right) }} \\ 
\text{{\small antiholomorphic:} }{\scriptsize \QTR{sc}{\Omega }_{3}^{\left(
0,3\right) }}%
\end{array}%
$ & \ \ \ - &  & {\small hyperkahler: }$\Omega ^{\left( ij\right) }$%
{\scriptsize $=$}$\left\{ 
\begin{array}{c}
{\scriptsize \QTR{sc}{\Omega }_{2}^{\left( 2,0\right) }} \\ 
{\scriptsize \QTR{sc}{\Omega }_{2}^{\left( 1,1\right) }} \\ 
{\scriptsize \QTR{sc}{\Omega }_{2}^{\left( 0,2\right) }}%
\end{array}%
\right. $ \\ 
&  &  &  \\ \hline
\end{tabular}%
\end{equation*}%
Notice that in the case of K3 compactification, Kahler and complex 2-form on
K3 are in the same cohomology class.\newline

\textbf{(2)} \emph{Kahler }$\rightarrow $\emph{\ hyperKahler and beyond}%
\newline
The Kahler and the \textit{complexified} Kahler structures of the Coulomb
branch of vacua in \emph{10D} type IIA on CY3 can be put in correspondence
with the hyperKahler and \textit{quaternionified} hyperKahler structure of $%
\boldsymbol{M}_{6D}^{N=2}$ as shown in the following table,

\begin{equation*}
\begin{tabular}{|llll|}
\hline
\underline{\emph{4D} $\mathcal{N}=2$ sugra} & $\longrightarrow \qquad $ &  & 
\underline{\emph{6D} $\mathcal{N}=2$ sugra} \\ 
&  &  &  \\ 
$\left\{ 
\begin{array}{c}
\text{{\small Kahler:}} \\ 
\text{$\Omega _{2}^{\left( 1,1\right) }\equiv K$\ }%
\end{array}%
\right. $ & \ \ \ - &  & $\left\{ 
\begin{array}{c}
\text{ {\small HyperKahler 2- form:}} \\ 
\Omega ^{\left( ij\right) }\text{\ }%
\end{array}%
\right. $ \\ 
&  &  &  \\ 
$\left\{ 
\begin{array}{c}
\text{ {\small Complexified Kahler:}} \\ 
J=B_{{\scriptsize NS}}+i\text{$K$}%
\end{array}%
\right. $ & \ \ \ - &  & $\left\{ 
\begin{array}{c}
\text{ {\small Quaternionic 2- form:}} \\ 
\boldsymbol{J}^{{\scriptsize ij}}=B_{{\scriptsize NS}}\varepsilon ^{%
{\scriptsize ij}}+\Omega ^{\left( ij\right) }%
\end{array}%
\right. $ \\ 
&  &  &  \\ \hline
\end{tabular}%
\end{equation*}%
where $\Omega _{2}^{\left( 1,1\right) }$ and $\Omega ^{\left( ij\right) }$
are as in previous tables and where $B_{{\scriptsize NS}}$ stands for the 
{\small NS-NS} B- field. Notice the remarkable role played by the $B$- field
in both cases.\newline

\textbf{(3)} \emph{Metrics building}\newline
As it is well known, the Kahler metric $g_{a\bar{b}}$ has a nice
interpretation in \emph{4D} $\mathcal{N}=1$ superspace. Denoting by $\Phi
^{a}$ a generic chiral superfield with leading component scalar field $z^{a}$%
; that is $\left( \Phi ^{a}\right) _{\theta =0}=z^{a}$, the metric $g_{a\bar{%
b}}$ can be obtained by integrating out the Grassmann variables $\theta $ in
the following superspace relation, 
\begin{equation}
\begin{tabular}{llll}
$L\left( x\right) $ & $=$ & $\int d^{4}\theta \mathcal{K}\left( \Phi ,\bar{%
\Phi}\right) $ & ,%
\end{tabular}%
\end{equation}%
where $\mathcal{K}\left( \Phi ,\bar{\Phi}\right) $ is the usual Kahler
(super)potential.\newline
HyperKahler metrics are engineered in a quite similar manner; but now by
using \emph{HSS}\ method. There, the Lagrangian density $L\left( x\right) $
is given by 
\begin{equation}
\begin{tabular}{llll}
$L\left( x\right) $ & $=$ & $\int_{S^{2}}du\left[ \int d^{4}\theta ^{+}%
\mathcal{L}^{+4}\left( \Phi ^{+},\tilde{\Phi}^{+}\right) \right] $ & ,%
\end{tabular}
\label{lp}
\end{equation}%
where $\Phi ^{+}$ is a hypermultiplet superfield and $\mathcal{L}^{+4}\left(
\Phi ^{+},\tilde{\Phi}^{+}\right) $ is the harmonic superspace Lagrangian
density as in eq(\ref{ls}). \newline
The hyperKahler metric associated with eq(\ref{lp}) is obtained as follows: 
\newline
\textbf{(i)} first integrate out the Grassmann variables $\theta ^{+}$ (\ref%
{lp}) to get the typical relation (\ref{aty}). \newline
\textbf{(ii)} Then eliminate the vector auxiliary fields B$_{\mu }^{-A}$
through their eqs of motion.\newline
\textbf{(iii)} Finally integrate out the harmonic variables $u^{\pm }$ and
end with eq(\ref{mm}).\newline
In the case of \emph{4D/6D} $\mathcal{N}=2$ supergravity theories embedded
in type IIA superstring compactifications, we have the following
correspondence 
\begin{equation*}
\begin{tabular}{|llllll|}
\hline
&  & \underline{\emph{4D} $\mathcal{N}=2$ sugra} & $\longrightarrow \qquad $
&  & \underline{\emph{6D} $\mathcal{N}=2$ sugra} \\ 
&  &  &  &  &  \\ 
{\small moduli} & $:$ & $z^{a}=\int_{C_{2}^{a}}J$ & \ \ \ - &  & $\phi
^{ijI}=\int_{C_{2}^{I}}\boldsymbol{J}^{ij}$ \\ 
&  &  &  &  &  \\ 
{\small prepotential} & $:$ & $F\left( z\right) =\left\{ 
\begin{array}{c}
\int_{CY3}J\wedge J\wedge J \\ 
=d_{abc}z^{a}z^{b}z^{c}%
\end{array}%
\right. $ & \ \ \ - &  & $\mathcal{F}^{ijkl}=\left\{ 
\begin{array}{c}
\int_{CY3}\boldsymbol{J}^{ij}\wedge \boldsymbol{J}^{kl} \\ 
=g_{IJ}\phi ^{ijI}\phi ^{klJ}%
\end{array}%
\right. $ \\ 
&  &  &  &  &  \\ \hline
\end{tabular}%
\end{equation*}%
where $d_{abc}$ are the 3-intersection numbers in the second homology of the
Calabi-Yau threefold and $g_{IJ}$ the 2-intersections of $H_{2}\left(
K3,R\right) $.\newline
Moreover, using the fact that Kahler 2-form can be also expressed as $K=%
\frac{1}{2}\left( J-\overline{J}\right) $, we have 
\begin{equation}
\begin{tabular}{llll}
$\int_{C_{2}^{a}}K$ & $=$ & $\frac{1}{2i}\left( z^{a}-\bar{z}^{a}\right) $ & 
.%
\end{tabular}
\label{kz}
\end{equation}%
Then, Kahler potential $\mathcal{K}=\mathcal{K}\left( \Phi ,\bar{\Phi}%
\right) $ reads in terms of the usual \emph{4D} $\mathcal{N}=1$ chiral
superfields $\Phi ^{a}$ and $\bar{\Phi}^{a}$ as,%
\begin{equation}
\begin{tabular}{llll}
$e^{\mathcal{K}}$ & $=$ & $\frac{i}{8}\sum\limits_{a,b,c}d_{abc}\left( \Phi
^{a}-\bar{\Phi}^{a}\right) \left( \Phi ^{b}-\bar{\Phi}^{b}\right) \left(
\Phi ^{c}-\bar{\Phi}^{c}\right) $ & .%
\end{tabular}%
\end{equation}%
The analogue of the above \emph{4D} $\mathcal{N}=1$ superspace (or
equivalently \emph{2D} $\mathcal{N}=2$) relations in six dimensional space
time can be also written down in \emph{6D} $\mathcal{N}=1$ \emph{HSS }or
equivalently\emph{\ 4D} $\mathcal{N}=2$ \emph{HSS}. There, the analogue of
eq(\ref{kz}) is obviously given by the hyperKahler isotriplet moduli%
\begin{equation}
\begin{tabular}{llll}
$\phi ^{\left( kl\right) I}$ & $=$ & $\int_{C_{2}^{I}}\Omega ^{\left(
kl\right) }$ & .%
\end{tabular}%
\end{equation}%
In harmonic superspace, this can be achieved by multiplying both sides of $%
\phi ^{ijI}=\int_{C_{2}^{I}}\boldsymbol{J}^{ij}$ by the harmonic variable
monomial $u_{i}^{+}u_{j}^{+}$ to end with, 
\begin{equation}
\begin{tabular}{lllllll}
$\phi ^{++I}$ & $=$ & $\int_{C_{2}^{I}}\Omega ^{++}$ & , & $B_{{\scriptsize %
NS}}\varepsilon ^{{\scriptsize ij}}u_{i}^{+}u_{j}^{+}$ & $=0$ & ,%
\end{tabular}%
\end{equation}%
and then the hyperKahler potential,%
\begin{equation}
\begin{tabular}{llll}
$\mathcal{L}_{\text{{\small HK}}}^{4+}$ & $=$ & $\sum%
\limits_{I,J}g_{IJ}T^{++I}T^{++J}$ & .%
\end{tabular}%
\end{equation}%
In the above relation the \emph{HSS} superfield $T^{++I}$ is given by the
following hypermultiplet composite 
\begin{equation}
\begin{tabular}{llll}
$T^{++I}$ & $=$ & $\frac{1}{i}Tr\left( \tilde{\Phi}^{+}\mathrm{H}^{I}\Phi
^{+}\right) $ & ,%
\end{tabular}%
\end{equation}%
where $\mathrm{H}^{I}$ are the Cartan generators of the $U\left( n\right) $
isotropy symmetry of the moduli space. The $T^{++I}$'s obey the \emph{HSS}
conservation laws%
\begin{equation}
\begin{tabular}{llll}
$D^{++}T^{++I}$ & $=$ & $0$ & ,%
\end{tabular}%
\end{equation}%
and are interpreted as the Noether super- currents in harmonic superspace.
This relation can be also stated as given by the sum of the two following
relation, 
\begin{equation}
\begin{tabular}{llll}
$\left( D^{++}-\lambda \sum\limits_{I}T_{I}^{++}H^{I}\right) \Phi ^{+}$ & $=$
& $0$ & $,$ \\ 
$\left( D^{++}+\lambda \sum\limits_{I}T_{I}^{++}H^{I}\right) \tilde{\Phi}%
^{+} $ & $=$ & $0$ & $.$%
\end{tabular}%
\end{equation}%
These eqs are precisely the superfield eqs of motion, from which we can read
the \emph{HSS}\ superfield action for the hypermultiplets.\newline
We end this study by noting that the analysis given in this paper can be
also used to deal with \emph{10D} type IIA on ALE spaces.

\begin{acknowledgement}
This research work is supported by the programme PROTARS\ D12/25/CNRST.
\end{acknowledgement}

\section{Appendix A: U$\left( 1\right) $ model}

To bring the superfield action (\ref{ac}) to the component fields form (\ref%
{mc}), we start by computing the hypermultiplet equations of motion. They
are given by%
\begin{equation}
\begin{tabular}{llll}
$D^{++}\mathrm{\Phi }^{+}-\lambda \left( \mathrm{\tilde{\Phi}}^{+}\mathrm{%
\Phi }^{+}\right) \mathrm{\Phi }^{+}$ & $=$ & $0$ & , \\ 
$D^{++}\mathrm{\tilde{\Phi}}+\lambda \left( \mathrm{\tilde{\Phi}}^{+}\mathrm{%
\Phi }^{+}\right) \mathrm{\tilde{\Phi}}^{+}$ & $=$ & $0$ & .%
\end{tabular}
\label{c11}
\end{equation}%
Then identify the conserved Noether \emph{HSS} current $T^{++}$ 
\begin{equation}
D^{++}T^{++}=0\text{ ,}  \label{csl}
\end{equation}%
associated with the symmetry (\ref{u1}). The super-current $T^{++}$ can
obtained by multiplying the first relation of the system (\ref{c11}) by $%
\mathrm{\tilde{\Phi}}^{+}$; and the second relation by $\mathrm{\Phi }^{+}$.
By adding both relations, we end with the \emph{HSS}\ conservation law $%
D^{++}\left( \mathrm{\tilde{\Phi}}^{+}\mathrm{\Phi }^{+}\right) =0$, from
which we learn:%
\begin{equation}
\begin{tabular}{llll}
$T^{++}=i\mathrm{\tilde{\Phi}}^{+}\mathrm{\Phi }^{+}$ & , & $\tilde{T}%
^{++}=T^{++}$ & .%
\end{tabular}
\label{ccs}
\end{equation}%
Moreover, substituting the superfields $\mathrm{\Phi }^{+}$ and $\mathrm{%
\tilde{\Phi}}^{+}\mathrm{\ }$by their $\theta ^{+}$- expansions (\ref{sp},%
\ref{ay}), we get the following component field relations:\newline
(\textbf{a}) the leading $\theta ^{+}$ component ($\theta ^{+}=0$) gives the
field eqs of motion of $\Delta ^{{\small ---}}$ and its conjugate $\tilde{%
\Delta}^{{\small ---}}$:%
\begin{equation}
\begin{tabular}{llll}
$\left[ \partial ^{++}\mathrm{q}^{+}-\lambda \left( \mathrm{\tilde{q}}^{+}%
\mathrm{q}^{+}\right) \right] \mathrm{q}^{+}$ & $=$ & $0$ & , \\ 
$\left[ \partial ^{++}\mathrm{\tilde{q}}^{+}+\lambda \left( \mathrm{\tilde{q}%
}^{+}\mathrm{q}^{+}\right) \right] \mathrm{\tilde{q}}^{+}$ & $=$ & $0$ & .%
\end{tabular}
\label{x1}
\end{equation}%
These are constraint eqs that fix the dependence of the scalar field $%
\mathrm{q}^{+}$ in the harmonic variables $u_{i}^{\pm }$; i.e:%
\begin{equation}
\begin{tabular}{llll}
$\mathrm{q}^{+}$ & $=$ & $\mathrm{q}^{+}\left( x,u^{\pm }\right) $ & .%
\end{tabular}%
\end{equation}%
(\textbf{b}) the $\theta ^{+}\sigma ^{\mu }\overline{\theta }^{+}$ component
of eq(\ref{csl}) gives the field eqs of motions of $\mathrm{B}_{\mu }^{-}$
and $\mathrm{\tilde{B}}_{\mu }^{-}$:%
\begin{equation}
\begin{tabular}{llll}
$\left[ \partial ^{++}-\lambda \left( \mathrm{\tilde{q}}^{+}\mathrm{q}%
^{+}\right) \right] \mathrm{B}_{\mu }^{-}-\lambda \left( \mathrm{\tilde{q}}%
^{+}\mathrm{B}_{\mu }^{-}+\mathrm{\tilde{B}}_{\mu }^{-}\mathrm{q}^{+}\right) 
\mathrm{q}^{+}$ & $=$ & $2\partial _{\mu }\mathrm{q}^{+}$ & , \\ 
$\left[ \partial ^{++}+\lambda \left( \mathrm{\tilde{q}}^{+}\mathrm{q}%
^{+}\right) \right] \mathrm{\tilde{B}}_{\mu }^{-}+\lambda \left( \mathrm{%
\tilde{q}}^{+}\mathrm{B}_{\mu }^{-}+\mathrm{\tilde{B}}_{\mu }^{-}\mathrm{q}%
^{+}\right) \mathrm{\tilde{q}}^{+}$ & $=$ & $2\partial _{\mu }\mathrm{\tilde{%
q}}^{+}$ & .%
\end{tabular}
\label{x2}
\end{equation}%
(\textbf{c}) The $\theta ^{+2}$ and $\overline{\theta }^{+2}$ components
give the eqs of motion of the auxiliary fields $F^{-}$ and $G^{-}$. But
these relations are irrelevant for the explicit computation the metric (\ref%
{mc}). They are rather needed for the determination of the scalar potential
that follows from the compactification from \emph{6D} down to \emph{4D}. 
\newline
The solution of eq(\ref{x1}) is given by,%
\begin{equation}
\begin{tabular}{llll}
$\mathrm{q}^{+}\left( x,u\right) $ & $=$ & $u_{i}^{+}e^{\lambda w}\mathrm{f}%
^{i}\left( x\right) $ & ,%
\end{tabular}
\label{x4}
\end{equation}%
with $w$ ($\tilde{w}=-w$) given by%
\begin{equation}
\begin{tabular}{ll}
$w=\frac{1}{2}\left( \mathrm{\tilde{f}}^{-}\mathrm{f}^{+}+\mathrm{\tilde{f}}%
^{+}\mathrm{f}^{-}\right) =\frac{1}{2}u_{(i}^{+}u_{j)}^{-}\overline{\mathrm{f%
}}^{(i}\mathrm{f}^{j)}$ & , \\ 
$\partial ^{++}w=\mathrm{\tilde{f}}^{+}\mathrm{f}^{+}$ & ,%
\end{tabular}
\label{x5}
\end{equation}%
where the complex doublets 
\begin{equation}
\begin{tabular}{llll}
$\mathrm{f}^{\pm }=u_{i}^{\pm }\mathrm{f}^{i}$ & , & $\mathrm{\tilde{f}}%
^{\pm }=u_{i}^{\pm }\mathrm{\bar{f}}^{i}$ & ,%
\end{tabular}%
\end{equation}%
are as in eqs(\ref{mc}). The solution of eq\ref{x2}) is given by, 
\begin{equation}
\begin{tabular}{llll}
$\mathrm{B}_{\mu }^{-}\left( x,u\right) $ & $=$ & $e^{\lambda w}\mathrm{C}%
_{\mu }^{-}\left( x\right) $ & ,%
\end{tabular}%
\end{equation}%
The field $\mathrm{C}_{\mu }^{-}$ is given by%
\begin{equation}
\begin{tabular}{llll}
$\mathrm{C}_{\mu }^{-}$ & $=$ & $2\partial _{\mu }\mathrm{f}^{-}+\lambda
\vartheta _{\mu }\mathrm{f}^{-}+\lambda \mathrm{f}^{+}\partial _{\mu }\left( 
\mathrm{\tilde{f}}^{-}\mathrm{f}^{-}\right) $ & ,%
\end{tabular}
\label{x6}
\end{equation}%
with%
\begin{equation}
\begin{tabular}{llll}
$\vartheta _{\mu }$ & $=$ & $\frac{-1}{1+\lambda \rho ^{2}}\left( \mathrm{%
\bar{f}}_{i}\partial _{\mu }\mathrm{f}^{i}-\mathrm{f}^{i}\partial _{\mu }%
\mathrm{\bar{f}}_{i}\right) $ & , \\ 
$\rho ^{2}$ & $=$ & $\mathrm{\bar{f}}_{i}\mathrm{f}^{i}$ & .%
\end{tabular}
\label{x7}
\end{equation}%
To get this result, we proceed in steps as follows:\newline
(\textbf{i}) compute the $\theta ^{+}$- expansion of $\mathrm{\tilde{\Phi}}%
^{+}\mathrm{\Phi }^{+}$ by using (\ref{sp}-\ref{ps}). We have,%
\begin{equation}
\begin{tabular}{llll}
$\mathrm{\tilde{\Phi}}^{+}\mathrm{\Phi }^{+}$ & $=$ & $\mathrm{w}%
^{++}+\theta ^{+2}\mathrm{M}+\overline{\theta }^{+2}\mathrm{N}+i\theta
^{+}\sigma ^{\mu }\overline{\theta }^{+}\mathrm{A}_{\mu }+\theta ^{+2}%
\overline{\theta }^{+2}\mathrm{P}^{--}$ & ,%
\end{tabular}%
\end{equation}%
with 
\begin{equation}
\begin{tabular}{llll}
$\mathrm{w}^{++}$ & $=$ & $\tilde{q}^{+}q^{+}$ & , \\ 
$\mathrm{M}$ & $=$ & $\left( \tilde{q}^{+}F^{-}-\tilde{G}^{-}q^{+}\right) $
& , \\ 
$\mathrm{\tilde{M}}$ & $=$ & $\left( \tilde{F}^{-}\phi ^{+}-\tilde{q}%
^{+}G^{-}\right) $ & , \\ 
$\mathrm{N}$ & $=$ & $\left( \tilde{q}^{+}G^{-}-\tilde{F}^{-}q^{+}\right) $
& , \\ 
$\mathrm{A}_{\mu }$ & $=$ & $\left( \tilde{q}^{+}B_{\mu }^{-}+\tilde{B}_{\mu
}^{-}q^{+}\right) $ & ,%
\end{tabular}%
\end{equation}%
and%
\begin{equation}
\begin{tabular}{llll}
$\mathrm{P}^{--}$ & $=$ & $\left( \tilde{q}^{+}\Delta ^{{\small -3}}-\tilde{%
\Delta}^{{\small -3}}\phi ^{+}-\tilde{F}^{-}F^{-}-\tilde{G}^{-}G^{-}-\tilde{B%
}_{\mu }^{-}B^{-\mu }\right) $ & .%
\end{tabular}%
\end{equation}%
We also have%
\begin{equation}
\begin{tabular}{llll}
$\mathrm{\tilde{w}}^{++}=-\mathrm{w}^{++}$ & , & $\mathrm{\tilde{A}}_{\mu }=%
\mathrm{A}_{\mu }$ & , \\ 
$\mathrm{\tilde{M}}=-\mathrm{N}$ & , & $\mathrm{\tilde{P}}^{--}=-\mathrm{P}%
^{--}$ & .%
\end{tabular}%
\end{equation}%
\textbf{(ii)} use the \emph{HSS} conservation law $D^{++}\left( \mathrm{%
\tilde{\Phi}}^{+}\mathrm{\Phi }^{+}\right) =0$, which leads, at the level of
the component fields, to:%
\begin{equation}
\begin{tabular}{llll}
$\partial ^{++}\left( \tilde{q}^{+}q^{+}\right) $ & $=$ & $0$ & , \\ 
$\partial ^{++}A_{\mu }$ & $=$ & $2\partial _{\mu }\left( \tilde{q}%
^{+}q^{+}\right) $ & .%
\end{tabular}
\label{x9}
\end{equation}%
These component fields conservation laws require the factorization (\ref{x4}%
); and allow to bring $A_{\mu }$ into the simple form%
\begin{equation}
\begin{tabular}{llll}
$A_{\mu }$ & $=$ & $\mathrm{\tilde{f}}^{+}\mathrm{C}_{\mu }^{-}+\mathrm{%
\tilde{C}}_{\mu }^{-}\mathrm{f}^{+}$ & , \\ 
$\partial ^{++}A_{\mu }$ & $=$ & $2\partial _{\mu }\left( \mathrm{\tilde{f}}%
^{+}\mathrm{f}^{+}\right) $ & .%
\end{tabular}
\label{x11}
\end{equation}%
The second relation of above eqs shows that $A_{\mu }$ can be decomposed like%
\begin{equation}
\begin{tabular}{llll}
$A_{\mu }$ & $=$ & $\vartheta _{\mu }+2\partial _{\mu }w$ & , \\ 
$\partial ^{++}\vartheta _{\mu }$ & $=$ & $0$ & ,%
\end{tabular}
\label{x12}
\end{equation}%
where 
\begin{equation}
\begin{tabular}{llll}
$w\left( x,u\right) $ & $=$ & $u_{(i}^{+}u_{j)}^{-}w^{\left( ij\right)
}\left( x\right) $ & ,%
\end{tabular}%
\end{equation}%
is as in eq(\ref{x5}). \newline
The extra term $\vartheta _{\mu }=\vartheta _{\mu }\left( x\right) $ is an
isosinglet; it is determined as follows:\newline
\emph{First} perform the change $\mathrm{B}_{\mu }^{-}=e^{\lambda w}\mathrm{C%
}_{\mu }^{-}$ to first bring eqs(\ref{x2}) into the simplest form%
\begin{equation}
\begin{tabular}{llll}
$\partial ^{++}\mathrm{C}_{\mu }^{-}-\lambda A_{\mu }\mathrm{f}^{+}$ & $=$ & 
$2\partial _{\mu }\mathrm{f}^{+}$ & , \\ 
$\partial ^{++}\mathrm{\tilde{C}}_{\mu }^{-}+\lambda A_{\mu }\mathrm{\tilde{f%
}}^{+}$ & $=$ & $2\partial _{\mu }\mathrm{\tilde{f}}^{+}$ & .%
\end{tabular}
\label{x13}
\end{equation}%
Substituting,%
\begin{equation}
\begin{tabular}{llll}
$A_{\mu }\mathrm{f}^{+}$ & $=$ & $\partial ^{++}\left( \vartheta _{\mu }%
\mathrm{f}^{-}+2\mathrm{f}^{+}\partial _{\mu }w^{--}\right) $ & , \\ 
$A_{\mu }\mathrm{\tilde{f}}^{+}$ & $=$ & $\partial ^{++}\left( \vartheta
_{\mu }\mathrm{\tilde{f}}^{-}+2\mathrm{\tilde{f}}^{+}\partial _{\mu
}w^{--}\right) $ & , \\ 
$\partial ^{++}w^{--}$ & $=$ & $w$ & , \\ 
$w^{--}$ & $=$ & $\frac{1}{2}\left( \mathrm{\tilde{f}}^{-}\mathrm{f}%
^{-}\right) $ & ,%
\end{tabular}
\label{x14}
\end{equation}%
we obtain,%
\begin{equation}
\begin{tabular}{llll}
$\mathrm{C}_{\mu }^{-}$ & $=$ & $2\partial _{\mu }\mathrm{f}^{-}+\lambda
\vartheta _{\mu }\mathrm{f}^{-}+2\lambda \mathrm{f}^{+}\left( \partial _{\mu
}\varphi ^{--}\right) $ & , \\ 
$\mathrm{\tilde{C}}_{\mu }^{-}$ & $=$ & $2\partial _{\mu }\mathrm{\tilde{f}}%
^{-}-\lambda \vartheta _{\mu }\mathrm{\tilde{f}}^{-}-2\lambda \mathrm{\tilde{%
f}}^{+}\left( \partial _{\mu }\varphi ^{--}\right) $ & .%
\end{tabular}
\label{x15}
\end{equation}%
Multiplying the first eq by $\mathrm{\tilde{f}}^{+}$ and the second by $%
\mathrm{f}^{+}$; we obtain%
\begin{equation}
\begin{tabular}{llll}
$\mathrm{\tilde{f}}^{+}\mathrm{C}_{\mu }^{-}$ & $=$ & $2\mathrm{\tilde{f}}%
^{+}\partial _{\mu }\mathrm{f}^{-}+\lambda \vartheta _{\mu }\mathrm{\tilde{f}%
}^{+}\mathrm{f}^{-}+2\lambda \mathrm{\tilde{f}}^{+}\mathrm{f}^{+}\left(
\partial _{\mu }w^{--}\right) $ & , \\ 
$\mathrm{f}^{+}\mathrm{\tilde{C}}_{\mu }^{-}$ & $=$ & $2\mathrm{f}%
^{+}\partial _{\mu }\mathrm{\tilde{f}}^{-}-\lambda \vartheta _{\mu }\mathrm{f%
}^{+}\mathrm{\tilde{f}}^{-}-2\lambda \mathrm{f}^{+}\mathrm{\tilde{f}}%
^{+}\left( \partial _{\mu }w^{--}\right) $ & .%
\end{tabular}
\label{x16}
\end{equation}%
Then adding both eqs, we get, 
\begin{equation}
\begin{tabular}{ll}
$\vartheta _{\mu }+2\partial _{\mu }w=2\left( \mathrm{\tilde{f}}^{+}\partial
_{\mu }\mathrm{f}^{-}+\mathrm{f}^{+}\partial _{\mu }\mathrm{\tilde{f}}%
^{-}\right) -\lambda \vartheta _{\mu }\left( \overline{\mathrm{f}}_{i}%
\mathrm{f}^{i}\right) $ & ,%
\end{tabular}
\label{x17}
\end{equation}%
from which we get the expression of $\vartheta _{\mu }$ (\ref{x7}).\newline
Notice that using $2w^{--}=\mathrm{\tilde{f}}^{-}\mathrm{f}^{-}$, we can
split the field $\mathrm{C}_{\mu }^{-}\left( x,u\right) $ (\ref{x15}) as the
sum of two irreducible components as follows%
\begin{equation}
\begin{tabular}{llll}
$\mathrm{C}_{\mu }^{-}\left( x,u\right) $ & $=$ & $u_{i}^{-}\mathcal{C}_{\mu
}^{i}\left( x\right) +u_{i}^{-}u_{j}^{-}u_{k}^{+}\mathcal{C}_{\mu
}^{(ijk)}\left( x\right) $ & ,%
\end{tabular}
\label{lk}
\end{equation}%
where%
\begin{equation}
\begin{tabular}{llll}
$\mathcal{C}_{\mu }^{i}$ & $=$ & $2\partial _{\mu }\mathrm{f}^{i}+\lambda
\vartheta _{\mu }\mathrm{f}^{i}+\frac{\lambda \rho ^{2}}{3}\partial _{\mu }%
\mathrm{f}^{i}+\frac{\lambda \overline{\mathrm{f}}^{i}\mathrm{f}^{k}}{3}%
\partial _{\mu }\mathrm{f}_{k}+\frac{\lambda \mathrm{f}^{i}\mathrm{f}^{k}}{3}%
\partial _{\mu }\overline{\mathrm{f}}_{k}$ & , \\ 
$\mathcal{C}_{\mu }^{(ijk)}$ & $=$ & $\lambda \mathrm{f}^{(i}\mathrm{f}%
^{j}\partial _{\mu }\overline{\mathrm{f}}^{k)}+\lambda \mathrm{f}^{(i}%
\overline{\mathrm{f}}^{j}\partial _{\mu }\mathrm{f}^{k)}$ & .%
\end{tabular}
\label{ci}
\end{equation}%
Similar relations can be written down for $\overline{\mathcal{C}}_{\mu i}$
and $\overline{\mathcal{C}}_{\mu (ijk)}$.

\emph{the Taub-NUT metric}\newline
By performing the integration of eq(\ref{tn}) with respect to the Grassmann
variables $\theta ^{+}$ and $\overline{\theta }^{+}$, we first get%
\begin{equation}
\begin{tabular}{llll}
$\mathcal{L}_{1}$ & $=$ & $\frac{1}{2}\int_{S^{2}}du\left( \mathrm{B}_{\mu
}^{-}\partial ^{\mu }\mathrm{\tilde{q}}^{+}-\mathrm{\tilde{B}}_{\mu
}^{-}\partial ^{\mu }\mathrm{q}^{+}\right) $ & .%
\end{tabular}%
\end{equation}%
Then using the solution (\ref{x4}), we can bring the above expression to%
\begin{equation}
\begin{tabular}{llll}
$\mathcal{L}_{1}$ & $=$ & $\frac{1}{2}\int_{S^{2}}du\left[ \left( \mathrm{C}%
_{\mu }^{-}\partial ^{\mu }\mathrm{\tilde{f}}^{+}-\mathrm{\tilde{C}}_{\mu
}^{-}\partial ^{\mu }\mathrm{f}^{+}\right) -\lambda A_{\mu }\partial ^{\mu }w%
\right] $ & ,%
\end{tabular}%
\end{equation}%
where we have used the identity $A_{\mu }=\mathrm{\tilde{f}}^{+}\mathrm{C}%
_{\mu }^{-}+\mathrm{\tilde{C}}_{\mu }^{-}\mathrm{f}^{+}$. \newline
Next integrating with respect to the harmonic variables u$^{\pm }$, we
obtain,%
\begin{equation}
\begin{tabular}{llll}
$\mathcal{L}_{1}$ & $=$ & $\frac{-1}{4}\left[ \left( \mathrm{C}_{\mu
}^{i}\partial ^{\mu }\overline{\mathrm{f}}_{i}-\overline{\mathrm{C}}_{\mu
}^{i}\partial ^{\mu }\mathrm{f}_{i}\right) +4\mathrm{a}\lambda \partial
_{\mu }\left( \overline{\mathrm{f}}_{(i}\mathrm{f}_{j)}\right) \partial
^{\mu }\left( \overline{\mathrm{f}}^{(i}\mathrm{f}^{j)}\right) \right] $ & ,%
\end{tabular}%
\end{equation}%
where $\mathrm{a}=\frac{1}{6}$; and where we have used%
\begin{equation}
\begin{tabular}{llll}
$\int_{S^{2}}du\left[ u^{+i}u^{-j}\right] $ & $=$ & $\frac{1}{2}\varepsilon
^{ij}$ & , \\ 
$\int_{S^{2}}du\left[ u^{+(i}u^{+j)}u_{(k}^{-}u_{l)}^{-}\right] $ & $=$ & $%
\frac{1}{3}\left( \delta _{k}^{i}\delta _{l}^{j}+\delta _{k}^{j}\delta
_{l}^{i}\right) $ & .%
\end{tabular}%
\end{equation}%
Substituting $\mathrm{C}_{\mu }^{i}$ by its expression (\ref{ci}), we get%
\begin{equation*}
\begin{tabular}{llll}
$\mathcal{L}_{1}$ & $=$ & $\frac{-1}{2}\left( \bar{g}_{ij}\partial _{\mu }%
\mathrm{f}^{i}\partial ^{\mu }\mathrm{f}^{j}+g^{_{ij}}\partial _{\mu }%
\overline{\mathrm{f}}_{i}\partial ^{\mu }\overline{\mathrm{f}}%
_{j}+2h_{i}^{j}\partial _{\mu }\mathrm{f}^{i}\partial ^{\mu }\overline{%
\mathrm{f}}_{j}\right) $ & ,%
\end{tabular}%
\end{equation*}%
with $\bar{g}_{ij}$, $g^{_{ij}}$ and $h_{i}^{j}$ as in eqs(\ref{tnm}).

\section{Appendix B: U$^{n}\left( 1\right) $ model, $n>1$}

In appendix B1, we determine the solution of eqs(\ref{del}-\ref{self}) and
in appendix B2, we derive the explicit expression on the component field of
the metric of the moduli space $\frac{SO\left( 4,20\right) }{SO\left(
4\right) \times SO\left( 20\right) }$.

\subsection{Solving the constraint eqs(\protect\ref{del}-\protect\ref{self})}

First we consider the solution of eq(\ref{del}). Then we deal with eq(\ref%
{self}).

\emph{Solving eq}(\ref{del})\newline
The solution of $\mathrm{q}^{+A}$, in terms of the fields doublets $\mathrm{f%
}^{\pm A}=u_{i}^{\pm }\mathrm{f}^{iA}$, $\mathrm{\tilde{f}}_{A}^{\pm
}=u_{i}^{\pm }\overline{\mathrm{f}}_{A}^{i}$ can be obtained by factorizing
it as follows%
\begin{equation}
\begin{tabular}{llll}
$\mathrm{q}^{+}$ & $=$ & $u_{i}^{\pm }\left( e^{\lambda w}\mathrm{f}%
^{i}\right) $ & , \\ 
$\mathrm{\tilde{q}}^{+}$ & $=$ & $u_{i}^{\pm }\left( \overline{\mathrm{f}}%
^{i}e^{-\lambda w}\right) $ & ,%
\end{tabular}%
\end{equation}%
where $w=w\left( \mathrm{f,\bar{f}}\right) $ is given by%
\begin{equation}
\begin{tabular}{llll}
$w$ & $=$ & $\dsum\limits_{I=1}^{n}w^{I}H_{I}$ & , \\ 
$w^{I}$ & $=$ & $\frac{1}{2}u_{k}^{+}u_{k}^{-}Tr\left( \overline{\mathrm{f}}%
^{k}H^{I}\mathrm{f}^{l}+\overline{\mathrm{f}}^{l}H^{I}\mathrm{f}^{k}\right) $
& .%
\end{tabular}
\label{wh}
\end{equation}%
Notice that $w^{I}$ has a particular dependence in the harmonic variables;
it captures an $SU_{R}\left( 2\right) $ isotriplet representation,%
\begin{equation}
\begin{tabular}{llll}
$w_{I}$ & $=$ & $\frac{1}{2}Tr\left( \overline{\mathrm{f}}^{+}H_{I}\mathrm{f}%
^{-}+\overline{\mathrm{f}}^{-}H_{I}\mathrm{f}^{+}\right) $ & .%
\end{tabular}%
\end{equation}%
Notice also%
\begin{equation}
\begin{tabular}{lllllll}
$\partial ^{++}w_{I}$ & $=$ & $w_{I}^{++}=Tr\left( \overline{\mathrm{f}}%
^{+}H_{I}\mathrm{f}^{+}\right) $ & , & $\left( \partial ^{++}\right)
^{2}w_{I}$ & $=0$ & , \\ 
$\partial ^{--}w_{I}$ & $=$ & $w_{I}^{--}=Tr\left( \overline{\mathrm{f}}%
^{-}H_{I}\mathrm{f}^{-}\right) $ & , & $\left( \partial ^{--}\right)
^{2}w_{I}$ & $=0$ & .%
\end{tabular}%
\end{equation}%
The solution of $\mathrm{q}^{+}=\mathrm{q}^{+}\left( x,u\right) $ in terms
of $\left( \mathrm{f}^{i},\overline{\mathrm{f}}_{i}\right) $ and the
harmonics u$_{i}^{\pm }$ reads therefore as%
\begin{equation}
\begin{tabular}{llll}
$\mathrm{q}^{+}$ & $=$ & $\left[ \exp \left( \frac{\lambda }{2}%
\sum\limits_{k,l=1}^{2}u_{(k}^{+}u_{l)}^{-}\sum\limits_{I=1}^{n}\left[
Tr\left( \overline{\mathrm{f}}^{k}H_{I}\mathrm{f}^{l}\right) \right]
H^{I}\right) \right] \mathrm{f}^{i}u_{i}^{+}$ & , \\ 
$\mathrm{\tilde{q}}^{+}$ & $=$ & $\left[ \exp \left( \frac{-\lambda }{2}%
\sum\limits_{k,l=1}^{2}u_{(k}^{+}u_{l)}^{-}\sum\limits_{I=1}^{n}\left[
Tr\left( \overline{\mathrm{f}}^{k}H_{I}\mathrm{f}^{l}\right) \right]
H^{I}\right) \right] \overline{\mathrm{f}}^{i}u_{i}^{+}$ & .%
\end{tabular}
\label{524}
\end{equation}%
In the limit $\lambda \longrightarrow 0$, we recover the free fields $%
\mathrm{q}^{+}=u_{i}^{+}\mathrm{f}^{i}\left( x\right) $.

\emph{Solving eq}(\ref{self})\newline
To get the solution of eq(\ref{self}), we need several steps:\newline
(\textbf{i}) \emph{Step 1}: We use the $U^{n}\left( 1\right) $ symmetry of
the Lagrangian density and the equations of motion to make the change of
field variables 
\begin{equation}
\begin{tabular}{llll}
$\mathrm{B}_{\mu }^{-}$ & $=$ & $e^{\lambda w}\mathrm{C}_{\mu }^{-}$ & , \\ 
$\mathrm{\tilde{B}}_{\mu }^{-}$ & $=$ & $e^{-\lambda w}\mathrm{\tilde{C}}%
_{\mu }^{-}$ & ,%
\end{tabular}
\label{525}
\end{equation}%
where $w$ is as in eq(\ref{wh}) and where $\mathrm{C}_{\mu }^{-}$ is the new
field to determine. This change of field variables allows us to bring eq(\ref%
{self}) into the form%
\begin{eqnarray}
\partial ^{++}\mathrm{C}_{\mu }^{-}-\lambda A_{\mu }\mathrm{f}^{+}
&=&2\nabla _{\mu }\mathrm{f}^{+},  \notag \\
\partial ^{++}\mathrm{C}_{\mu }^{-}+\lambda A_{\mu }\overline{\mathrm{f}}%
^{+} &=&2\overline{\nabla }_{\mu }\overline{\mathrm{f}}^{+},  \label{cp}
\end{eqnarray}%
where we have set%
\begin{equation}
\begin{tabular}{llll}
$\nabla _{\mu }\mathrm{f}^{+}$ & $=$ & $\left[ \partial _{\mu }+\lambda
\left( \partial _{\mu }w\right) \right] \mathrm{f}^{+}$ & , \\ 
$\overline{\nabla }_{\mu }\overline{\mathrm{f}}^{+}$ & $=$ & $\left[
\partial _{\mu }-\lambda \left( \partial _{\mu }w\right) \right] \overline{%
\mathrm{f}}^{+}$ & .%
\end{tabular}%
\end{equation}%
We also have 
\begin{equation}
\begin{tabular}{llll}
$A_{\mu }$ & $=$ & $\sum_{I=1}^{n}A_{\mu }^{I}H_{I}$ & , \\ 
$A_{\mu }^{I}$ & $=$ & $\mathrm{\tilde{C}}_{\mu }^{-}H^{I}\mathrm{f}^{+}+%
\mathrm{\tilde{f}}^{+}H^{I}\mathrm{C}_{\mu }^{-}$ & ,%
\end{tabular}
\label{jc}
\end{equation}%
which satisfy%
\begin{equation}
\begin{tabular}{llll}
$\partial ^{++}A_{\mu }^{I}$ & $=$ & $2\left[ \partial _{\mu }\left( \mathrm{%
\tilde{f}}^{+}H^{I}\mathrm{f}^{+}\right) \right] $ & , \\ 
& $=$ & $2\partial ^{++}\left( \partial _{\mu }w^{I}\right) $ & .%
\end{tabular}
\label{dp}
\end{equation}%
Eq(\ref{dp}) implies in turns $\left( \partial ^{++}\right) ^{2}A_{\mu
}^{I}=0;$ and so can be solved as follows,%
\begin{equation}
A_{\mu }^{I}=\vartheta _{\mu }^{I}+2\partial _{\mu }w^{I}.  \label{at}
\end{equation}%
The $\vartheta _{\mu }^{I}$'s are isosinglets, 
\begin{equation}
\begin{tabular}{llll}
$\partial ^{++}\vartheta _{\mu }^{I}$ & $=$ & $0$ & , \\ 
$\vartheta _{\mu }$ & $=$ & $\sum_{I=1}^{n}\vartheta _{\mu }^{I}H_{I}$ & ,%
\end{tabular}%
\end{equation}%
they will be determined in terms of the dynamical scalars $\mathrm{f}^{iA}$.%
\newline
(ii) \emph{Step 2}: We use the identity%
\begin{equation}
\lambda A_{\mu }\mathrm{f}^{+}=\lambda \partial ^{++}\left( \vartheta _{\mu }%
\mathrm{f}^{-}+\partial _{\mu }w^{--}\mathrm{f}^{+}\right)
\end{equation}%
to solve eq(\ref{cp}) like,%
\begin{equation}
\begin{tabular}{llll}
$\mathrm{C}_{\mu }^{-}$ & $=$ & $2\partial _{\mu }\mathrm{f}^{-}+\lambda
\vartheta _{\mu }\mathrm{f}^{-}+\lambda \left( \partial _{\mu }w^{--}\right) 
\mathrm{f}^{+}$ & , \\ 
$\mathrm{\tilde{C}}_{\mu }^{-}$ & $=$ & $2\partial _{\mu }\mathrm{\tilde{f}}%
^{-}-\lambda \vartheta _{\mu }\mathrm{\tilde{f}}^{-}-\lambda \left( \partial
_{\mu }w^{--}\right) \mathrm{\tilde{f}}^{+}$ & .%
\end{tabular}
\label{pa}
\end{equation}%
To determine $\vartheta _{\mu }$, we compute $\left( \mathrm{\tilde{f}}%
^{+}H^{I}\mathrm{C}_{\mu }^{-}+\mathrm{\tilde{C}}_{\mu }^{-}H^{I}\mathrm{f}%
^{+}\right) $ by using eqs(\ref{pa}) and derive a constraint equation that
allows us to fix $\vartheta _{\mu }$. We have%
\begin{equation}
\begin{tabular}{llll}
$\mathrm{\tilde{f}}^{+}H^{I}\mathrm{C}_{\mu }^{-}$ & $=$ & $2\mathrm{\tilde{f%
}}^{+}H^{I}\partial _{\mu }\mathrm{f}^{-}+\lambda \vartheta _{\mu J}\left( 
\mathrm{\tilde{f}}^{+}H^{I}H^{J}\mathrm{f}^{-}\right) +2\lambda \partial
_{\mu }w_{J}^{--}\left( \mathrm{\tilde{f}}^{+}H^{I}H^{J}\mathrm{f}%
^{+}\right) $ & , \\ 
$\mathrm{\tilde{C}}_{\mu }^{-}H^{I}\mathrm{f}^{+}$ & $=$ & $2\partial _{\mu }%
\mathrm{\tilde{f}}^{-}H^{I}\mathrm{f}^{+}-\lambda \vartheta _{\mu J}\left( 
\mathrm{\tilde{f}}^{-}H^{J}H^{I}\mathrm{f}^{+}\right) -2\lambda \partial
_{\mu }w_{J}^{--}\left( \mathrm{\tilde{f}}^{+}H^{J}H^{I}\mathrm{f}%
^{+}\right) $ & .%
\end{tabular}
\label{nv}
\end{equation}%
Before going ahead, it is convenient to simplify a little bit the above
relations by using the following conventional notations: 
\begin{equation}
\begin{tabular}{llllll}
$\mathrm{Q}_{I}^{\pm A}$ & $=$ & $u_{i}^{\pm }\mathrm{Q}_{I}^{iA}$ & $\equiv 
$ & $\left( \mathrm{f}^{\pm }H_{I}\right) ^{A}$ & , \\ 
$\mathrm{\tilde{Q}}_{B}^{\pm I}$ & $=$ & $u_{i}^{\pm }\overline{\mathrm{Q}}%
_{B}^{iI}$ & $\equiv $ & $\left( \mathrm{\tilde{f}}^{\pm }H^{I}\right) _{B}$
& ,%
\end{tabular}
\label{qq}
\end{equation}%
with 
\begin{equation}
\begin{tabular}{llll}
$\mathrm{Q}_{I}^{iA}$ & $=$ & $\left( H_{I}\right) _{C}^{A}\mathrm{f}^{iC}$
& , \\ 
$\overline{\mathrm{Q}}_{jB}^{I}$ & $=$ & $\overline{\mathrm{f}}_{jD}\left(
H^{I}\right) _{B}^{D}$ & , \\ 
$R_{B}^{A}$ & $=$ & $\overline{\mathrm{Q}}_{iB}^{I}\mathrm{Q}_{I}^{iA}$ & .%
\end{tabular}
\label{qqb}
\end{equation}%
From these fields, one can build the following quantities%
\begin{equation}
\begin{tabular}{llll}
$\overline{\mathrm{Q}}_{iB}^{I}\mathrm{Q}_{I}^{iA}$ & , & $\mathrm{Q}%
_{I}^{iA}\overline{\mathrm{Q}}_{jB}^{J}$ & , \\ 
$\mathrm{Q}_{I}^{iA}\mathrm{Q}_{J}^{kC}$ & , & $\overline{\mathrm{Q}}%
_{jB}^{I}\overline{\mathrm{Q}}_{lD}^{J}$ & ,%
\end{tabular}
\label{Q}
\end{equation}%
Notice that for $n=1$, we have $H_{1}=I$, and the fields $\mathrm{Q}%
_{I}^{iA} $\ reduce to $\mathrm{f}^{i}$ and eqs(\ref{Q}) to%
\begin{equation}
\begin{tabular}{llll}
$\overline{\mathrm{Q}}_{iB}^{I}\mathrm{Q}_{I}^{iA}\rightarrow \overline{%
\mathrm{f}}_{i}\mathrm{f}^{i}$ & , & $\mathrm{Q}_{I}^{iA}\overline{\mathrm{Q}%
}_{jB}^{J}\rightarrow \mathrm{f}^{i}\overline{\mathrm{f}}_{j}$ & , \\ 
$\mathrm{Q}_{I}^{iA}\mathrm{Q}_{J}^{kC}\rightarrow \mathrm{f}^{i}\mathrm{f}%
^{k}$ & , & $\overline{\mathrm{Q}}_{jB}^{I}\overline{\mathrm{Q}}%
_{lD}^{J}\rightarrow \overline{\mathrm{f}}_{j}\overline{\mathrm{f}}_{l}$ & .%
\end{tabular}
\label{QQ}
\end{equation}%
Using these new field moduli, we can rewrite eqs(\ref{nv}) like,%
\begin{equation}
\begin{tabular}{llll}
$\mathrm{\tilde{Q}}^{+I}\mathrm{C}_{\mu }^{-}$ & $=$ & $2\mathrm{\tilde{Q}}%
^{+I}\partial _{\mu }\mathrm{f}^{-}+\lambda \vartheta _{\mu J}\left( \mathrm{%
\tilde{Q}}^{+I}\mathrm{Q}^{-J}\right) +\lambda \partial _{\mu
}w_{J}^{--}\left( \mathrm{\tilde{Q}}^{+I}\mathrm{Q}^{+J}\right) $ & , \\ 
$\mathrm{\tilde{C}}_{\mu }^{-}\mathrm{Q}^{+I}$ & $=$ & $2\partial _{\mu }%
\mathrm{\tilde{f}}^{-}\mathrm{Q}^{+I}-\lambda \vartheta _{\mu J}\left( 
\mathrm{\tilde{Q}}^{-J}\mathrm{Q}^{+I}\right) -2\lambda \partial _{\mu
}\varphi _{J}^{--}\left( \mathrm{\tilde{Q}}^{+J}\mathrm{Q}^{+I}\right) $ & .%
\end{tabular}%
\end{equation}%
Next, adding these two relations and using eq(\ref{jc}), we end with 
\begin{equation}
\begin{tabular}{llll}
$\vartheta _{\mu }^{I}$ & $=$ & $\left( \mathrm{Q}^{iAI}\partial _{\mu }%
\overline{\mathrm{f}}_{iA}-\overline{\mathrm{Q}}_{iA}^{I}\partial _{\mu }%
\mathrm{f}^{iA}\right) -\lambda \vartheta _{\mu }^{J}\left( \overline{%
\mathrm{Q}}_{iAJ}\mathrm{Q}^{iAI}\right) $ & ,%
\end{tabular}%
\end{equation}%
which can be also rewritten as 
\begin{equation}
\mathcal{E}_{J}^{I}\vartheta _{\mu }^{J}=\upsilon _{\mu }^{I},  \label{t1}
\end{equation}%
with%
\begin{equation}
\begin{tabular}{llll}
$\upsilon _{\mu }^{I}$ & $=$ & $\left( \mathrm{Q}^{iAI}\partial _{\mu }%
\overline{\mathrm{f}}_{iA}-\overline{\mathrm{Q}}_{iA}^{I}\partial _{\mu }%
\mathrm{f}^{iA}\right) $ & , \\ 
$\mathcal{E}_{J}^{I}$ & $=$ & $\left[ \delta _{J}^{I}+\lambda \overline{%
\mathrm{Q}}_{iAJ}\mathrm{Q}^{iAI}\right] $ & .%
\end{tabular}
\label{t2}
\end{equation}%
Using eq(\ref{qq}), these relations can be also put in the equivalent form%
\begin{equation}
\begin{tabular}{llll}
$\upsilon _{\mu }^{I}$ & $=$ & $\left( \mathrm{f}^{i}H^{I}\partial _{\mu }%
\overline{\mathrm{f}}_{i}-\overline{\mathrm{f}}_{i}H^{I}\partial _{\mu }%
\mathrm{f}^{i}\right) $ & , \\ 
$\mathcal{E}_{J}^{I}$ & $=$ & $\left[ \delta _{J}^{I}+\lambda \overline{%
\mathrm{Q}}_{iAJ}\mathrm{Q}^{iAI}\right] $ & .%
\end{tabular}%
\end{equation}%
Thus, the solution of $\vartheta _{\mu }^{I}$ reads as, 
\begin{equation}
\begin{tabular}{llll}
$\vartheta _{\mu }^{J}=\mathcal{F}_{I}^{J}v_{\mu }^{I}$ & , & $\mathcal{F}%
_{I}^{J}\mathcal{E}_{K}^{I}=\delta _{K}^{I}$ & .%
\end{tabular}
\label{t3}
\end{equation}%
Notice that for $n=1$, eqs(\ref{t2}-\ref{t3}) reduce to%
\begin{equation}
\begin{tabular}{llll}
$\upsilon _{\mu }^{I}$ & $\rightarrow $ & $\upsilon _{\mu }=\left( \mathrm{f}%
^{i}\partial _{\mu }\overline{\mathrm{f}}_{i}-\overline{\mathrm{f}}%
_{i}\partial _{\mu }\mathrm{f}^{i}\right) $ & , \\ 
$\mathcal{E}_{J}^{I}$ & $\rightarrow $ & $\mathcal{E}=\left[ 1+\lambda 
\overline{\mathrm{f}}_{i}\mathrm{f}^{i}\right] $ & , \\ 
$\mathcal{F}_{I}^{J}$ & $\rightarrow $ & $\mathcal{F}=\frac{1}{\left[
1+\lambda \overline{\mathrm{f}}_{i}\mathrm{f}^{i}\right] }$ & , \\ 
$\mathcal{EF}$ & $=$ & $1$ & .%
\end{tabular}%
\end{equation}%
Notice moreover that because of the property,%
\begin{equation}
\begin{tabular}{llll}
$\overline{\mathrm{f}}_{i}H_{J}H^{I}\mathrm{f}^{i}$ & $=$ & $\overline{%
\mathrm{f}}_{i}H^{I}H_{J}\mathrm{f}^{i}$ & ,%
\end{tabular}%
\end{equation}%
we have the identity%
\begin{equation}
\begin{tabular}{llll}
$\overline{\mathrm{Q}}_{iAJ}\mathrm{Q}^{iAI}$ & $=$ & $\overline{\mathrm{Q}}%
_{iA}^{I}\mathrm{Q}_{J}^{iA}$ & .%
\end{tabular}%
\end{equation}%
The solution $\mathrm{C}_{\mu }^{-A}\left( x,u\right) $ and $\mathrm{\tilde{C%
}}_{\mu B}^{-}\left( x,u\right) $ read, in terms of $\mathrm{Q}_{J}^{\pm }$,
as follows:%
\begin{equation}
\begin{tabular}{lll}
$\mathrm{C}_{\mu }^{-A}=$ & $2\partial _{\mu }\mathrm{f}^{-A}+\lambda 
\mathcal{F}_{I}^{J}\upsilon _{\mu }^{I}\mathrm{Q}_{J}^{-A}+\lambda \mathrm{Q}%
_{J}^{+A}\mathrm{\tilde{Q}}_{B}^{-J}\left( \partial _{\mu }\mathrm{f}%
^{-B}\right) $ &  \\ 
& $+\lambda \mathrm{Q}_{J}^{+A}\mathrm{Q}^{-BJ}\left( \partial _{\mu }%
\mathrm{\tilde{f}}_{B}^{-}\right) $ & ,%
\end{tabular}
\label{ca}
\end{equation}%
\begin{equation}
\begin{tabular}{lll}
$\mathrm{\tilde{C}}_{\mu A}^{-}=$ & $2\partial _{\mu }\mathrm{\tilde{f}}%
_{A}^{-}-\lambda \mathcal{F}_{I}^{J}\upsilon _{\mu }^{I}\mathrm{\tilde{Q}}%
_{AJ}^{-}-\lambda \mathrm{\tilde{Q}}_{AJ}^{+}\mathrm{Q}^{-BJ}\left( \partial
_{\mu }\mathrm{\tilde{f}}_{B}^{-}\right) $ &  \\ 
& $-\lambda \mathrm{\tilde{Q}}_{AJ}^{+}\mathrm{\tilde{Q}}_{B}^{-J}\left(
\partial _{\mu }\mathrm{f}^{-B}\right) $ & .%
\end{tabular}
\label{cb}
\end{equation}%
Like in eq(\ref{lk}), these fields obey $\left( \partial ^{--}\right) ^{2}%
\mathrm{C}_{\mu }^{-A}=0$; they can be then decomposed in quite similar
manner like 
\begin{equation}
\begin{tabular}{llll}
$\mathrm{C}_{\mu }^{-A}\left( x,u\right) $ & $=$ & $u_{i}^{-}\mathcal{C}%
_{\mu }^{iA}\left( x\right) +u_{(i}^{-}u_{j}^{-}u_{k)}^{+}\mathcal{C}_{\mu
}^{\left( ijk\right) A}\left( x\right) $ & ,%
\end{tabular}%
\end{equation}%
with%
\begin{equation}
\begin{tabular}{lll}
$\mathcal{C}_{\mu }^{iA}=$ & $2\partial _{\mu }\mathrm{f}^{iA}+\lambda 
\mathcal{F}_{I}^{J}\upsilon _{\mu }^{I}\mathrm{Q}_{J}^{iA}$ &  \\ 
& $+\frac{\lambda }{3}\mathrm{Q}_{J}^{jA}\overline{\mathrm{Q}}%
_{jB}^{J}\left( \partial _{\mu }\mathrm{f}^{iB}\right) +\frac{\lambda }{3}%
\overline{\mathrm{Q}}_{B}^{iJ}\mathrm{Q}_{J}^{jA}\left( \partial _{\mu }%
\mathrm{f}_{j}^{B}\right) $ &  \\ 
& $+\frac{\lambda }{3}\mathrm{Q}_{J}^{jA}\mathrm{Q}_{j}^{BJ}\left( \partial
_{\mu }\overline{\mathrm{f}}_{B}^{i}\right) +\frac{\lambda }{3}\mathrm{Q}%
^{iBJ}\mathrm{Q}_{J}^{jA}\left( \partial _{\mu }\overline{\mathrm{f}}%
_{jB}\right) $ & ,%
\end{tabular}%
\end{equation}%
and%
\begin{equation}
\begin{tabular}{llll}
$\overline{\mathcal{C}}_{\mu iA}$ & $=$ & $2\partial _{\mu }\overline{%
\mathrm{f}}_{iA}-\lambda \mathcal{F}_{I}^{J}\upsilon _{\mu }^{I}\overline{%
\mathrm{Q}}_{iAJ}$ &  \\ 
&  & $-\frac{\lambda }{3}\overline{\mathrm{Q}}_{AJ}^{j}\mathrm{Q}%
_{j}^{BJ}\left( \partial _{\mu }\overline{\mathrm{f}}_{iB}\right) -\frac{%
\lambda }{3}\mathrm{Q}_{i}^{BJ}\overline{\mathrm{Q}}_{AJ}^{j}\left( \partial
_{\mu }\overline{\mathrm{f}}_{jB}\right) $ &  \\ 
&  & $-\frac{\lambda }{3}\overline{\mathrm{Q}}_{AJ}^{j}\overline{\mathrm{Q}}%
_{jB}^{J}\left( \partial _{\mu }\mathrm{f}_{i}^{B}\right) -\frac{\lambda }{3}%
\overline{\mathrm{Q}}_{iB}^{J}\overline{\mathrm{Q}}_{AJ}^{j}\left( \partial
_{\mu }\mathrm{f}_{j}^{B}\right) $ & .%
\end{tabular}%
\end{equation}%
Similar relations may be written down for $\mathcal{C}_{\mu }^{\left(
ijk\right) A}$ and $\overline{\mathcal{C}}_{\mu \left( ijk\right) A}$.

\subsection{Deriving the metric (\protect\ref{mc})}

Performing the integration of eq(\ref{lan}) with respect to the Grassmann
variables $\theta ^{+}$ and $\overline{\theta }^{+}$, we obtain the
following action,%
\begin{equation}
\mathcal{S}_{n}=\frac{1}{2}\int d^{4}x\left[ \int_{S^{2}}du\left( \mathrm{B}%
_{\mu }^{-A}\partial ^{\mu }\mathrm{\tilde{q}}_{A}^{+}-\mathrm{\tilde{B}}%
_{\mu A}^{-}\partial ^{\mu }\mathrm{q}^{+A}\right) \right] .  \label{ty}
\end{equation}%
To get the space time field action,%
\begin{equation}
\mathcal{S}_{n}=\frac{1}{2}\int d^{4}x\left( 2h_{iA}^{jB}\partial _{\mu }%
\mathrm{f}^{iA}\partial ^{\mu }\overline{\mathrm{f}}_{jB}+\overline{g}%
_{iAjB}\partial _{\mu }\mathrm{f}^{iA}\partial ^{\mu }\mathrm{f}%
^{jB}+g^{_{iAjB}}\partial _{\mu }\overline{\mathrm{f}}_{iA}\partial ^{\mu }%
\overline{\mathrm{f}}_{jB}\right) \text{ },
\end{equation}%
we have to integrate with respect the harmonic variables $u^{\pm }$. \newline
To that purpose, we start by substituting $\mathrm{B}_{\mu }^{-A}$ and $%
\mathrm{q}^{+A}$ by of their expressions in terms of $\mathrm{C}_{\mu }^{-A}$
and $\mathrm{f}^{+A}$ (\ref{524}-\ref{525},\ref{ca}). Doing this, we can
bring $\mathcal{S}_{n}$ to the form%
\begin{equation}
\mathcal{S}=\frac{1}{2}\int d^{4}x\left[ L_{n1}\left( x\right) +L_{n2}\left(
x\right) \right]
\end{equation}%
where we have set%
\begin{equation}
\begin{tabular}{llll}
$L_{n1}\left( x\right) $ & $=$ & $\int_{S^{2}}du\left( \mathrm{C}_{\mu
}^{-A}\partial ^{\mu }\mathrm{\tilde{f}}_{A}^{+}-\mathrm{\tilde{C}}_{\mu
A}^{-}\partial ^{\mu }\mathrm{f}^{+A}\right) $ & , \\ 
$L_{n2}\left( x\right) $ & $=$ & $-\lambda \int_{S^{2}}du\left[ \left( 
\mathrm{\tilde{Q}}_{IA}^{+}\mathrm{C}_{\mu }^{-A}+\mathrm{\tilde{C}}_{\mu
A}^{-}\mathrm{Q}_{I}^{+A}\right) \left( \partial ^{\mu }w^{I}\right) \right] 
$ & ,%
\end{tabular}
\label{su}
\end{equation}%
with $w^{I}$ as in eq(\ref{x5}). Notice that using eq(\ref{jc}), we also have%
\begin{equation}
L_{n2}=-\lambda \int_{S^{2}}du\left( \sum_{I=1}^{n}A_{\mu I}\partial ^{\mu
}w^{I}\right) .  \label{l2}
\end{equation}%
As the integration with respect to the harmonic variables is technical, let
us give details regarding the explicit calculations of $L_{n1}$ and $L_{n2}$.

\textbf{(1) Computing} $L_{n1}$\newline
Substituting $\mathrm{C}_{\mu }^{-A}$ and $\mathrm{\tilde{C}}_{\mu A}^{-}$
by their expressions (\ref{ca}-\ref{cb}) in terms of the dynamical fields $%
\mathrm{f}_{A}^{\pm }$ and $\mathrm{\tilde{f}}_{A}^{\pm }$, we can determine 
$L_{n1}$. The calculations are lengthy, we shall then proceed by steps.
Setting%
\begin{equation}
\begin{tabular}{llll}
$\mathcal{A}_{1}=\mathrm{C}_{\mu }^{-A}\partial ^{\mu }\mathrm{\tilde{f}}%
_{A}^{+}$ & , & $\widetilde{\mathcal{A}}_{1}=\mathrm{\tilde{C}}_{\mu
A}^{-}\partial ^{\mu }\mathrm{f}^{+A}$ & ,%
\end{tabular}
\label{a1a}
\end{equation}%
we first compute their explicit expression in terms of the dynamical fields $%
\mathrm{f}^{\pm }$ and $\mathrm{\tilde{f}}^{\pm },$%
\begin{equation}
\begin{tabular}{llll}
$\mathcal{A}_{1}=\mathcal{A}_{1}\left( \mathrm{f}^{\pm },\mathrm{\tilde{f}}%
^{\pm }\right) $ & , & $\widetilde{\mathcal{A}}_{1}=\widetilde{\mathcal{A}}%
_{1}\left( \mathrm{f}^{\pm },\mathrm{\tilde{f}}^{\pm }\right) $ & .%
\end{tabular}%
\end{equation}%
Then we integrate with respect to the harmonic variables.

\textbf{(i)} \emph{Computing} $\mathcal{A}_{1}$ and $\widetilde{\mathcal{A}}%
_{1}$\newline
Putting eqs(\ref{ca}-\ref{cb}) back into eqs(\ref{a1a}), we obtain 
\begin{equation}
\begin{tabular}{llll}
$\mathcal{A}_{1}$ & $=$ & $2\partial _{\mu }\mathrm{f}^{-A}\partial ^{\mu }%
\mathrm{\tilde{f}}_{A}^{+}+\lambda \mathcal{F}_{I}^{J}\upsilon _{\mu }^{I}%
\mathrm{Q}_{J}^{-A}\partial ^{\mu }\mathrm{\tilde{f}}_{A}^{+}$ &  \\ 
&  & $+\lambda \mathrm{Q}_{J}^{+A}\text{ }\mathrm{\tilde{Q}}%
_{B}^{-J}\partial _{\mu }\mathrm{f}^{-B}\partial ^{\mu }\mathrm{\tilde{f}}%
_{A}^{+}+\lambda \mathrm{Q}_{J}^{+A}\mathrm{Q}^{-BJ}\partial _{\mu }\mathrm{%
\tilde{f}}_{B}^{-}\partial ^{\mu }\mathrm{\tilde{f}}_{A}^{+}$ & ,%
\end{tabular}%
\end{equation}%
and 
\begin{equation}
\begin{tabular}{llll}
$\widetilde{\mathcal{A}}_{1}$ & $=$ & $2\partial _{\mu }\widetilde{\mathrm{f}%
}_{A}^{-}\partial ^{\mu }\mathrm{f}^{+A}-\lambda \mathcal{F}_{I}^{J}\upsilon
_{\mu }^{I}\widetilde{\mathrm{Q}}_{AJ}^{-}\partial ^{\mu }\mathrm{f}^{+A}$ & 
\\ 
&  & $-\lambda \widetilde{\mathrm{Q}}_{AJ}^{+}\mathrm{Q}^{-BJ}\partial _{\mu
}\widetilde{\mathrm{f}}_{B}^{-}\partial ^{\mu }\mathrm{f}^{+A}-\lambda 
\widetilde{\mathrm{Q}}_{AJ}^{+}\widetilde{\mathrm{Q}}_{B}^{-J}\partial _{\mu
}\mathrm{f}^{-B}\partial ^{\mu }\mathrm{f}^{+A}$ & .%
\end{tabular}%
\end{equation}%
In these relations, $\mathcal{F}_{I}^{J}$ and $\upsilon _{\mu }^{I}$ are
given by (\ref{t1}-\ref{t3}) and $\mathrm{\tilde{Q}}_{AJ}^{+}$ and $\mathrm{Q%
}^{-BJ}$ are as in eqs(\ref{qq}-\ref{Q}).

\textbf{(ii)} \emph{Integration over }$S^{2}$\emph{\ }\newline
The integration of the above eqs with respect to the harmonic variables
gives 
\begin{equation}
\begin{tabular}{llll}
$\int_{S^{2}}du\mathcal{A}_{1}=\mathcal{C}_{\mu }^{iA}\partial ^{\mu }%
\overline{\mathrm{f}}_{iA}$ & , & $\int_{S^{2}}du\widetilde{\mathcal{A}}_{1}=%
\overline{\mathcal{C}}_{\mu iA}\partial ^{\mu }\mathrm{f}^{iA}$ & ,%
\end{tabular}%
\end{equation}%
where%
\begin{equation*}
\begin{tabular}{lll}
$\mathcal{C}_{\mu }^{iA}\partial ^{\mu }\overline{\mathrm{f}}_{iA}=$ & $%
-\partial _{\mu }\mathrm{f}^{iA}\partial ^{\mu }\overline{\mathrm{f}}_{iA}-%
\frac{\lambda }{2}\mathcal{F}_{I}^{J}\upsilon _{\mu }^{I}\mathrm{Q}%
_{J}^{iA}\partial ^{\mu }\overline{\mathrm{f}}_{iA}$ &  \\ 
& $+\mathrm{\xi }\lambda \widetilde{\mathrm{Q}}_{(iB}^{J}\partial _{\mu }%
\mathrm{f}_{j)}^{B}\mathrm{Q}_{J}^{(iA}\partial ^{\mu }\overline{\mathrm{f}}%
_{A}^{j)}$ &  \\ 
& $+\mathrm{\xi }\lambda \mathrm{Q}_{(i}^{BJ}\partial _{\mu }\overline{%
\mathrm{f}}_{j)B}\mathrm{Q}_{J}^{(iA}\partial ^{\mu }\overline{\mathrm{f}}%
_{A}^{j)}$ & ,%
\end{tabular}%
\end{equation*}%
and%
\begin{equation}
\begin{tabular}{lll}
$\overline{\mathcal{C}}_{\mu iA}\partial ^{\mu }\mathrm{f}^{iA}=$ & $%
\partial _{\mu }\overline{\mathrm{f}}_{iA}\partial ^{\mu }\mathrm{f}^{iA}-%
\frac{\lambda }{2}\mathcal{F}_{I}^{J}\upsilon _{\mu }^{I}\overline{\mathrm{Q}%
}_{iAJ}\partial ^{\mu }\mathrm{f}^{iA}$ &  \\ 
& $-\mathrm{\xi }\lambda \mathrm{Q}_{(i}^{BJ}\partial _{\mu }\overline{%
\mathrm{f}}_{j)B}\overline{\mathrm{Q}}_{AJ}^{(i}\partial ^{\mu }\mathrm{f}%
^{j)A}$ &  \\ 
& $-\mathrm{\xi }\lambda \overline{\mathrm{Q}}_{(iB}^{J}\partial _{\mu }%
\mathrm{f}_{j)}^{B}\overline{\mathrm{Q}}_{AJ}^{(i}\partial ^{\mu }\mathrm{f}%
^{j)A}$ & .%
\end{tabular}%
\end{equation}%
For later use it is also interesting to rewrite these relations as follows:%
\begin{equation}
\begin{tabular}{lll}
$\mathcal{C}_{\mu }^{iA}\partial ^{\mu }\overline{\mathrm{f}}_{iA}=$ & $%
-\partial _{\mu }\mathrm{f}^{iA}\partial ^{\mu }\overline{\mathrm{f}}_{iA}-%
\frac{\lambda }{2}\left( \mathcal{F}_{I}^{J}\mathrm{Q}_{J}^{iA}\right)
\upsilon _{\mu }^{I}\partial ^{\mu }\overline{\mathrm{f}}_{iA}$ &  \\ 
& $+2\mathrm{\xi }\lambda \left[ \left( \overline{\mathrm{Q}}_{C}^{lJ}%
\mathrm{Q}_{kJ}^{D}\right) -\left( \overline{\mathrm{Q}}_{iC}^{J}\mathrm{Q}%
_{J}^{iD}\right) \delta _{k}^{l}\right] \partial _{\mu }\mathrm{f}%
^{kC}\partial ^{\mu }\overline{\mathrm{f}}_{lD}$ &  \\ 
& $+2\mathrm{\xi }\lambda \left[ \left( \mathrm{Q}_{i}^{CJ}\mathrm{Q}%
_{J}^{iD}\right) \varepsilon ^{kl}-\left( \mathrm{Q}^{lCJ}\mathrm{Q}%
_{J}^{kD}\right) \right] \partial _{\mu }\overline{\mathrm{f}}_{kC}\partial
^{\mu }\overline{\mathrm{f}}_{lD}$ & .%
\end{tabular}%
\end{equation}%
with $\mathrm{\xi =}\frac{1}{8}$, and%
\begin{equation}
\begin{tabular}{lll}
$\overline{\mathcal{C}}_{\mu iA}\partial ^{\mu }\mathrm{f}^{iA}=$ & $%
\partial _{\mu }\overline{\mathrm{f}}_{iA}\partial ^{\mu }\mathrm{f}^{iA}-%
\frac{\lambda }{2}\mathcal{F}_{I}^{J}\upsilon _{\mu }^{I}\overline{\mathrm{Q}%
}_{iAJ}\partial ^{\mu }\mathrm{f}^{iA}$ &  \\ 
& $-2\mathrm{\xi }\lambda \left[ \left( \mathrm{Q}_{k}^{DJ}\overline{\mathrm{%
Q}}_{CJ}^{l}\right) +\left( \mathrm{Q}_{i}^{DJ}\overline{\mathrm{Q}}%
_{CJ}^{i}\right) \delta _{k}^{l}\right] \partial ^{\mu }\mathrm{f}%
^{kC}\partial _{\mu }\overline{\mathrm{f}}_{lD}$ &  \\ 
& $+2\mathrm{\xi }\lambda \left[ \left( \overline{\mathrm{Q}}_{kD}^{J}%
\overline{\mathrm{Q}}_{lCJ}\right) -\left( \overline{\mathrm{Q}}_{iD}^{J}%
\overline{\mathrm{Q}}_{CJ}^{i}\right) \varepsilon _{kl}\right] \partial
_{\mu }\mathrm{f}^{lD}\partial ^{\mu }\mathrm{f}^{kC}$ & .%
\end{tabular}%
\end{equation}%
Subtracting the two terms as in (\ref{su}), we obtain%
\begin{equation}
\begin{tabular}{lll}
$L_{n1}=$ & $-2\partial _{\mu }\mathrm{f}^{iA}\partial ^{\mu }\overline{%
\mathrm{f}}_{iA}-\frac{\lambda }{2}\mathcal{F}_{I}^{J}\upsilon _{\mu
}^{I}\upsilon _{J}^{\mu }$ &  \\ 
& $+4\mathrm{\xi }\lambda \left[ \left( \overline{\mathrm{Q}}_{C}^{lJ}%
\mathrm{Q}_{kJ}^{D}\right) -\left( \overline{\mathrm{Q}}_{iC}^{J}\mathrm{Q}%
_{J}^{iD}\right) \delta _{k}^{l}\right] \partial _{\mu }\mathrm{f}%
^{kC}\partial ^{\mu }\overline{\mathrm{f}}_{lD}$ &  \\ 
& $-2\mathrm{\xi }\lambda \left[ \left( \mathrm{Q}^{lCJ}\mathrm{Q}%
_{J}^{kD}\right) -\left( \mathrm{Q}_{i}^{CJ}\mathrm{Q}_{J}^{iD}\right)
\varepsilon ^{kl}\right] \partial _{\mu }\overline{\mathrm{f}}_{kC}\partial
^{\mu }\overline{\mathrm{f}}_{lD}$ &  \\ 
& $-2\mathrm{\xi }\lambda \left[ \left( \overline{\mathrm{Q}}_{kD}^{J}%
\overline{\mathrm{Q}}_{lCJ}\right) -\left( \overline{\mathrm{Q}}_{iD}^{J}%
\overline{\mathrm{Q}}_{CJ}^{i}\right) \varepsilon _{kl}\right] \partial
_{\mu }\mathrm{f}^{lD}\partial ^{\mu }\mathrm{f}^{kC}$ & .%
\end{tabular}%
\end{equation}%
Using eqs(\ref{t1}-\ref{t3}) and the identity%
\begin{equation}
\begin{tabular}{llll}
$\mathcal{F}_{I}^{J}\upsilon _{\mu }^{I}\upsilon _{J}^{\mu }$ & $=$ & $%
\mathcal{F}_{I}^{J}\mathrm{Q}_{J}^{kC}\mathrm{Q}^{lDI}\partial ^{\mu }%
\overline{\mathrm{f}}_{kC}\partial _{\mu }\overline{\mathrm{f}}_{lD}-%
\mathcal{F}_{I}^{J}\overline{\mathrm{Q}}_{kCJ}\mathrm{Q}^{lDI}\partial ^{\mu
}\mathrm{f}^{kC}\partial _{\mu }\overline{\mathrm{f}}_{lD}$ &  \\ 
&  & $-\mathcal{F}_{I}^{J}\mathrm{Q}_{J}^{lD}\overline{\mathrm{Q}}%
_{kC}^{I}\partial _{\mu }\mathrm{f}^{kC}\partial ^{\mu }\overline{\mathrm{f}}%
_{lD}+\mathcal{F}_{I}^{J}\overline{\mathrm{Q}}_{lDJ}\overline{\mathrm{Q}}%
_{kC}^{I}\partial _{\mu }\mathrm{f}^{kC}\partial ^{\mu }\mathrm{f}^{lD}$ & ,%
\end{tabular}%
\end{equation}%
we can put $L_{1n}$ like 
\begin{equation}
\begin{tabular}{llll}
$L_{n1}$ & $=$ & $-2\partial _{\mu }\mathrm{f}^{iA}\partial ^{\mu }\overline{%
\mathrm{f}}_{iA}+\frac{\lambda }{2}\mathcal{N}_{kC}^{lD}\partial _{\mu }%
\mathrm{f}^{kC}\partial ^{\mu }\overline{\mathrm{f}}_{lD}$ &  \\ 
&  & $-\frac{\lambda }{2}\mathcal{U}^{kClD}\partial _{\mu }\overline{\mathrm{%
f}}_{kC}\partial ^{\mu }\overline{\mathrm{f}}_{lD}-\frac{\lambda }{2}%
\overline{\mathcal{U}}_{kClD}\partial _{\mu }\mathrm{f}^{lD}\partial ^{\mu }%
\mathrm{f}^{kC}$ & ,%
\end{tabular}
\label{la1}
\end{equation}%
where we have set%
\begin{equation}
\begin{tabular}{llll}
$\mathcal{N}_{kC}^{lD}$ & $=$ & $\mathcal{F}_{I}^{J}\mathrm{Q}_{J}^{lD}%
\overline{\mathrm{Q}}_{kC}^{I}+\mathcal{F}_{I}^{J}\overline{\mathrm{Q}}_{kCJ}%
\mathrm{Q}^{lDI}$ &  \\ 
&  & $+8\mathrm{\xi }\left( \overline{\mathrm{Q}}_{C}^{lJ}\mathrm{Q}%
_{kJ}^{D}\right) -8\mathrm{\xi }\left( \overline{\mathrm{Q}}_{iC}^{J}\mathrm{%
Q}_{J}^{iD}\right) \delta _{k}^{l}$ & , \\ 
$\overline{\mathcal{U}}_{kC,lD}$ & $=$ & $\mathcal{F}_{I}^{J}\overline{%
\mathrm{Q}}_{lDJ}\overline{\mathrm{Q}}_{kC}^{I}$ &  \\ 
&  & $+4\mathrm{\xi }\left( \overline{\mathrm{Q}}_{kD}^{J}\overline{\mathrm{Q%
}}_{lCJ}\right) -4\mathrm{\xi }\left( \overline{\mathrm{Q}}_{iD}^{J}%
\overline{\mathrm{Q}}_{CJ}^{i}\right) \varepsilon _{kl}$ & , \\ 
$\mathcal{U}^{kC,lD}$ & $=$ & $\mathcal{F}_{I}^{J}\mathrm{Q}_{J}^{kC}\mathrm{%
Q}^{lDI}$ &  \\ 
&  & $+4\mathrm{\xi }\left( \mathrm{Q}^{lCJ}\mathrm{Q}_{J}^{kD}\right) -4%
\mathrm{\xi }\left( \mathrm{Q}_{i}^{CJ}\mathrm{Q}_{J}^{iD}\right)
\varepsilon ^{kl}$ & .%
\end{tabular}
\label{al1}
\end{equation}%
In the particular case where $n=1$, these quantities reduce to%
\begin{equation}
\begin{tabular}{llll}
$\mathcal{N}_{k}^{l}$ & $=$ & $\frac{2\mathrm{f}^{l}\overline{\mathrm{f}}_{k}%
}{1+\lambda \overline{\mathrm{f}}\mathrm{f}}+8\mathrm{\xi }\left( \overline{%
\mathrm{f}}^{l}\mathrm{f}_{k}-\delta _{k}^{l}\left( \overline{\mathrm{f}}_{i}%
\mathrm{f}^{i}\right) \right) $ & , \\ 
$\overline{\mathcal{U}}_{kl}$ & $=$ & $\frac{\overline{\mathrm{f}}_{l}%
\overline{\mathrm{f}}_{k}}{1+\lambda \overline{\mathrm{f}}\mathrm{f}}+4%
\mathrm{\xi }\left( \overline{\mathrm{f}}_{k}\overline{\mathrm{f}}%
_{l}\right) $ & , \\ 
$\mathcal{U}^{kl}$ & $=$ & $\frac{\mathrm{f}^{k}\mathrm{f}^{l}}{1+\lambda 
\overline{\mathrm{f}}\mathrm{f}}+4\mathrm{\xi }\left( \mathrm{f}^{l}\mathrm{f%
}^{k}\right) $ & ,%
\end{tabular}
\label{nu}
\end{equation}%
where $\overline{\mathrm{f}}\mathrm{f}$ stands for $\overline{\mathrm{f}}_{i}%
\mathrm{f}^{i}$ and $\mathrm{\xi =}\frac{1}{8}$.

\textbf{(2) Computing} $L_{n2}$ eq(\ref{l2})\newline
Using eqs(\ref{ca}-\ref{cb}), we have 
\begin{equation}
\begin{tabular}{llll}
$\mathrm{\tilde{Q}}_{AI}^{+}\mathrm{C}_{\mu }^{-A}$ & $=$ & $2\mathrm{\tilde{%
Q}}_{AI}^{+}\partial _{\mu }\mathrm{f}^{-A}+\lambda \mathcal{F}%
_{I}^{J}\upsilon _{\mu }^{I}\mathrm{\tilde{Q}}_{AI}^{+}\mathrm{Q}%
_{J}^{-A}+\lambda \mathrm{\tilde{Q}}_{AI}^{+}\mathrm{Q}_{J}^{+a}\mathrm{%
\tilde{Q}}_{B}^{-J}\left( \partial _{\mu }\mathrm{f}^{-B}\right) $ &  \\ 
&  & $+\lambda \mathrm{\tilde{Q}}_{AI}^{+}\mathrm{Q}_{J}^{+A}\mathrm{Q}%
^{-BJ}\left( \partial _{\mu }\mathrm{\tilde{f}}_{B}^{-}\right) $ & ,%
\end{tabular}%
\end{equation}%
and%
\begin{equation}
\begin{tabular}{llll}
$\mathrm{\tilde{C}}_{\mu A}^{-}\mathrm{Q}_{I}^{+A}$ & $=$ & $2\partial _{\mu
}\mathrm{\tilde{f}}_{A}^{-}\mathrm{Q}_{I}^{+A}-\lambda \mathcal{F}%
_{I}^{J}\upsilon _{\mu }^{I}\mathrm{\tilde{Q}}_{AJ}^{-}\mathrm{Q}%
_{I}^{+A}-\lambda \mathrm{Q}_{I}^{+A}\mathrm{\tilde{Q}}_{AJ}^{+}\mathrm{Q}%
^{-bJ}\left( \partial _{\mu }\mathrm{\tilde{f}}_{B}^{-}\right) $ &  \\ 
&  & $-\lambda \mathrm{Q}_{I}^{+A}\mathrm{\tilde{Q}}_{AJ}^{+}\mathrm{\tilde{Q%
}}_{B}^{-J}\left( \partial _{\mu }\mathrm{f}^{-B}\right) $ & .%
\end{tabular}%
\end{equation}%
Putting these relations back into $A_{\mu I}=\left( \mathrm{\tilde{Q}}%
_{AI}^{+}\mathrm{C}_{\mu }^{-A}+\mathrm{\tilde{C}}_{\mu A}^{-}\mathrm{Q}%
_{I}^{+A}\right) $, we obtain 
\begin{equation}
\begin{tabular}{llll}
$A_{\mu I}$ & $=$ & $2\left( \mathrm{\tilde{Q}}_{AI}^{+}\partial _{\mu }%
\mathrm{f}^{-A}+\mathrm{Q}_{I}^{+A}\partial _{\mu }\widetilde{\mathrm{f}}%
_{A}^{-}\right) $ &  \\ 
&  & $+\lambda \mathcal{F}_{I}^{J}\upsilon _{\mu }^{I}\left( \mathrm{\tilde{Q%
}}_{AI}^{+}\mathrm{Q}_{J}^{-A}-\mathrm{\tilde{Q}}_{AJ}^{-}\mathrm{Q}%
_{I}^{+A}\right) $ &  \\ 
&  & $+\lambda \left( \mathrm{\tilde{Q}}_{AI}^{+}\mathrm{Q}_{J}^{+A}-\mathrm{%
\tilde{Q}}_{AJ}^{+}\mathrm{Q}_{I}^{+A}\right) \mathrm{\tilde{Q}}%
_{B}^{-J}\left( \partial _{\mu }\mathrm{f}^{-B}\right) $ &  \\ 
&  & $+\lambda \left( \mathrm{\tilde{Q}}_{AI}^{+}\mathrm{Q}_{J}^{+A}-\mathrm{%
\tilde{Q}}_{AJ}^{+}\mathrm{Q}_{I}^{+A}\right) \mathrm{Q}^{-AJ}\left(
\partial _{\mu }\widetilde{\mathrm{f}}_{A}^{-}\right) $ & .%
\end{tabular}%
\end{equation}%
Moreover, using the following identities%
\begin{equation}
\begin{tabular}{llll}
$\mathrm{\tilde{Q}}_{AJ}^{-}\mathrm{Q}_{I}^{+A}$ & $=$ & $\mathrm{\tilde{Q}}%
_{AI}^{-}\mathrm{Q}_{J}^{+A}$ & , \\ 
$\mathrm{\tilde{Q}}_{AJ}^{+}\mathrm{Q}_{I}^{+A}$ & $=$ & $\mathrm{\tilde{Q}}%
_{AI}^{+}\mathrm{Q}_{J}^{+A}$ & , \\ 
$\mathrm{\tilde{Q}}_{AJ}^{+}\mathrm{Q}_{I}^{+A}$ & $=$ & $\mathrm{\tilde{Q}}%
_{AI}^{+}\mathrm{Q}_{J}^{+A}$ & ,%
\end{tabular}%
\end{equation}%
the above expression of $A_{\mu I}$ gets reduced to%
\begin{equation}
\begin{tabular}{llll}
$A_{\mu I}$ & $=$ & $2\left( \mathrm{\tilde{Q}}_{AI}^{+}\partial _{\mu }%
\mathrm{f}^{-A}+\mathrm{Q}_{I}^{+A}\partial _{\mu }\widetilde{\mathrm{f}}%
_{A}^{-}\right) -\lambda \mathcal{F}_{I}^{J}\upsilon _{\mu }^{I}\left( 
\overline{\mathrm{Q}}_{iAJ}\mathrm{Q}_{I}^{iA}\right) $ & .%
\end{tabular}%
\end{equation}%
Now using eq(\ref{l2}), and substituting $A_{\mu I}$, we have%
\begin{equation}
\begin{tabular}{llll}
$L_{n2}$ & $=$ & $-\lambda \int_{S^{2}}du\left[ \sum_{I=1}^{n}2\left( 
\mathrm{\tilde{Q}}_{AI}^{+}\partial _{\mu }\mathrm{f}^{-A}+\mathrm{Q}%
_{I}^{+A}\partial _{\mu }\widetilde{\mathrm{f}}_{A}^{-}\right) \partial
^{\mu }w^{I}\right] $ & .%
\end{tabular}%
\end{equation}%
The term $\lambda \mathcal{F}_{I}^{J}\upsilon _{\mu }^{I}\left( \overline{%
\mathrm{Q}}_{iAJ}\mathrm{Q}_{I}^{iA}\right) \partial ^{\mu }w^{I}$ drops out
because of the property $\int_{S^{2}}du$ $u_{(i}^{+}u_{j)}^{-}=0$. By
integration by parts, we can also put $L_{n2}$ in the form%
\begin{equation}
\begin{tabular}{llll}
$L_{n2}$ & $=$ & $+\lambda \int_{S^{2}}du\left[ \sum_{I=1}^{n}\left( \mathrm{%
\tilde{Q}}_{AI}^{+}\partial _{\mu }\mathrm{f}^{+A}+\mathrm{Q}%
_{I}^{+A}\partial _{\mu }\widetilde{\mathrm{f}}_{A}^{+}\right) \partial
^{\mu }w^{--I}\right] $ & .%
\end{tabular}%
\end{equation}%
Substituting $w^{--I}=\mathrm{\tilde{Q}}^{-I}\mathrm{f}^{-}=\mathrm{\tilde{f}%
}^{-}\mathrm{Q}^{-I}$, we get the following 
\begin{equation}
\begin{tabular}{lll}
$L_{n2}=$ & $\frac{-\lambda }{2}\left( \widehat{\mathcal{U}}^{kC,lD}\partial
_{\mu }\overline{\mathrm{f}}_{kC}\partial ^{\mu }\overline{\mathrm{f}}_{lD}+%
\widehat{\overline{\mathcal{U}}}_{kC,lD}\partial _{\mu }\mathrm{f}%
^{lD}\partial ^{\mu }\mathrm{f}^{kC}\right) $ &  \\ 
& $+\frac{\lambda }{2}\mathcal{\hat{N}}_{kC}^{lD}\partial _{\mu }\mathrm{f}%
^{kC}\partial ^{\mu }\overline{\mathrm{f}}_{lD}$ & ,%
\end{tabular}
\label{la2}
\end{equation}%
where we have set%
\begin{equation}
\begin{tabular}{llll}
$\mathcal{\hat{N}}_{kC}^{lD}$ & $=$ & $8\lambda \mathrm{\xi }\left( 
\overline{\mathrm{Q}}_{CI}^{l}\mathrm{Q}_{k}^{DI}-\overline{\mathrm{Q}}_{iCI}%
\mathrm{Q}^{iDI}\delta _{k}^{l}\right) $ & , \\ 
$\widehat{\overline{\mathcal{U}}}_{kC,lD}$ & $=$ & $4\lambda \mathrm{\xi }%
\left( \overline{\mathrm{Q}}_{lCI}\overline{\mathrm{Q}}_{kD}^{I}-\overline{%
\mathrm{Q}}_{CI}^{i}\overline{\mathrm{Q}}_{iD}^{I}\varepsilon _{kl}\right) $
& , \\ 
$\widehat{\mathcal{U}}^{kC,lD}$ & $=$ & $4\lambda \mathrm{\xi }\left( 
\mathrm{Q}_{I}^{lC}\mathrm{Q}^{kDI}-\mathrm{Q}_{iI}^{C}\mathrm{Q}%
^{iDI}\varepsilon ^{kl}\right) $ & .%
\end{tabular}
\label{al2}
\end{equation}%
In the case $n=1$, these tensors reduce to%
\begin{equation}
\begin{tabular}{llll}
$\mathcal{\hat{N}}_{k}^{l}$ & $=$ & $8\lambda \mathrm{\xi }\left( \overline{%
\mathrm{f}}^{l}\mathrm{f}_{k}-\overline{\mathrm{f}}\mathrm{f}\delta
_{k}^{l}\right) $ & , \\ 
$\widehat{\overline{\mathcal{U}}}_{kl}$ & $=$ & $4\lambda \mathrm{\xi }%
\overline{\mathrm{f}}_{l}\overline{\mathrm{f}}_{k}$ & , \\ 
$\widehat{\mathcal{U}}^{kl}$ & $=$ & $4\lambda \mathrm{\xi f}^{l}\mathrm{f}%
^{k}$ & .%
\end{tabular}%
\end{equation}

\emph{the U}$^{n}\left( 1\right) $\emph{\ hyperKahler metric}\newline
Adding eqs(\ref{la1}-\ref{al1}) and eqs(\ref{la2}-\ref{al2}), we get the
total Lagrangian density%
\begin{equation}
\begin{tabular}{llll}
$L_{n}$ & $=$ & $+g^{kc,ld}\partial _{\mu }\overline{\mathrm{f}}%
_{kc}\partial ^{\mu }\overline{\mathrm{f}}_{ld}+\overline{g}_{kc,ld}\partial
_{\mu }\mathrm{f}^{ld}\partial ^{\mu }\mathrm{f}^{kc}$ &  \\ 
&  & $+2h_{kc}^{ld}\partial _{\mu }\mathrm{f}^{kc}\partial ^{\mu }\overline{%
\mathrm{f}}_{ld}$ & ,%
\end{tabular}%
\end{equation}%
with $g^{kc,ld}$, $\overline{g}_{kc,ld}$ and $h_{kc}^{ld}$ as in eqs(\ref{a}-%
\ref{c}).

\end{document}